\documentclass[a4paper, 12pt]{article}

\usepackage{amsmath, enumerate, amsthm, amssymb, mathrsfs, graphicx, multirow, listings, lipsum, color, hyperref, cite, natbib, adjustbox, algpseudocode, ulem, multicol, lscape, parskip, tablefootnote, comment}
\usepackage[table]{xcolor}

\newcommand{\ind}{\perp\!\!\!\!\perp} 
\newcommand{\tr}{\textcolor{black}}

\begin{document}

\title{\textbf{Predictive Performance Test based on the Exhaustive Nested Cross-Validation for High-dimensional data}}
\author{
Iris Ivy Gauran$^{1}$, 
Hernando Ombao$^{1}$, 
Zhaoxia Yu$^{2}$}
\vspace{-3em}
\date{\small 
$^{1}$ Statistics Program, King Abdullah University of Science and Technology (KAUST)\\ 
$^{2}$ Department of Statistics, University of California, Irvine}
\maketitle

\begin{abstract}%
It is crucial to assess the predictive performance of a model to establish its practicality and relevance in real-world scenarios, particularly for high-dimensional data analysis.  Among data splitting or resampling methods, cross-validation (CV) is extensively used for several tasks such as estimating the prediction error, tuning the regularization parameter, and selecting the most suitable predictive model among competing alternatives.  The $K$-fold cross-validation is a popular CV method but its limitation is that the risk estimates are highly dependent on the partitioning of the data (for training and testing). Here, the issues regarding the reproducibility of the $K$-fold CV estimator are demonstrated in hypothesis testing wherein different partitions lead to notably disparate conclusions.  This study presents a novel predictive performance test and valid confidence intervals based on \underline{exhaustive} nested cross-validation for determining the difference in prediction error between two model-fitting algorithms.  A naive implementation of the exhaustive nested cross-validation is computationally costly. Here, we address concerns regarding computational complexity by devising a computationally tractable closed-form expression for the proposed cross-validation estimator. Our study also investigates strategies aimed at enhancing statistical power within high-dimensional scenarios while controlling the Type I error rate.  \tr{Through comprehensive numerical experiments, we demonstrate that Ridge-based methods using bias to measure uncertainty of CV estimates and adaptive hyperparameter selection provide the most reliable approach for high-dimensional predictive performance testing.}  To illustrate the practical utility of our method, we apply it to an RNA sequencing study and demonstrate its effectiveness in the context of biological data analysis. 
\end{abstract}
\vspace{0.5em}

% -------------------------------------------------
\section{Introduction}
\label{sec:Introduction}
% -------------------------------------------------

In high-throughput omics studies, including genomics, metabolomics, transcriptomics, proteomics, and epigenomics, scientists \tr{routinely} generate massive datasets \tr{containing millions of features}. \tr{These datasets may include millions of single nucleotide polymorphisms (SNPs) in genomics studies, or differentially methylated probes in epigenomics research.} \tr{Similarly, neuroimaging studies produce data with hundreds of thousands of voxels in functional magnetic resonance images and dozens to hundreds of electroencephalogram channels.  The central inferential question across these domains --- whether changes in disease-related biological metrics can be predicted from genetic information or potential biomarkers --- underpins feature selection, biomarker discovery, and therapeutic target identification.  To address this challenge, our primary objective is to investigate how models fitted on training data generalize to unseen test data.}

% -------------------------------------------------
\subsection{Limitations of Current Cross-Validation Methods}
\label{sec:LimitationsCV}
% -------------------------------------------------

Cross-validation (CV) has emerged as the standard technique for evaluating predictive performance \citep{stone1974cross, geisser1975predictive, stone1977asymptotics, stone1978cross}, with $K$-fold CV gaining widespread adoption due to its conceptual simplicity. While CV has been widely recognized as a reliable tool for estimating generalization error \citep{nadeau1999inference, nadeau2003inference}, recent research reveals critical issues that undermine its reliability for statistical inference. Contrary to common assumption, CV estimates the average prediction error of models fit on other training sets from the same population, not the prediction error for the model fitted using training data \citep{bates2023cross}. Furthermore, cross-validated LASSO exhibits marked instability across different test-train partitions \citep{lund2013instability}, and CV point estimates show high variability, especially for small samples \citep{varma2006bias, krstajic2014cross}. This variability creates substantial uncertainty about true predictive performance and makes it difficult to draw reliable conclusions about whether specific features significantly improve prediction.

\tr{Moreover, the variability in cross-validation point estimates necessitates proper uncertainty quantification through confidence intervals, yet constructing valid intervals presents significant methodological gaps. \citet{bengio2003no} established} that a universal unbiased estimator of the standard error of $K$-fold CV point estimates does not exist, \tr{prompting decades of research into alternative approaches. This theoretical limitation pushed researchers to impose} additional assumptions or modifications for standard error estimation \citep{jiang2008calculating}. \tr{Recent advances have explored asymptotic properties of cross-validation \citep{dudoit2005asymptotics, bayle2020cross} and developed various confidence intervals for generalization errors \citep{nadeau1999inference, bayle2020cross, jiang2008calculating, bates2023cross, rajanala2022confidence}. However, substantial challenges persist in high-dimensional settings where traditional asymptotic assumptions may not hold and finite sample guarantees are essential for reliable inference.}

\tr{Besides theoretical limitations, cross-validation suffers from a critical crisis in reproducibility that becomes particularly acute in high-dimensional contexts. The core issue lies in dependence on random partitioning schemes, which introduces substantial variability in performance estimates. Several methods have been proposed to address this problem, including \citet{lei2020cross} through bootstrap-based confidence sets that quantify uncertainty in model comparison, \citet{cui2018test} through improved variance estimation via refitted CV, \citet{yu2014modified} through modified CV criteria for penalized high-dimensional regression models, and \citet{simon2011using} through practical guidance for evaluating predictive accuracy in high-dimensional genomic studies. While these methods enhance inference from cross-validation, there is no existing approach that completely eliminates the partition dependency problem, where different random seeds or partitioning schemes produce inherently unreproducible testing results.}

\tr{To address these methodological limitations, this study develops a novel framework for high-dimensional predictive performance testing using exhaustive nested cross-validation to eliminate partition dependency while maintaining computational tractability.  To motivate our proposed framework, we first demonstrate the severity of these theoretical concerns, and provide empirical evidence for the partition dependency problem.}

% -------------------------------------------------
\subsection{Empirical Evidence of the Reproducibility Crisis}
\label{sec:ReproducibilityCrisis}
% -------------------------------------------------

\tr{In this section, we demonstrate the reproducibility crisis inherent in high-dimensional hypothesis testing based on $K$-fold cross-validation estimators using} the liver toxicity dataset by \citet{bushel2007simultaneous} \tr{as our} motivating example. This high-dimensional \tr{dataset} contains gene expression measurements of 3116 genes for 64 rats exposed to \tr{varying} doses of acetaminophen in a controlled experiment. \tr{This dataset serves as an ideal case study because it has been previously utilized to investigate the problem of overfitting in high-dimensional settings \citep{kobak2020optimal, gauran2022ridge}.}

\tr{The fundamental flaw with $K$-fold cross-validation is due to its dependence on a {\it single}  random partitioning of the data. Each sampling unit is assigned to exactly one fold, and the choice of random seed critically affects reproducibility despite being typically arbitrary and unreported. To quantify this instability, we conducted an experiment using 500 different random seeds to predict Alkaline Phosphatase (ALP) from liver mRNA expression data. The seeds are organized into 20 groups of 25 each and we computed test statistics using both $K$ = 5 and $K$ = 10 fold CV estimators. The description of the test and the rigorous calculation of these test statistics are presented in Appendix \ref{sec:KFoldCVTest}.}

\begin{center}
    \begin{figure}[ht]
    \includegraphics[width=\linewidth]{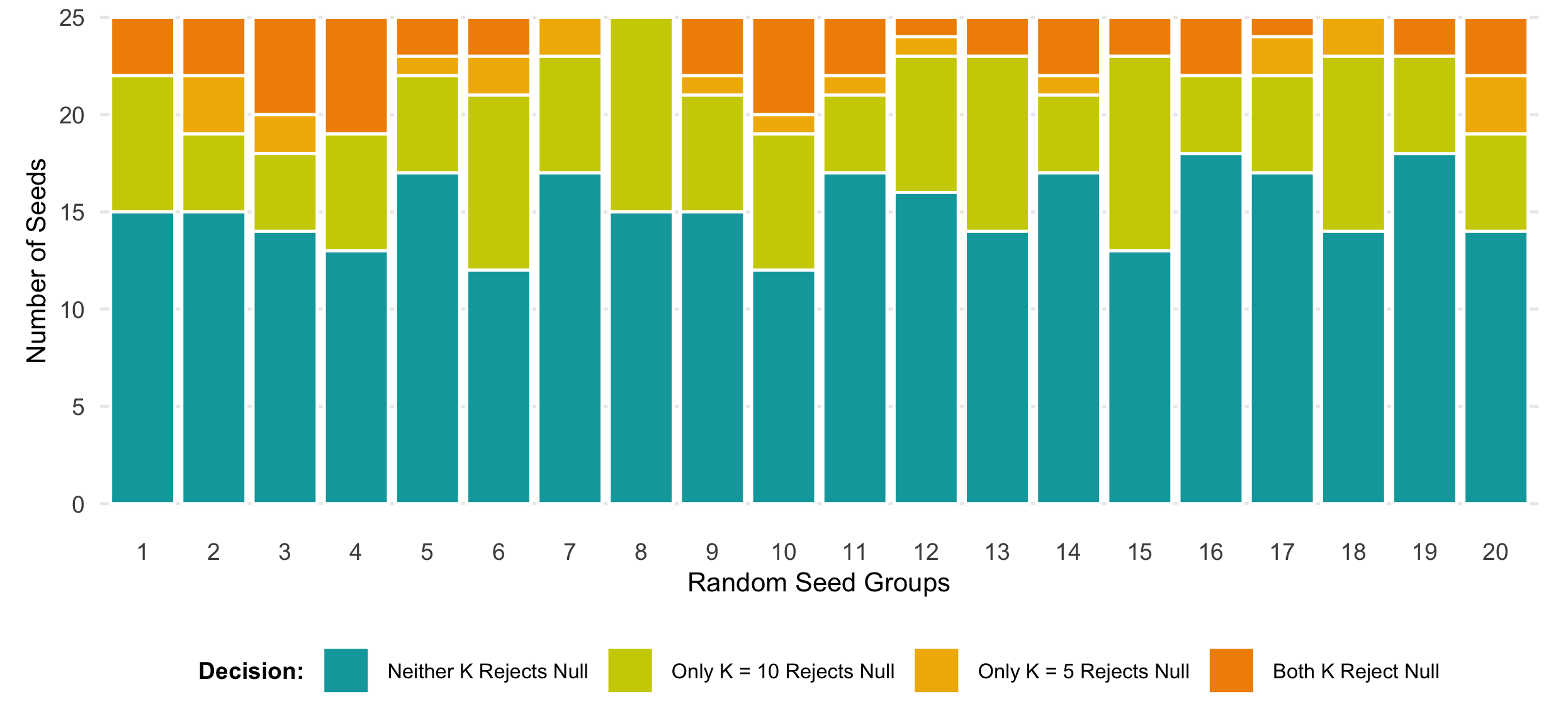}
    \caption{\tr{Illustration of the problem of reproducibility in hypothesis testing based on $K$-fold CV demonstrated using the response variable Alkaline Phosphatase (ALP). Results from 500 different random seeds (8892 to 9391) are grouped into 20 consecutive seed groups of 25 seeds each. The stacked bars show the distribution of statistical decisions across seeds within each group.}}
    \label{fig:RationaleLiverToxicity}
\end{figure}
\end{center}

\tr{Figure \ref{fig:RationaleLiverToxicity} illustrates this reproducibility crisis. The stacked bars show the distribution of statistical decisions across the 25 seeds within each group, revealing four possible outcomes: both $K = 5$ and $K = 10$ reject the null hypothesis (orange), only $K = 5$ rejects the null hypothesis (yellow), only $K = 10$ rejects the null hypothesis (green), or neither $K = 5$ and $K = 10$ rejects the null hypothesis (teal).  The results reveal substantial variation both across and within seed groups. Different seed groups produce dramatically different patterns of statistical decisions, with some groups showing rejections only for $K = 5$, others only for $K = 10$, and still others showing mixed or no rejections. Even within individual seed groups, considerable variation persists, demonstrating the problem of instability, where the results fluctuate even under mild perturbation of the seeds. The comparison between $K = 5$ and $K = 10$ reveals additional inconsistencies, where identical data with different $K$ values leads to opposite statistical conclusions.}

\tr{This problem of reproducibility   extends beyond ALP to other clinical measurements in the dataset, demonstrating that the problem represents a systemic limitation of the $K$-fold CV approach rather than a variable-specific artifact (see Appendix \ref{sec:KFoldCVTest} for results on other clinical measurements). The nature of this instability suggests that any scientific conclusion drawn from $K$-fold CV hypothesis testing carries substantial risk of irreproducibility.}

\tr{Practitioners typically use one value of $K$ with a single random seed---often without even reporting their choice. But this investigation reveals a troubling instability: a researcher conducting the same analysis with a different, equally valid seed could reach opposite conclusions about statistical significance. This variation across seeds exposes a serious problem with current $K$-fold CV approaches. Results lack consistency, threatening the reproducibility that's essential for advancing scientific knowledge in high-dimensional settings. These findings suggest the need for a different approach to cross-validation.}

% -------------------------------------------------
\subsection{Our Approach and Contributions}
\label{sec:Contributions}
% -------------------------------------------------

\tr{In this study, we develop a novel framework for high-dimensional predictive performance testing. Our key contributions are the following. First, we develop a computationally feasible exhaustive cross-validation method that considers all possible data divisions. While exhaustive CV methods can provide complete enumeration of validation outcomes \citep{arlot2010survey}, they are often computationally impractical. Our approach makes exhaustive methods tractable while eliminating the partition dependency and reproducibility issues demonstrated in Figure \ref{fig:RationaleLiverToxicity}.}

\tr{Second, our proposed method  incorporates a nested structure within the exhaustive framework. While nested cross-validation has demonstrated excellent performance, particularly in datasets with limited samples \citep{aliferis2006challenges,tsamardinos2015performance}, existing implementations remain highly sensitive to the partition.  We developed an exhaustive approach that eliminates this limitation while properly handling both model selection and assessment.}

\tr{Third, the scalability problem of permutation tests is addressed by characterizing the null distribution of our test statistic analytically, avoiding the high computational cost of generating samples from the null distribution.}

\tr{Finally, we derived computationally feasible closed-form expressions for the exhaustive nested cross-validation estimator, building on scalable approaches for CV \citep{rad2020scalable, wang2018approximate}. This mathematical development addresses the computational burden of naive approaches while providing valid confidence intervals for prediction error differences between models with and without hypothesized features.}

The remainder of this paper is organized as follows.  Section \tr{\ref{sec:PredictivePerformanceInference} establishes a hypothesis testing framework that focuses on algorithm improvement through prediction error reduction.}  We present several main results in Section \ref{sec:MainResults} and develop the proposed method in \ref{sec:NestedExhaustiveCVTest}. Numerical studies that investigate the performance of the proposed method are discussed in Section \ref{sec:NumericalExperiments}, and the application of the proposed method to an RNA sequencing data set are presented in Section \ref{sec:RealDataAnalysis}.  Novel findings were highlighted in this section including the significant improvement in the prediction of proteins that are highly implicated biomarkers in the progression of Alzheimer's disease.  Finally, conclusions and open problems are discussed in Section \ref{sec:Conclusions}.

% -------------------------------------------------
\section{Predictive Performance Inference}
\label{sec:PredictivePerformanceInference}
% -------------------------------------------------

\tr{Consider the supervised learning setting with dataset $\mathcal{D} = \{\mathbf{D}_n\}_{n = 1}^N$, where each observation $\mathbf{D}_n = (\mathbf{X}_n, Y_n) \in \mathbb{R}^P \times \mathbb{R}$ represents the $n$th sampling unit for $n \in [N] = \{1, 2, \ldots, N\}$. Each response variable $Y_n$ corresponds to a feature vector $\mathbf{X}_n^\top = (X_{n0}, X_{n1}, \ldots, X_{nP^*})$, with $X_{n0} = 1$ serving as the intercept term, and $P^* = P-1$ denoting the number of predictive features $\{X_{np}\}_{p=1}^{P^*}$.  The modeling framework employs the response vector $\mathbf{y} = (Y_1, Y_2, \ldots, Y_N)^\top \in \mathbb{R}^N$ and the design matrix $\mathbf{X} = (\mathbf{1}_N, \mathbf{X}^\star) \in \mathbb{R}^{N \times P}$ where $\mathbf{X}^\star = (\mathbf{X}_1^\top, \cdots, \mathbf{X}_n^\top, \cdots, \mathbf{X}_N^\top)^\top$ represents the matrix of covariates.}

\tr{Given this data structure, we consider models of the form $\mathbf{y} = f(\mathbf{X}, \boldsymbol{\beta}) + \boldsymbol{\varepsilon}$, parameterized by $\boldsymbol{\beta} \in \boldsymbol{\Theta}$ for a closed convex set $\boldsymbol{\Theta} \subseteq \mathbb{R}^{P}$. Here $f$ is a linear predictive function of the design matrix $\mathbf{X}$, and $\boldsymbol{\varepsilon}$ represents random noise with zero mean, variance $\sigma_\varepsilon^2$, and independence from $\mathbf{X}$. Under this specification, suppose $(\mathbf{X}_n, Y_n) \sim \mathcal{Q}(Y_n \mid f(\mathbf{X}_n, \boldsymbol{\beta})) \mathcal{P}(\mathbf{X}_n)$ for $n \in [N]$, where $\mathcal{Q}(\cdot \mid \cdot)$ represents the conditional distribution of the response given the predictive function $f(\mathbf{X}_n, \boldsymbol{\beta})$, and $\mathcal{P}(\cdot)$ denotes the marginal distribution of the features.}

% -------------------------------------------------
\subsection{Regularized Estimation and Prediction Error Quantification}
\label{sec:RegularizedEstimationHD}
% -------------------------------------------------

\tr{The fundamental question in model assessment becomes how well a fitted model generalizes to new data.  In high-dimensional settings, our main goal then is to evaluate whether a model-fitting method $\mathcal{M}$ can yield $\widehat{f}$ that accurately predict the target variable for newly observed data points $(\mathbf{X}_{\text{new}}, Y_{\text{new}})$ that are independent of the training dataset $\mathcal{D}$ and drawn from the same underlying distribution $\mathcal{Q}(Y_n \mid f(\mathbf{X}_n, \boldsymbol{\beta})) \mathcal{P}(\mathbf{X}_n)$.}

\tr{To formalize this evaluation, consider a penalized estimation method $\mathcal{M}$ that takes $N$ training observations as input and produces a prediction rule $\widehat{f}$.  When applied to the linear model $\mathbf{y} = \mathbf{X} \boldsymbol{\beta} + \boldsymbol{\varepsilon}, ~\boldsymbol{\varepsilon} \sim N(\boldsymbol{0}, \sigma^2_{\varepsilon} \mathbf{I}_N)$, $N < P$, we can define the general framework:}
\begin{eqnarray}
    \label{eqn:PR_ModelFitting}
    \mathcal{M} &:& \mathbb{R}^{N \times P} \times \mathbb{R}^N \rightarrow \mathbb{R}^{N} \hspace{2.75em}
    \mathcal{M}(\{\mathbf{D}_n\}_{n = 1}^N) 
    = \widehat{f}(\mathbf{X}, \widehat{\boldsymbol{\beta}}(\lambda)) = \mathbf{X}\widehat{\boldsymbol{\beta}}(\lambda) \\
    \label{eqn:PR_LossFunction}
    \mathcal{L} 
    &:& \mathbb{R}^{N} \times \mathbb{R}^{N} \rightarrow \mathbb{R}^{\dagger} \cup \{0\} \hspace{1.5em}
    \mathcal{L}(\mathbf{y}, ~\mathbf{X}\widehat{\boldsymbol{\beta}}(\lambda)) = \lVert\mathbf{y} - \mathbf{X}\widehat{\boldsymbol{\beta}}(\lambda)\rVert^2_2 
    = \mathbf{r}^\top(\lambda) \mathbf{r}(\lambda)
\end{eqnarray}
\tr{where $\widehat{\boldsymbol{\beta}}(\lambda)$ estimates the unknown parameter $\boldsymbol{\beta}$ by minimizing a composite objective function}
\begin{equation}
\label{eqn:PR_Estimator}
\widehat{\boldsymbol{\beta}}(\lambda) = \underset{\mathbf{b} \in \mathbb{R}^{P} }{\text{arg min}} \left\{ \sum\limits_{n = 1}^N \mathcal{L}(Y_n, f(\mathbf{X}_n, \mathbf{b})) + \lambda \mathcal{R}(\mathbf{b}) \right\}.
\end{equation}
\tr{In this formulation, we utilize a convex loss function $\mathcal{L}: \mathbb{R}^2 \rightarrow \mathbb{R}$ described in \eqref{eqn:PR_LossFunction} and \eqref{eqn:PR_Estimator}, while the regularizer in \eqref{eqn:PR_Estimator} denoted by $\mathcal{R}: \mathbb{R}^{P} \rightarrow \mathbb{R}^+$ encodes prior beliefs about desirable parameter properties. The tuning parameter $\lambda \in \boldsymbol{\Lambda}$ in \eqref{eqn:PR_Estimator} controls the relative importance of data fitting versus regularization, thereby determining model complexity and enabling theoretical guarantees on estimator performance \citep{wainwright2014structured}.}

\tr{Moreover, this framework captures the essential components of predictive performance evaluation: the method $\mathcal{M}$ in \eqref{eqn:PR_ModelFitting} transforms training data into predictions, while the loss function $\mathcal{L}$ in \eqref{eqn:PR_LossFunction} quantifies prediction accuracy by comparing these predictions to observed responses.  Some prediction error terminologies and formulations available in previous works are presented in Table \ref{tab:PredictionErrorReview}. As shown in the table, various terminologies exist for this prediction task, each reflecting different perspectives on how to formalize and measure predictive performance.}

% -------------------------------------------------
\subsection{Hypothesis Testing for Predictive Improvement}
\label{sec:HypothesisTestingPredictiveImprovement}
% -------------------------------------------------

\tr{Rather than testing traditional parameter significance $\boldsymbol{\beta} = \mathbf{0}$ in the linear model fixed-effects setting, we frame the problem in terms of predictive improvement. This paradigm shift addresses a more practical question: \textit{Do the hypothesized features actually improve our ability to predict the response}? To answer this, we compare two models: ($i$) $\mathbf{y} = \mathbf{1}_N\beta_0 + \boldsymbol{\varepsilon}$, an intercept-only model and ($ii$) $\mathbf{y} = \mathbf{X}\boldsymbol{\beta} + \boldsymbol{\varepsilon}$, the full model with features $\mathbf{X} = (\mathbf{1}_N, \mathbf{X}^\star)$. This corresponds to two model-fitting approaches where $\mathcal{M}_0$ produces $\widehat{f}_0$ as a function of $\mathbf{y}$ only, while $\mathcal{M}_1$ yields $\widehat{f}_1$ as a function of $\mathbf{X}$ and $\mathbf{y}$.}

\begin{landscape}
\begin{table}[h!]
    \centering
    \caption{Some Prediction Error Terminologies and Formulations available in Literature}
    \small{
    \begin{tabular}{lll}
    \hline\hline
    \textbf{Terminology} & \textbf{Formula for Prediction Error} & \textbf{References} \\
    \hline
    \vspace{0.15em} & & \\
    Generalization Error & $_n\mu := \mathbb{E}\left[\mathcal{L}(Y_{\text{new}}, \widehat{f}_{\mathcal{D}}(\mathbf{X}_{\text{new}}) \right]$ & \citet{nadeau1999inference, nadeau2003inference}\tablefootnote{\scriptsize$\widehat{f}_{\mathcal{D}}$ is the output of the learning algorithm when training the algorithm on $\mathcal{D}$}\\
    \vspace{0.15em} & & \\
    \hline
    \vspace{0.15em} & & \\
    Extra-sample Error/
    & \multirow{2}*{$\text{Err} := \mathbb{E}_{\mathbf{D}_{\text{new}}}\left[\mathcal{L}(Y_{\text{new}}, \widehat{f}_{\mathcal{D}}(\mathbf{X}_{\text{new}})) \mid \mathcal{D}, \widehat{f}_{\mathcal{D}}~\right]$} & \multirow{5}*{\citet{efron1997improvements}\tablefootnote{\scriptsize$\widehat{f}_{\mathcal{D}}$ is the value of the prediction rule on the basis of $\mathcal{D}$}}\\
    True Error Rate & & \\
    \vspace{0.15em} & & \\
    Expected True Error & $\mu := \mathbb{E}_{\mathcal{D}}\left[\text{Err}\right] = \mathbb{E}_{\mathcal{D}}\left[\mathbb{E}_{\mathbf{D}_{\text{new}}}\left(\mathcal{L}(Y_{\text{new}}, \widehat{f}_{\mathcal{D}}(\mathbf{X}_{\text{new}})) \mid \mathcal{D}, \widehat{f}_{\mathcal{D}}\right)\right]$ & \\
    \vspace{0.15em} & & \\
    \hline
    \vspace{0.15em} & & \\
    Prediction Error of $\widehat{f}$ & $\text{Err}(\widehat{f}_{\mathcal{D}}) = \mathbb{E}_{\mathbf{X}_{\text{new}}} \left \{\mathbb{E}_{Y_{\text{new}} \mid \mathbf{X}_{\text{new}}} \left[ \mathcal{L}(Y_{\text{new}}, \widehat{f}_{\mathcal{D}}(\mathbf{X}_{\text{new}})) \mid \mathbf{X}_{\text{new}}, \widehat{f}_{\mathcal{D}}\right]\right\}$ & \multirow{4}*{\citet{borra2010measuring}\tablefootnote{\scriptsize$\widehat{f}_{\mathcal{D}}$ is the estimate of the general unknown function of $\mathbf{X}$ using the training set $\mathcal{D}$ where $\mathbf{y} = f(\mathbf{X}) + \boldsymbol{\varepsilon}$}}\\
    \vspace{0.15em} & & \\
    Expected Prediction Error & $\overline{\text{Err}} 
    = \mathbb{E}_{\mathcal{D}} \left[\text{Err}(\widehat{f}_{\mathcal{D}})\right]$ & \\
    \vspace{0.15em} & & \\
    \hline
    \vspace{0.15em} & & \\
    \multirow{2}*{Out-of-sample Error} & \multirow{2}*{$\text{Err}_{\text{XY}} := \mathbb{E}_{\mathbf{D}_{\text{new}}}\left[\mathcal{L}(Y_{\text{new}}, \widehat{f}_{\mathbf{X}, \mathbf{y}}(\mathbf{X}_{\text{new}}, \widehat{\boldsymbol{\beta}})) \mid \mathbf{X}, \mathbf{y}\right]$} & \citet{bates2023cross,patil2021uniform}\tablefootnote{\scriptsize $\widehat{f}_{\mathbf{X}, \mathbf{y}}(\cdot, \widehat{\boldsymbol{\beta}})$ is the function that predicts $Y_{\text{new}}$ from $\mathbf{X}_{\text{new}}$ using the model with parameter $\boldsymbol{\beta}; \widehat{\boldsymbol{\beta}}$ is the fitted value of the parameter based on the training data $(\mathbf{X}, \mathbf{y})$}\\
    && \citet{rad2020scalable}\\
    \vspace{0.15em} & & \\
    \hline
    \vspace{0.15em} & & \\
    In-sample Error of $\widehat{f}$ & $\text{Err}^{(\text{in})}_{\mathbf{X}} = \displaystyle\frac{1}{N}\sum\limits_{n = 1}^N \mathbb{E}_{Y_n \mid \mathbf{X}_n = \mathbf{x}_n} \left[ \mathcal{L}(Y_n, \widehat{f}_{\mathcal{D}}(\mathbf{x}_n)) \mid \widehat{f}_{\mathcal{D}}, \mathbf{X}\right]$ & \citet{efron2004estimation}\tablefootnote{\scriptsize $\widehat{f}_{\mathcal{D}}$ is the estimate of the general unknown function of $\mathbf{X}$ using the training set $\mathcal{D}$ where $\mathbf{y} = f(\mathbf{X}) + \boldsymbol{\varepsilon}$; covariates are fixed at observed values $\mathbf{x}_n$}\\
    \vspace{0.15em} & & \\
    \hline\hline
    \end{tabular}}
    \label{tab:PredictionErrorReview}
\end{table}
\end{landscape}

\tr{For high-dimensional testing, we propose to use the expected in-sample prediction error difference as the measure to compare the predictive performance with vs. without the hypothesized features.  We test the null hypothesis $H_0: \overline{\text{Err}}^{(0)}_{\mathbf{X}} - \overline{\text{Err}}^{(1)}_{\mathbf{X}} \leq 0$ versus the alternative $H_1: \overline{\text{Err}}^{(0)}_{\mathbf{X}} - \overline{\text{Err}}^{(1)}_{\mathbf{X}} > 0$.  Equivalently, this can be expressed as the percent change in prediction error, denoted by $\Delta$,
\begin{equation}
\label{eqn:PercentChange_PE}
   \Delta = \displaystyle \frac{\overline{\text{Err}}^{(0)}_{\mathbf{X}} - \overline{\text{Err}}^{(1)}_{\mathbf{X}}}{\overline{\text{Err}}^{(0)}_{\mathbf{X}}} \times 100\%, ~~ \text{provided}~\overline{\text{Err}}^{(0)}_{\mathbf{X}} > 0,
\end{equation}
where large values of $\Delta$ in \eqref{eqn:PercentChange_PE} provide evidence against the null hypothesis.  This aligns with the interpretation that hypothesized features do not yield significant additional improvement in predictive performance if we fail to reject $H_0$. Conversely, rejecting $H_0$ suggests that the feature set $\mathbf{X}^\star$ produces meaningful improvement in predicting $\mathbf{y}$.}

\tr{This hypothesis testing framework offers several important advantages over traditional model comparison approaches. First, it directly addresses the question practitioners actually care about: rather than asking whether parameters are statistically different from zero, we ask whether features provide meaningful predictive value. This distinction is crucial in high-dimensional settings where statistical significance may not translate to practical utility. In contrast, the percent change metric $\widehat{\Delta}$ provides immediate practical interpretation, allowing practitioners to assess whether the added complexity of including features $\mathbf{X}^\star$ justifies the predictive gains.  Second, our approach establishes a universal baseline by comparing model-fitting algorithms against $\mathcal{M}_0$ rather than against each other --- if a method cannot outperform this simple baseline, further comparisons become futile.  This scales naturally with the modern machine learning landscape because by establishing a single baseline, methods are efficiently screened to identify those that  merit further investigation.}

\tr{To estimate the prediction errors $\overline{\text{Err}}^{(0)}_{\mathbf{X}}$ and $\overline{\text{Err}}^{(1)}_{\mathbf{X}}$ needed for our test statistic, we employ the Leave-$N_0$-out Cross-Validation (L$N_0$OCV) estimator. This approach allows us to compute prediction errors for both the intercept-only model and the full model while addressing the reproducibility concerns in standard $K$-fold cross-validation methods.  The formal definitions and estimation procedure are detailed in the next section.}

% -------------------------------------------------
\section[Main Results on Leave-N0-out Cross-validation Estimator]{Main Results on  Leave-\(N_0\)-out Cross-validation Estimator}
\label{sec:MainResults}
% -------------------------------------------------

\tr{In practice, truly independent new data $(\mathbf{X}_{\text{new}}, Y_{\text{new}})$ described in the previous section are not always available during model development. Therefore, we measure generalization performance by partitioning the available dataset.  Data splitting procedures partition $\mathcal{D}$ into disjoint subsets called folds. For a standard train-test split, we use $N_1 = N - N_0$ observations to train the model and $N_0$ held-out observations to evaluate predictive performance.}

\tr{Train-test partitions are inherently non-unique, creating a critical issue in cross-validation methodology. For $N$ observations with testing set size $N_0$, there exist $L = \binom{N}{N_0}$ distinct ways to select the testing observations. To formalize this enumeration, we define the testing index set for the $\ell$th partition, where $\ell \in [L] = \{1, 2, \ldots, L\}$, as $\mathcal{T}_{0\ell} = \{\ell(1), \ell(2), \ldots, \ell(N_0)\}$ where $\ell(h) < \ell(h + 1)$ for all $h \in \{1, \ldots, N_0 - 1\}$. The corresponding training index set is given by $\mathcal{T}_{1\ell} = [N] \setminus \mathcal{T}_{0\ell} = \{\ell(N_0+1), \ell(N_0+2), \ldots, \ell(N)\}$, where the ordering constraint $\ell(h) < \ell(h + 1)$ holds for all $h \in \{N_0+1, \ldots, N - 1\}$.  For every $n \in [N]$, the $n$th original observation $\mathbf{D}_n = (\mathbf{X}_n, Y_n)$ can be equivalently expressed as the $h$th sampling unit in the $\ell$th partition, denoted by $\mathbf{D}_{\ell(h)} = (\mathbf{X}_{\ell(h)}, Y_{\ell(h)})$.  This systematic indexing framework enables comprehensive enumeration of all possible train-test splits and facilitates rigorous analysis of the statistical properties of cross-validation estimators across different partitioning schemes.}

\tr{Utilizing this partitioning framework, each train-test split can be represented using matrices. For the $\ell$th partition and $g \in \{0, 1\}$, we define $\mathbf{y}_{\mathcal{T}_{g\ell}}$ as the $N_g \times 1$ vector of observed responses and $\mathbf{X}_{\mathcal{T}_{g\ell}}$ as the corresponding $N_g \times P$ covariate matrix.  Here, each row of $\mathbf{X}_{\mathcal{T}_{g\ell}}$ contains features for one sampling unit in the corresponding subset.  The training set consists of the design matrix $\mathbf{X}_{\mathcal{T}_{1\ell}} := (\mathbf{X}^\top_{\ell(N_0 + 1)}, \ldots, \mathbf{X}^\top_{\ell(N)})^\top$ and response vector $\mathbf{y}_{\mathcal{T}_{1\ell}} := (Y_{\ell(N_0 + 1)}, \ldots, Y_{\ell(N)})^\top$. Similarly, the testing set comprises the design matrix $\mathbf{X}_{\mathcal{T}_{0\ell}} := (\mathbf{X}^\top_{\ell(1)}, \ldots, \mathbf{X}^\top_{\ell(N_0)})^\top$ and response vector $\mathbf{y}_{\mathcal{T}_{0\ell}} := (Y_{\ell(1)}, \ldots, Y_{\ell(N_0)})^\top$. Both response vectors are assumed to be drawn from the same underlying distribution.}

\tr{As mentioned in Section \ref{sec:HypothesisTestingPredictiveImprovement}, let $\mathcal{M}_0$ and $\mathcal{M}_1$ denote the model-fitting methods for the intercept-only and full models, respectively. Using this matrix representation, $\mathcal{M}_1$ takes the training data $(\mathbf{X}_{\mathcal{T}_{1\ell}}, \mathbf{y}_{\mathcal{T}_{1\ell}})$ as input to obtain $\widehat{\boldsymbol{\beta}}_{\mathcal{T}_{1\ell}}$ and construct the predictive function $\widehat{f}_1$. The fitted model generates testing predictions $\widehat{\mathbf{y}}_{\mathcal{T}_{0\ell}} = \widehat{f}_1(\mathbf{X}_{\mathcal{T}_{0\ell}}, \widehat{\boldsymbol{\beta}}_{\mathcal{T}_{1\ell}})$. The model-fitting mapping and loss function are defined as
\begin{eqnarray}
\label{eqn:PR_ModelFitting_Test}
    \mathcal{M}_1(\mathbf{X}_{\mathcal{T}_{0\ell}}, \mathbf{X}_{\mathcal{T}_{1\ell}}, \mathbf{y}_{\mathcal{T}_{1\ell}}) = \widehat{f}_1(\mathbf{X}_{\mathcal{T}_{0\ell}}, \widehat{\boldsymbol{\beta}}_{\mathcal{T}_{1\ell}}) = \mathbf{X}^\top_{\mathcal{T}_{0\ell}} \widehat{\boldsymbol{\beta}}_{\mathcal{T}_{1\ell}}
    &:& \mathbb{R}^{N_0 \times P} \times \mathbb{R}^{N_1 \times P} \times \mathbb{R}^{N_1} \rightarrow \mathbb{R}^{N_0}  \\
    \mathcal{L}(\mathbf{y}_{\mathcal{T}_{0\ell}}, \widehat{\mathbf{y}}_{\mathcal{T}_{0\ell}}) = \lVert\mathbf{y}_{\mathcal{T}_{0\ell}} - \mathbf{X}^\top_{\mathcal{T}_{0\ell}} \widehat{\boldsymbol{\beta}}_{\mathcal{T}_{1\ell}}\rVert^2_2
    &:& \mathbb{R}^{N_0} \times \mathbb{R}^{N_0} \rightarrow \mathbb{R}^{\dagger} \cup \{0\}.
\label{eqn:PR_LossFunction_Test}\nonumber
\end{eqnarray}}
\tr{On the other hand, $\mathcal{M}_0$ and its associated $\mathcal{L}$ can be characterized as}
\begin{eqnarray}
    \label{eqn:Null_ModelFitting}
    \mathcal{M}_0(\mathbf{y}_{\mathcal{T}_{0\ell}}) 
    = \widehat{f}_0(\overline{Y}_{\mathcal{T}_{1\ell}}, \mathbf{1}_{N_0}) = \overline{Y}_{\mathcal{T}_{1\ell}}\mathbf{1}_{N_0} &:& \mathbb{R}^{N_1} \rightarrow \mathbb{R}^{N_0} 
    \\
    \label{eqn:Null_LossFunction}
    \mathcal{L}(\mathbf{y}_{\mathcal{T}_{0\ell}}, \widehat{\mathbf{y}}_{\mathcal{T}_{0\ell}}) = \lVert\mathbf{y}_{\mathcal{T}_{0\ell}} - \overline{Y}_{\mathcal{T}_{1\ell}}\mathbf{1}_{N_0}\rVert^2_2
    &:& \mathbb{R}^{N_0} \times \mathbb{R}^{N_0} \rightarrow \mathbb{R}^{\dagger} \cup \{0\}.\nonumber
\end{eqnarray}
\tr{Understanding how the fitted models $\mathcal{M}_0$ and $\mathcal{M}_1$ generalize to the testing data addresses the classical prediction problem and provides the foundation for the testing procedure.}

% -------------------------------------------------
\subsection[Computationally Efficient Leave-N0-out Cross-validation Estimator]{Computationally Efficient Leave-\(N_0\)-out Cross-validation Estimator}
\label{sec:ComputationallyEfficientLN0OCVEstimator}
% -------------------------------------------------

\tr{Having established our data-splitting approach, we now define the prediction errors that form the foundation of our test statistic. To assess the incremental value of $\mathbf{X}^\star$ for the prediction of $\mathbf{y}$, we define the baseline and full model expected in-sample prediction errors $\overline{\text{Err}}_{\mathbf{X}}^{(v)}$, $v \in \{0, 1\}$, as}
\begin{equation}
\label{eqn:InSampleError}
    \overline{\text{Err}}_{\mathbf{X}}^{(0)} = \displaystyle \frac{1}{L}\sum\limits_{\ell = 1}^L \mathbb{E}\left[\mathcal{L}(\mathbf{y}_{\mathcal{T}_{0\ell}}, \widehat{f}_0(\mathbf{y}_{\mathcal{T}_{1\ell}})) \right] ~\text{and}~ 
    \overline{\text{Err}}_{\mathbf{X}}^{(1)} = \displaystyle \frac{1}{L}\sum\limits_{\ell = 1}^L \mathbb{E}\left[\mathcal{L}(\mathbf{y}_{\mathcal{T}_{0\ell}}, \widehat{f}_1(\mathbf{X}, \mathbf{y}_{\mathcal{T}_{1\ell}})) \mid \mathbf{X}\right].
\end{equation} 
\tr{Since these theoretical quantities in \eqref{eqn:InSampleError} are unknown, we estimate them using the Leave-$N_0$-out Cross-Validation (L$N_0$OCV) estimator. We select this estimator because it resolves reproducibility concerns in standard $K$-fold CV methods by exhaustively evaluating all possible data partitions.  The L$N_0$OCV estimator for model-fitting methods $\mathcal{M}_v$, $v \in \{0, 1\}$, in \eqref{eqn:PR_ModelFitting_Test} and \eqref{eqn:Null_ModelFitting} is 
\begin{equation}
\label{eqn:PR_LN0OCV}
    \text{L}N_0\text{OCV}^{(v)} = 
    \begin{cases}
        \displaystyle \frac{1}{L N_0} \sum \limits_{\ell = 1}^{L} \sum \limits_{h = 1}^{N_0} \left(Y_{\ell(h)} - \mathbf{X}_{\ell(h)}^\top \widehat{\boldsymbol{\beta}}_{\mathcal{T}_{1\ell}}(\lambda)\right)^2, & v = 1\\
        \displaystyle \frac{1}{L N_0} \sum \limits_{\ell = 1}^{L} \sum \limits_{h = 1}^{N_0} \left(Y_{\ell(h)} - \overline{Y}_{\mathcal{T}_{1\ell}}\right)^2,& v = 0
    \end{cases}
\end{equation}
provided $N_0 > 0$. When $N_0 = 0$, the quantity in \eqref{eqn:PR_LN0OCV} reduces to the standard mean squared error using all available data.}

\tr{Direct implementation of L$N_0$OCV is computationally expensive since it requires computing $\widehat{\boldsymbol{\beta}}_{\mathcal{T}_{1\ell}}$ or $\overline{Y}_{\mathcal{T}_{1\ell}}$ across $L$ different training sets.  Under appropriate regularity conditions, however, we can derive tractable closed-form expressions. Ridge regression proves particularly suitable for this analysis due to its analytical properties \citep{tikhonov1943stability}. Combining a linear predictor with quadratic loss and $L_2$ regularization yields $\widehat{\boldsymbol{\beta}}_{\text{R}}(\lambda) = (\mathbf{X}^\top\mathbf{X} + \lambda \mathbf{I}_{P})^{\dagger} \mathbf{X}^\top \mathbf{y}$ where $\mathbf{A}^{\dagger}$ denotes the Moore-Penrose generalized inverse of a matrix $\mathbf{A}$. This formulation offers several key advantages, with two particularly important features. First, both the loss function $\mathcal{L}(\mathbf{y}, ~\mathbf{X}\mathbf{b}) = \left\|\mathbf{y} - \mathbf{X}\mathbf{b}\right\|^2_2$ and regularizer $\mathcal{R}(\mathbf{b}) = \left\|\mathbf{b}\right\|^2_2$ are twice differentiable, ensuring well-behaved optimization. Second, the quadratic structure provides an explicit solution, enabling direct theoretical analysis without iterative methods \citep{hoerl1962applications, hoerl1975ridge, hoerl1970ridge}.}

\tr{These analytical properties of ridge regression enable us to dramatically reduce the computational burden of L$N_0$OCV. Rather than refitting $\widehat{\boldsymbol{\beta}}_{\mathcal{T}_{1\ell}}$ across all $L$ training sets, we can obtain the ridge estimate $\widehat{\boldsymbol{\beta}}_{\text{R}}(\lambda)$ once using the full dataset, then compute appropriately weighted residuals across all $L$ terms. This reformulation maintains the statistical properties of the original estimator while achieving substantial computational savings compared to the standard implementation. The following main result establishes this efficient formulation.}

\noindent
{\bf Lemma 1:} {
\label{lemma:RR_LN0OCV_lemma}
\tr{Let $\{(\mathbf{X}_n, Y_n)\}_{n = 1}^N$ be independent and identically distributed draws from $\mathcal{Q}(Y_n \mid f(\mathbf{X}_n, \boldsymbol{\beta}))\mathcal{P}(\mathbf{X}_n)$.  Consider the linear model $\mathbf{y} = \mathbf{X}\boldsymbol{\beta} + \boldsymbol{\varepsilon}~$ where $\mathbf{y} \in \mathbb{R}^N, \mathbf{X} \in \mathbb{R}^{N \times P}$, $\boldsymbol{\beta} \in \mathbb{R}^{P}$, $\mathbb{E}(\boldsymbol{\varepsilon}) = \mathbf{0}$, $\mathbb{V}(\boldsymbol{\varepsilon}) = \sigma^2_{\varepsilon} \mathbf{I}_N$, and $\boldsymbol{\varepsilon} \ind \mathbf{X}$.  For a fixed $\lambda \in \boldsymbol{\Lambda}$ and $N_0 > 0$, the L$N_0$OCV$^{(1)} := ~$L$N_0$OCV$^{(1)}(\lambda)$ estimator simplifies to
\begin{eqnarray}
\label{eqn:RR_LN0OCV_lemma}
    \text{L$N_0$OCV}^{(1)}(\lambda)
    = \frac{1}{L N_0} \displaystyle \sum \limits_{\ell = 1}^{L} \mathbf{r}^\top_{\mathcal{T}_{0\ell}}(\lambda) [\mathbf{I}_{N_0} - \mathbf{H}_{\mathcal{T}_{0\ell}}(\lambda)]^{-2}\mathbf{r}_{\mathcal{T}_{0\ell}}(\lambda) = \frac{1}{LN_0} \displaystyle \sum \limits_{\ell = 1}^{L} \mathbf{y}^\top [\mathbf{W}_{\mathcal{T}_{0\ell}}(\lambda)]\mathbf{y}
\end{eqnarray}
where $\mathbf{r}_{\mathcal{T}_{0\ell}}(\lambda)$ is the $N_0 \times 1$ subvector of original residuals indexed by $\mathcal{T}_{0\ell}$, $\mathbf{H}_{\mathcal{T}_{0\ell}}(\lambda)$ is the $N_0 \times N_0$ principal submatrix of $\mathbf{H}(\lambda) = \mathbf{X}(\mathbf{X}^\top \mathbf{X} + \lambda \mathbf{I}_{P})^{\dagger} \mathbf{X}^\top$ with rows and columns indexed by $\mathcal{T}_{0\ell}$, and $\mathbf{W}_{\mathcal{T}_{0\ell}}(\lambda)$ is an $N \times N$ matrix derived from $\mathbf{I}_N - \mathbf{H}(\lambda)$.  The proof of Lemma 1 is provided in Appendix \ref{sec:ProofLemma1}.}}

\tr{Lemma 1 reveals a key insight about ridge regularized cross-validation methods: they share a unified computational structure despite employing different sampling strategies. To illustrate this connection, consider standard $K$-fold CV where each test set $\mathcal{T}_k = \{k(j)\}_{j = 1}^{N_k}$ contains $N_0 := N_k = \lfloor N/K \rfloor$ sampling units.  The $K$-fold CV estimator can be expressed using an identical matrix formulation with \eqref{eqn:RR_LN0OCV_lemma} given by
\begin{equation*}
    \text{$K$-fold CV}^{(1)}(\lambda)
    = \frac{1}{K} \displaystyle \sum \limits_{k = 1}^{K} \frac{\mathbf{r}^\top_{\mathcal{T}_{k}}(\lambda) [\mathbf{I}_{N_k} - \mathbf{H}_{\mathcal{T}_{k}}(\lambda)]^{-2}\mathbf{r}_{\mathcal{T}_{k}}(\lambda)}{N_k}, ~~\text{where} \sum\limits_{k = 1}^K N_k = N.
\end{equation*}
The mathematical similarity becomes clear: both methods compute weighted sums of squared residuals following the same matrix structure. The crucial difference lies in their sampling approaches. Unlike the exhaustive L$N_0$OCV, $K$-fold CV employs non-exhaustive, non-overlapping partitions with typically $K \ll L = \binom{N}{N_0}$. This unified framework demonstrates that the computational complexity differences between these methods stem from their enumeration strategies rather than fundamentally different mathematical operations.}

\noindent
{\bf Corollary 1:} {
\label{corollary:Null_LN0OCV}
Let $Y_1, Y_2, \ldots, Y_N$ be independent and identically distributed draws from some distribution $\mathcal{P}$. Suppose the null model corresponds to $\mathbf{y} = \beta_0\mathbf{1}_N + \boldsymbol{\varepsilon}$ where $\varepsilon_n$ is the error term with zero mean and variance $\sigma^2_{\varepsilon}, n \in [N]$.  The estimator of L$N_0$OCV$^{(0)}$ when $N_0 > 0$ can be simplified as
\begin{equation}
\text{L}N_0\text{OCV}^{(0)} = \displaystyle \frac{1}{L N_0} \sum\limits_{\ell = 1}^{L} \sum\limits_{h = 1}^{N_0}  (Y_{\ell(h)} - \overline{Y}_{\mathcal{T}_{1\ell}})^2 = \left(1 + \displaystyle \frac{1}{N - N_0}\right) S_Y^2
\label{eqn:Null_LN0OCV_Variance}
\end{equation}
where $S_Y^2$ is the sample variance of the entire response vector $\mathbf{y}$. The proof of Corollary 1 is available in Appendix \ref{sec:ProofCorollary1}.}

\tr{According to \eqref{eqn:Null_LN0OCV_Variance}, $\text{L}N_0\text{OCV}^{(0)}$ can be approximated by the sample variance $S_Y^2$ of the observed response vector when the number of sampling units in the training data is sufficiently large, i.e.,  $\text{L}N_0\text{OCV}^{(0)} - S_Y^2$ converges in probability to 0 as $N \rightarrow \infty$, while $N_0/N \rightarrow 0$.}

\tr{The computational advantages extend beyond ridge regression to include regularized methods with non-smooth penalties like LASSO. \citet{rad2020scalable} developed a scalable Approximate Leave-One-Out (ALO) framework that achieves dramatic speedups while preserving statistical benefits. \citet{wang2018approximate} extended this to non-differentiable regularizers including LASSO and elastic net, while \citet{stephenson2020approximate} identified that ALO methods can deteriorate in high-dimensional settings unless the parameter structure exhibits sparsity. These developments demonstrate that while exact solutions remain limited to cases like ridge regression, approximate methods can deliver substantial computational gains across diverse regularized learning problems.}

% -------------------------------------------------
\subsection[Asymptotic Results for Leave-N0-out Cross-validation Estimator]{Asymptotic Results for Leave-\(N_0\)-out Cross-validation Estimator}
\label{sec:AsymptoticResultsLN0OCVEstimator}
% -------------------------------------------------

\tr{The hypothesis testing framework for predictive improvement introduced in Section \ref{sec:HypothesisTestingPredictiveImprovement} centers on using cross-validation to reliably determine whether hypothesized features actually improve our ability to predict the response.  To provide the theoretical foundation for this comparison between model-fitting methods, we first investigate the asymptotic behavior of the $\text{LN}_0\text{OCV}^{(1)}$ estimator, provided $N_0/N \to 0$. In this analysis, we will interchangeably use the notations $\text{LN}_0\text{OCV}^{(v)} \equiv \text{LN}_0\text{OCV}(\mathcal{M}_v), v \in \{0, 1\}$ to emphasize that the method $\mathcal{M}_v$ is used.}

\tr{By adding and subtracting $f(\mathbf{X}_{\ell(h)}) = \mathbb{E}[\widehat{f}(\mathbf{X}_{\ell(h)} , \widehat{\boldsymbol{\beta}}_{\mathcal{T}_{1\ell}}) \mid \mathbf{X}]$, we decompose $\text{LN}_0\text{OCV}(\mathcal{M}_1) = \mathcal{C}_1(\mathcal{M}_1) + \mathcal{C}_2(\mathcal{M}_1) + 2\mathcal{C}_3(\mathcal{M}_1)$ where
\begin{eqnarray*}
\mathcal{C}_1(\mathcal{M}_1) &=& \frac{1}{L N_0} \sum\limits_{\ell = 1}^{L} \sum\limits_{h = 1}^{N_0} (Y_{\ell(h)} - f(\mathbf{X}_{\ell(h)}))^2 \\
\mathcal{C}_2(\mathcal{M}_1)&=& \frac{1}{L N_0} \sum\limits_{\ell = 1}^{L} \sum\limits_{h = 1}^{N_0} (f(\mathbf{X}_{\ell(h)}) - \widehat{f}(\mathbf{X}_{\ell(h)}, \widehat{\boldsymbol{\beta}}_{\mathcal{T}_{1\ell}}))^2\\
\mathcal{C}_3(\mathcal{M}_1)&=& \frac{1}{L N_0} \sum\limits_{\ell = 1}^{L} \sum\limits_{h = 1}^{N_0} (Y_{\ell(h)} - f(\mathbf{X}_{\ell(h)}))(f(\mathbf{X}_{\ell(h)}) - \widehat{f}(\mathbf{X}_{\ell(h)}, \widehat{\boldsymbol{\beta}}_{\mathcal{T}_{1\ell}})).
\end{eqnarray*}
From this decomposition, $\mathcal{C}_1(\mathcal{M}_1)$ represents the irreducible error component, $\mathcal{C}_2(\mathcal{M}_1)$ captures the excess error of the fitted predictive rule, and $\mathcal{C}_3(\mathcal{M}_1)$ is the cross-term between noise and estimation error \citep{rosset2019fixed, wager2020cross}. Notably, $\mathcal{C}_2(\mathcal{M}_1)$ and $\mathcal{C}_3(\mathcal{M}_1)$ depend on the training data through $\widehat{f}(\mathbf{X}_{\ell(h)}, \widehat{\boldsymbol{\beta}}_{\mathcal{T}_{1\ell}})$, while $\mathcal{C}_1(\mathcal{M}_1)$ depends only on the test observations $\{(\mathbf{X}_{\ell(h)}, Y_{\ell(h)})\}_{h = 1}^{N_0}$.  Counting techniques reveal that each sampling unit $(\mathbf{X}_n, Y_n)$ appears $\binom{N-1}{N_0-1}$ times across all testing sets, yielding
\begin{eqnarray}
\label{eqn:CV1_Simplification}
    \mathcal{C}_1(\mathcal{M}_1) = \frac{1}{L N_0} \dbinom{N - 1}{N_0 - 1}\sum\limits_{n = 1}^{N} (Y_{n} - f(\mathbf{X}_{n}))^2 = \frac{1}{N} \sum\limits_{n = 1}^{N} (Y_{n} - f(\mathbf{X}_{n}))^2 = \mathcal{C}_1.
\end{eqnarray}
Thus $\mathcal{C}_1$ in \eqref{eqn:CV1_Simplification} is no longer a function of the model-fitting algorithm $\mathcal{M}_1$ and can be viewed as the training error of the optimally fitted predictor.}

\noindent
{\bf Corollary 2:} {
\label{corollary:RR_Asymptotics}
Consider a sequence of independent and identically distributed samples $(\mathbf{X}_n, Y_n) \in \mathbb{R}^P \times \mathbb{R}$ which satisfy the following assumptions:
\begin{enumerate}
    \item[{A.1}] $\mathbb{E}(Y_n^4) < \infty$ and $\mathbb{V}(Y_n \mid \mathbf{X}_n) \leq \boldsymbol{\Sigma}$, $\left\|\boldsymbol{\Sigma}\right\|_\infty < \infty$ 
    \item[{A.2}] There are constants $0 < c_{-} \leq c_+ < \infty$ and $0.25 < \delta < 0.5$ such that the excess risk of $\widehat{f}(\mathbf{X}_{\ell(h)}, \widehat{\boldsymbol{\beta}}_{\mathcal{T}_{1\ell}})$ evaluated on a testing data point scales as
\begin{eqnarray*}
\lim \limits_{N \rightarrow \infty} \mathbb{P}\left[N_1^\delta \mathbb{E}\left[ \left(\widehat{f}(\mathbf{X}_{\ell(h)}, \widehat{\boldsymbol{\beta}}_{\mathcal{T}_{1\ell}}) - f(\mathbf{X}_{\ell(h)})\right)^2 \mid \{\mathbf{X}_{\ell(h)}, Y_{\ell(h)}\}_{h = N_0 + 1}^N \right]^{\frac{1}{2}} \leq c_{-}\right] &=& 0,\\
\lim \limits_{N \rightarrow \infty} \mathbb{P}\left[N_1^\delta \mathbb{E}\left[ \left(\widehat{f}(\mathbf{X}_{\ell(h)}, \widehat{\boldsymbol{\beta}}_{\mathcal{T}_{1\ell}}) - f(\mathbf{X}_{\ell(h)})\right)^2 \mid \{\mathbf{X}_{\ell(h)}, Y_{\ell(h)}\}_{h = N_0 + 1}^N \right]^{\frac{1}{2}} \leq c_{+}\right] &=& 1
\end{eqnarray*}
for a given method $\mathcal{M}_1$, where $\mathbf{X}_{\ell(h)}$ is the set of features from a randomly selected observation in the testing set, drawn independently from the collection of observations belonging to the training set $\{\mathbf{X}_{\ell(h)}, Y_{\ell(h)}\}_{h = N_0 + 1}^N$.
\end{enumerate}
It follows that
\begin{equation}
\label{eqn:RR_LN0OCV_Distribution}
    \sqrt{N}(\text{L}N_0\text{OCV}(\mathcal{M}_1) - \text{Err}^{(1)}) \overset{d}{\rightarrow} \mathcal{N}(0, \sigma^2_{\mathcal{C}_1}), ~\text{as} ~N \rightarrow \infty
\end{equation}
where $\mathbb{E}(\mathcal{C}_1) = \text{Err}^{(1)}$ and $\mathbb{V}(\mathcal{C}_1) = \frac{\sigma^2_{\mathcal{C}_1}}{N}$.  The proof of Corollary 2 is detailed in Appendix \ref{sec:ProofCorollary2}.}

\tr{The asymptotic result in Corollary 2 reveals that the first-order behavior of $\text{LN}_0\text{OCV}(\mathcal{M}_1)$ is dominated by $\mathcal{C}_1$, which depends only on the irreducible error and is independent of the model-fitting algorithm. This algorithm-independence has important implications for model comparison: when evaluating two methods $\mathcal{M}_0$ and $\mathcal{M}_1$, the dominant term $\mathcal{C}_1$ cancels out, leaving only the algorithm-dependent components:
\begin{equation}
\text{LN}_0\text{OCV}(\mathcal{M}_0) - \text{LN}_0\text{OCV}(\mathcal{M}_1) = (\mathcal{C}_2(\mathcal{M}_0) - \mathcal{C}_2(\mathcal{M}_1)) + 2(\mathcal{C}_3(\mathcal{M}_0) - \mathcal{C}_3(\mathcal{M}_1))
\end{equation}
This cancellation property enables reliable model selection. Consider two methods $\mathcal{M}_0$ and $\mathcal{M}_1$ that satisfy the assumptions in Corollary 2 with excess risk scaling parameters $(\delta, c_{-}, c_{+})$ and $(\delta^\star, c^\star_{-}, c^\star_{+})$, respectively. These parameters characterize how each method's prediction error decreases as the training sample size grows, where $\delta$ controls the rate of convergence and the constants $c_{-}, c_{+}$ bound the scaled excess risk. When method $\mathcal{M}_1$ exhibits superior convergence properties---specifically when $\delta < \delta^\star$ (faster convergence rate) or $\delta = \delta^\star$ and $c_{+} > c^\star_{-}$ (same convergence rate but consistently lower error bounds)---the cross-validation difference converges to the true difference in excess errors:
\begin{equation*}
\frac{\text{LN}_0\text{OCV}(\mathcal{M}_0) - \text{LN}_0\text{OCV}(\mathcal{M}_1)}{\mathcal{C}_2(\mathcal{M}_0) - \mathcal{C}_2(\mathcal{M}_1)} \xrightarrow{p} 1 ~\text{and}~\lim\limits_{N \rightarrow \infty} \mathbb{P}[\text{LN}_0\text{OCV}(\mathcal{M}_0) > \text{LN}_0\text{OCV}(\mathcal{M}_1)] = 1,
\end{equation*}
meaning cross-validation will asymptotically show that $\mathcal{M}_0$ has higher error than $\mathcal{M}_1$, correctly identifying the better performing method. These results establish that cross-validation enables asymptotically perfect model selection, as demonstrated by \citet{wager2020cross}.}

\tr{Building on the theoretical requirement that $N_0/N \rightarrow 0$ established in this section, we now develop high-dimensional statistical tests by choosing $N_0$ to be small (specifically $N_0 = 1$ or 2), ensuring this key asymptotic condition is satisfied in practice.}

% -------------------------------------------------
\section{Nested Exhaustive Cross-validation Tests for High-dimensional data}
\label{sec:NestedExhaustiveCVTest}
% -------------------------------------------------

\tr{This section presents a statistical testing framework for comparing competing methods $\mathcal{M}_0$ and $\mathcal{M}_1$ in high-dimensional settings. The framework addresses a persistent challenge in statistical methodology: achieving both computational efficiency and statistical rigor when evaluating predictive performance. Leveraging the computational results of Lemma 1 and Corollary 1, and the asymptotic properties established in Corollary 2, our approach centers on the L$N_0$OCV estimator to enable precise method comparison through direct matched-pair analysis.  This paired design is theoretically advantageous, as it substantially enhances statistical power for detecting meaningful performance differences \citep{schrauf2021comparing}.}

\tr{However, a critical methodological challenge emerges from the dependence of $\text{L}{N_0}\text{OCV}^{(1)}$ from \eqref{eqn:PR_LN0OCV} on the regularization parameter $\lambda$.  This dependence introduces two distinct sources of bias that can compromise model evaluation, each requiring different solutions.  According to \citet{varma2006bias}, parameter selection bias arises when standard single-level cross-validation violates the principle of independence between hyperparameter selection and predictive performance evaluation. When the same data informs both hyperparameter choice and error estimation, the resulting performance metrics become systematically optimistic and underestimate the true error rates.}

\begin{figure}[ht]
\scriptsize
\centering
    \begin{tabular}{ccc}
    \includegraphics[width=0.3\linewidth]{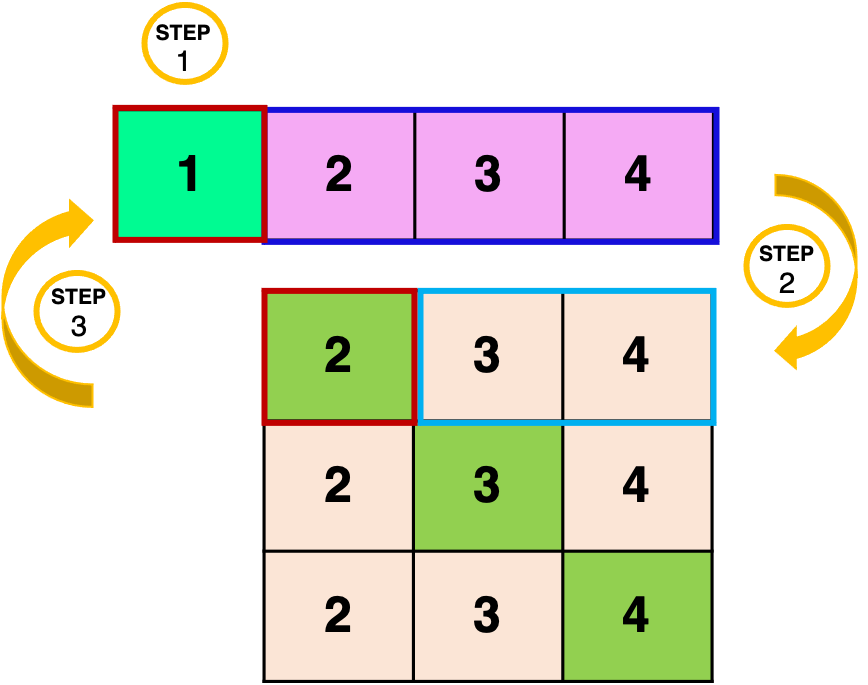} &
    \includegraphics[width=0.3\linewidth]{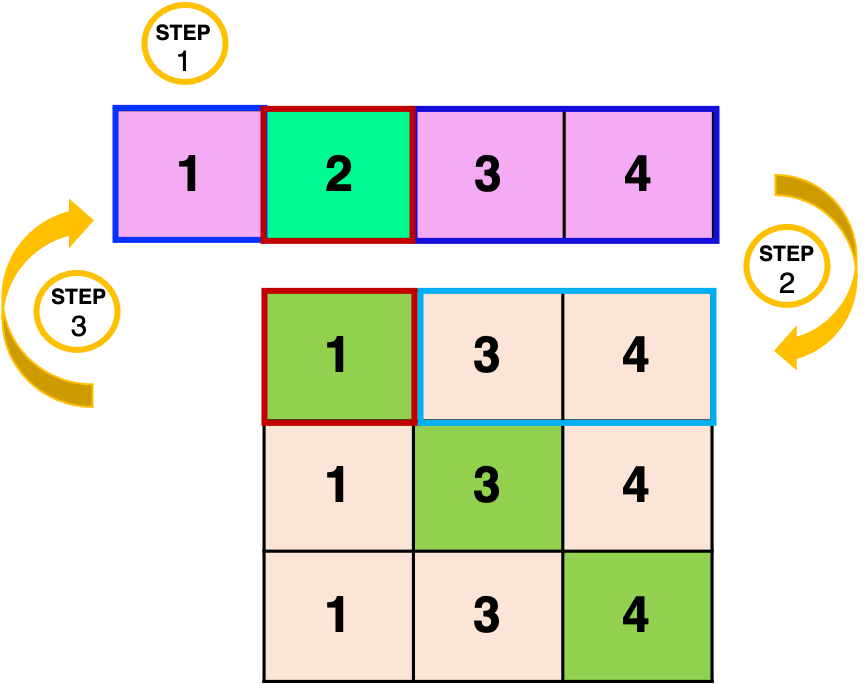} & \includegraphics[width=0.3\linewidth]{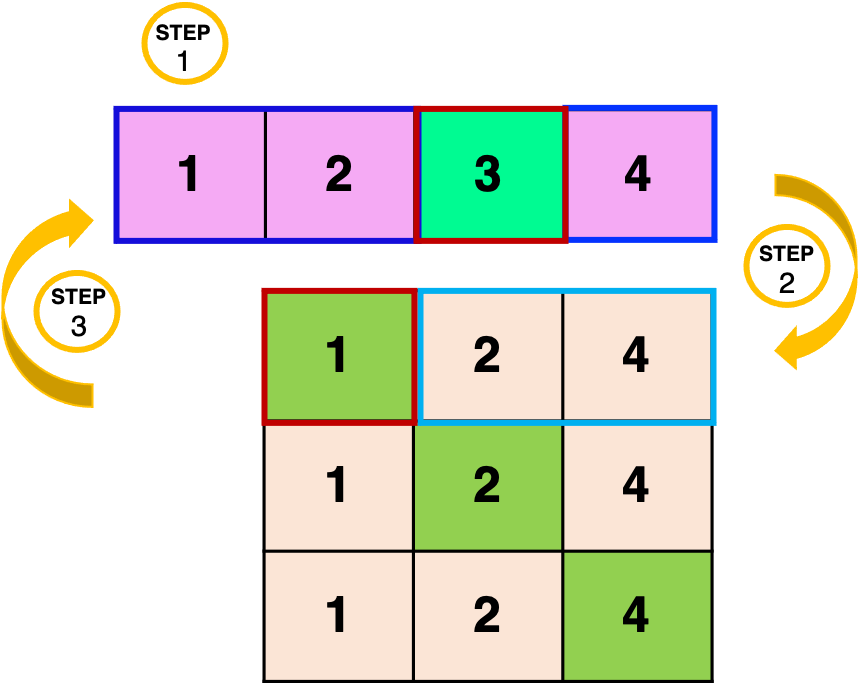} \\
    $T^{(0)}_{1} = (Y_1 - \overline{Y}_{-1})^2$ & $T^{(0)}_{2} = (Y_2 - \overline{Y}_{-2})^2$ & $T^{(0)}_{3} = (Y_3 - \overline{Y}_{-3})^2$\\
    $T^{(1)}_{1} = (Y_1 - \mathbf{X}_1^\top \widehat{\boldsymbol{\beta}}_{-1}(\widehat{\lambda}_1))^2$ & $T^{(1)}_{2} = (Y_2 - \mathbf{X}_2^\top \widehat{\boldsymbol{\beta}}_{-2}(\widehat{\lambda}_2))^2$ & $T^{(1)}_{3} = (Y_3 - \mathbf{X}_3^\top \widehat{\boldsymbol{\beta}}_{-3}(\widehat{\lambda}_3))^2$\\
    \includegraphics[width=0.3\linewidth]{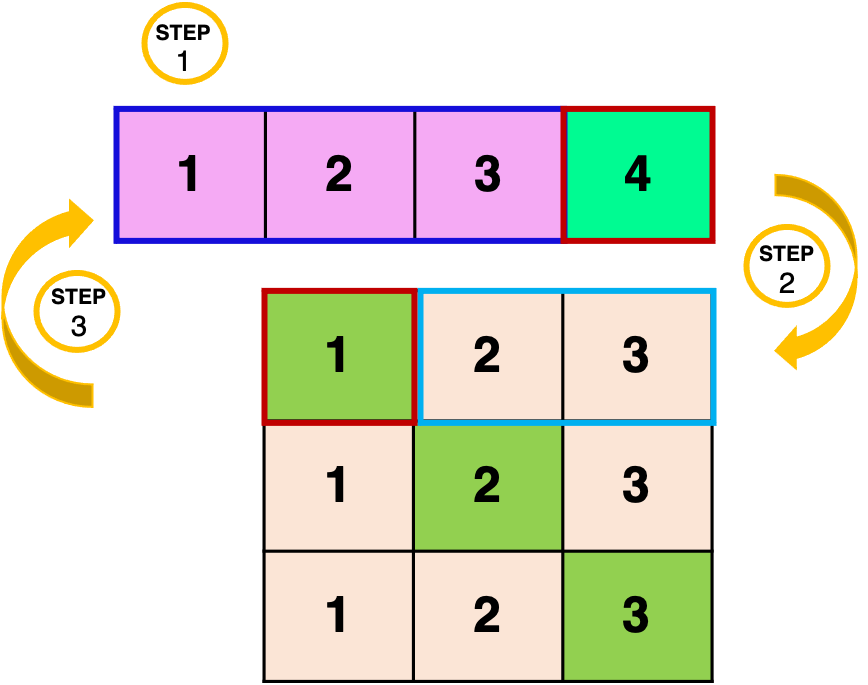} & \multicolumn{2}{c}{\includegraphics[width=0.6\linewidth]{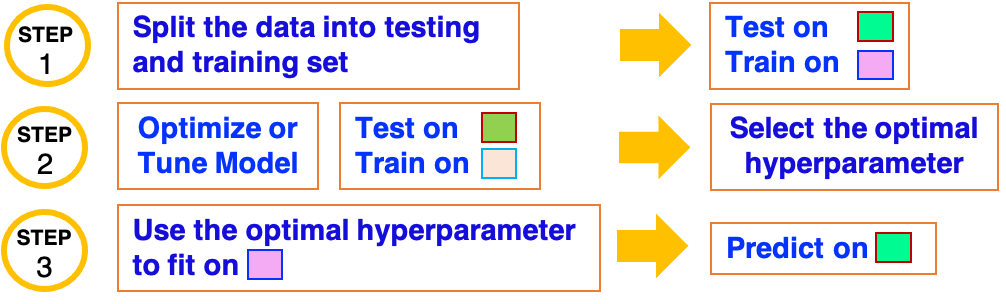}}\\
    $T^{(0)}_{4} = (Y_4 - \overline{Y}_{-4})^2$\\
    $T^{(1)}_{4} = (Y_4 - \mathbf{X}_4^\top \widehat{\boldsymbol{\beta}}_{-4}(\widehat{\lambda}_4))^2$\\
    \end{tabular}
    \caption{Nested Leave-one-out Cross-validation procedure for Testing Predictive Improvement with $N = 4$. Each panel shows one observation is held out for testing while the remaining three form the training set. Within each training set, nested cross-validation is performed to select the optimal regularization parameter $\widehat{\lambda}_\ell$ before computing the prediction errors $T^{(v)}_{\ell}, v = 0, 1$ for $\ell = 1, 2, 3, 4$.}
    \label{fig:IllustrationNCV}
\end{figure}

\tr{The nested cross-validation framework addresses parameter selection bias through a two-level architecture: an inner optimization loop identifies optimal tuning parameters using only training data, while an outer evaluation loop provides predictive performance estimates on completely independent held-out data. This rigorous separation between selection and evaluation phases eliminates the underestimation characteristic of single-level approaches.}

\tr{The methodology employs a three-tier partitioning scheme that maximizes data utilization. For the $\ell$th partition, the primary split divides data into testing ($\mathcal{T}_{0\ell} = \{\ell(h)\}_{h=1}^{N_0}$ with $N_0$ indices) and training ($\mathcal{T}_{1\ell} = \{\ell(h)\}_{h=N_0+1}^N$ with remaining $N - N_0$ indices) components.  The critical step lies in the mutually exclusive nested decomposition of the testing set where $\mathcal{T}_{0\ell} = \mathcal{R}_{O\ell} \cup \mathcal{R}_{I\ell}$. The outer testing set $\mathcal{R}_{O\ell} = \{\ell(h)\}_{h=1}^{R_O}$ contains $R_O$ ``reserved" observation indices, serving exclusively as an evaluation benchmark to ensure unbiased predictive performance assessment. The inner testing set $\mathcal{R}_{I\ell} = \{\ell(h)\}_{h=R_O+1}^{N_0}$ comprises the remaining $R_I = N_0 - R_O$ testing indices used specifically for hyperparameter optimization.  Figure \ref{fig:IllustrationNCV} demonstrates this procedure for the simple case $R_O = R_I = 1$.}

\tr{While this nested partitioning scheme provides the conceptual foundation, the practical implementation requires addressing the computational complexity of the inner and outer cross-validation stages.}

% -------------------------------------------------
\subsection{Computationally Efficient Inner CV for Hyperparameter Optimization}
\label{sec:ComputationallyEfficientInnerCV}
% -------------------------------------------------

\tr{In this study, our proposed inner cross-validation stage represents both a computational advancement and a theoretical contribution. The advantage is that, rather than computing the penalized regression coefficients $\widehat{\boldsymbol{\beta}}_{\mathcal{T}_{1\ell}}$,  $\mathcal{T}_{1\ell} = [N] \backslash (\mathcal{R}_{O\ell} \cup \mathcal{R}_{I\ell})$, across all $L_I = \binom{N-R_O}{R_I}$ possible data partitions, Lemma 1 provides computationally efficient closed-form expressions that make exhaustive parameter selection tractable for large datasets.  The computational efficiency comes from recognizing that the exhaustive inner cross-validation optimization
\begin{equation}
\widehat{\lambda}_\ell = \underset{\lambda_b \in \boldsymbol{\Lambda}}{\text{arg min}}~\frac{1}{L_I R_I} \sum_{\ell'=1}^{L_I} \sum_{h=R_O+1}^{N_0} \left(Y_{\ell'(h)} - \mathbf{X}_{\ell'(h)}^\top \widehat{\boldsymbol{\beta}}_{[N] \backslash \mathcal{T}_{0\ell}}(\lambda_b) \right)^2 
\end{equation}
can be expressed using the hat matrix $\mathbf{H}(\lambda_b)$ and residual vector $\mathbf{r}(\lambda_b)$, eliminating the need to explicitly enumerate all partition combinations.}

\tr{For the practically important Nested Leave-one-out case with $R_O = R_I = 1$, $L_O = N$ and $L_I = N-1$, Lemma 1 yields the efficient form:
\begin{equation}
\widehat{\lambda}_n = \underset{\lambda_b \in \boldsymbol{\Lambda}}{\text{arg min}} ~\frac{1}{N-1} \sum_{m \neq n} \left[\frac{\mathcal{H}_{0n}(\lambda_b)r_m(\lambda_b) + [\mathbf{H}(\lambda_b)]_{mn}r_n(\lambda_b)}{\mathcal{H}_{1mn}(\lambda_b)}\right]^2
\end{equation}
where $\mathcal{H}_{0a}(\lambda_b) = 1 - [\mathbf{H}(\lambda_b)]_{aa}$, $\mathcal{H}_{1mn}(\lambda_b) = \mathcal{H}_{0n}(\lambda_b)\mathcal{H}_{0m}(\lambda_b) - [\mathbf{H}(\lambda_b)]_{mn}^2$, and $r_n(\lambda_b)$ represents the $n$th residual.}

\tr{Similarly, for the Nested Leave-two-out case with $R_O = 2$, $R_I = 1$, $L_O = N(N-1)/2$ and $L_I = N-2$, Lemma 1 provides:
\begin{equation*}
\widehat{\lambda}_{mn} = \underset{\lambda_b \in \boldsymbol{\Lambda}}{\text{arg min}} ~\frac{1}{N-2} \sum_{l \neq m,n} \left[\frac{\mathcal{H}_{1mn}(\lambda_b) r_l(\lambda_b) + \mathcal{H}_{2lm}^{(n)}(\lambda_b)r_m(\lambda_b) + \mathcal{H}_{2ln}^{(m)}(\lambda_b)r_n(\lambda_b)}{\mathcal{H}_{3mn}^{(l)}(\lambda_b)}\right]^2
\end{equation*}
where the auxiliary functions are
\begin{eqnarray*}
    \mathcal{H}^{(m)}_{2ln}(\lambda_b) 
    &=& \mathcal{H}_{0m}(\lambda_b)[\mathbf{H}(\lambda_b)]_{ln}+[\mathbf{H}(\lambda_b)]_{mn} [\mathbf{H}(\lambda_b)]_{lm}\\
    \mathcal{H}^{(n)}_{2lm}(\lambda_b) 
    &=& \mathcal{H}_{0n}(\lambda_b)[\mathbf{H}(\lambda_b)]_{lm}+[\mathbf{H}(\lambda_b)]_{mn} [\mathbf{H}(\lambda_b)]_{ln} ~~\text{and}\\
    \mathcal{H}^{(l)}_{3mn}(\lambda_b)
    &=& \mathcal{H}_{0l}(\lambda_b)\mathcal{H}_{1mn}(\lambda_b) -  [\mathbf{H}(\lambda)]_{ln}\mathcal{H}^{(m)}_{2ln}(\lambda_b) -[\mathbf{H}(\lambda)]_{lm}\mathcal{H}^{(n)}_{2lm}(\lambda_b). 
\end{eqnarray*}
This computational breakthrough essentially transforms nested CV from an intractable combinatorial problem into a tractable linear algebra solution. By expressing all possible inner CV configurations through closed-form matrix operations computed once per candidate $\lambda_b$, we eliminate the need to explicitly fit models across exponentially many data partitions. The resulting algorithm scales polynomially rather than exponentially with dataset size, making exhaustive parameter space exploration computationally feasible even for large-scale applications where traditional nested CV would be prohibitive. This efficiency gain unlocks rigorous model selection for previously inaccessible problem domains, enabling researchers to achieve optimal hyperparameter tuning without computational compromise.}

% -------------------------------------------------
\subsection{Computationally Efficient Outer CV for Prediction Error Estimation}
\label{sec:ComputationallyEfficientOuterCV}
% -------------------------------------------------

\tr{Having rigorously selected hyperparameters $\widehat{\lambda}_{\ell}$ through the nested cross-validation framework, we obtain prediction error estimates for each outer partition under both model-fitting approaches. The nested structure naturally leads to a decomposition that separates the average prediction error, inherent bias, and sampling variance components that collectively determine the quality of our predictive performance comparison.}

\tr{For each outer partition $\ell \in [L_O]$, we obtain two $L_O R_O \times 1$ vectors $\mathbf{T}^{(v)} = (\mathbf{T}^{(v)}_1, \ldots, \mathbf{T}^{(v)}_{L_O})^\top$ where each block $\mathbf{T}^{(v)}_\ell = (T^{(v)}_{\ell(1)}, \ldots, T^{(v)}_{\ell(R_O)})^\top$ contains the prediction error estimates for outer partition $\ell$ under model-fitting approach $v \in \{0, 1\}$. For Nested Leave-one-out cross-validation, we obtain two $N \times 1$ vectors with elements $T^{(v)}_{n}$ as follows
\begin{equation}
\label{eqn:T_L1OCV}
    T^{(v)}_{n} =
    \begin{cases}
        \left(Y_{n} - \mathbf{X}_{n}^\top \widehat{\boldsymbol{\beta}}_{-n}(\widehat{\lambda}_n)\right)^2 = \displaystyle\left(\frac{Y_{n} - \mathbf{X}_{n}^\top \widehat{\boldsymbol{\beta}}(\widehat{\lambda}_{n})}{1 - [\mathbf{H}({\widehat{\lambda}_{n}})]_{nn}} \right)^2, & v = 1\\
        \left(Y_{n} - \overline{Y}_{-n}\right)^2 = \displaystyle\left(\frac{N}{N - 1}\right)^2 (Y_{n} - \overline{Y})^2,& v = 0.
    \end{cases}
\end{equation}
These expressions follow from applying Lemma 1 and Corollary 1. While the formulation of $T^{(v)}_{n}$ in \eqref{eqn:T_L1OCV} is derived for ridge regularization, approximate leave-one-out for LASSO follows a similar form \citep{rad2020scalable}. The key difference is that the hat matrix $\mathbf{H}_\mathcal{E}$ is defined in terms of $\mathbf{X}_\mathcal{E}$, the submatrix that include only columns of $\mathbf{X}$ in the equicorrelation set $\mathcal{E} = \{p \in \{1, 2, \ldots, P\}: |\mathbf{x}_p^\top(\mathbf{y} - \mathbf{X}\widehat{\boldsymbol{\beta}})| = \lambda\}$, where $\mathbf{x}_p \in \mathbb{R}^N$ is the $p$th column of $\mathbf{X}$.  Another special case is Nested Leave-two-out Cross-validation with
\begin{eqnarray*}
\label{eqn:L2OCV_T}
    T^{(v)}_{mn} =
    \begin{cases}
        \left(Y_{m} - \mathbf{X}_{m}^\top \widehat{\boldsymbol{\beta}}_{-mn}(\widehat{\lambda}_{mn})\right)^2 = \displaystyle \left[\frac{\mathcal{H}_{0n}(\widehat{\lambda}_{mn})r_m(\widehat{\lambda}_{mn}) + [\mathbf{H}(\widehat{\lambda}_{mn})]_{mn}r_n(\widehat{\lambda}_{mn})}{\mathcal{H}_{1mn}(\widehat{\lambda}_{mn})}\right]^2, & v = 1\\
        \left(Y_{m} - \overline{Y}_{-mn}\right)^2 = \displaystyle \left[\left(\frac{N - 1}{N - 2}\right)(Y_{m} - \overline{Y}) + \frac{Y_{n} - \overline{Y}}{N - 2}\right]^2,& v = 0
    \end{cases}
\end{eqnarray*}
where $\mathcal{H}_{0a}(\widehat{\lambda}_{mn}) = 1 - [\mathbf{H}(\widehat{\lambda}_{mn})]_{aa}$, and $\mathcal{H}_{1mn}(\widehat{\lambda}_{mn}) = \mathcal{H}_{0n}(\widehat{\lambda}_{mn})\mathcal{H}_{0m}(\widehat{\lambda}_{mn}) - [\mathbf{H}(\widehat{\lambda}_{mn})]_{mn}^2$.}

\tr{Consequently, the difference in the average prediction error simplifies to the more computationally tractable form
\begin{equation}
    \overline{T}_{R_O} = \frac{1}{L_O R_O} \sum \limits_{\ell = 1}^{L_O} \sum \limits_{h = 1}^{R_O} (T^{(0)}_{\ell(h)} - T^{(1)}_{\ell(h)}) = \left(1 + \frac{1}{N - R_O}\right)S_Y^2 - \frac{1}{L_O}\sum\limits_{\ell = 1}^{L_O} \frac{\mathbf{y}^\top [\mathbf{W}^{(1)}_{\mathcal{R}_{O\ell}}(\lambda)]\mathbf{y}}{R_O}
\end{equation}
using Lemma 1 and Corollary 1, provided $R_O > 0$.}

\tr{While the nested structure eliminates parameter selection bias through architectural separation, inherent bias remains due to the training set size asymmetry between inner and outer procedures \citep{varma2006bias}. Following the bias quantification methodology of Bates et al. (2023), we quantify this deviation as
\begin{equation}
\label{eqn:TN0CV_Bias}
    \widehat{\mathbb{B}}(\overline{T}_{R_O}) = (\overline{T}^{(0)}_{R_I} - \overline{T}^{(0)}_{R_O})^2 + (\overline{T}^{(1)}_{R_I} - \overline{T}^{(1)}_{R_O})^2.
\end{equation}
This captures the squared difference between what the inner CV process optimizes for (performance on training sets of size $N - R_I - R_O$) versus what the outer CV process evaluates (performance on training sets of size $N - R_O$). The computational approach also follows from \citet{arlot2010survey}, who demonstrated that nested cross-validation procedures inherently introduce dependencies between the inner optimization and outer evaluation phases.}

\tr{On the other hand, the variance component captures sampling variability across different cross-validation partitions.  Specifically, 
\begin{eqnarray}
\label{eqn:TN0CV_Variance}
    \widehat{\mathbb{V}}(\overline{T}_{R_O})
    &=& 
    \begin{cases}
    \varphi_1 \mathcal{V}_1 + \varphi_2 \mathcal{V}_2 + \varphi_3 \mathcal{V}_3, & R_O > 1\\
    \varphi_1 \mathcal{V}_1 + \varphi_3 \mathcal{V}_3, & R_O = 1\\
    \end{cases}
\end{eqnarray}
where $\mathcal{V}_1 = \displaystyle\sum\limits_{\ell = 1}^{L_O} \sum\limits_{h = 1}^{R_O} T_{\ell(h)}^2, \mathcal{V}_2 = \sum\limits_{\ell = 1}^{L_O} \sum\limits_{h = 1}^{R_O}\sum \limits_{h \neq h^\star} T_{\ell(h)} T_{\ell(h^\star)}$, and $\mathcal{V}_3 = \sum\limits_{\ell = 1}^{L_O}  \sum \limits_{\ell^\star \neq \ell} \sum\limits_{h = 1}^{R_O}\sum \limits_{h^\star = 1}^{R_O}T_{\ell(h)} T_{\ell^\star(h^\star)}$.}

\tr{The three variance components capture distinct sources of variability in the cross-validation procedure. The first component, $\mathcal{V}_1$, represents the sum of squared individual prediction errors across all observations and partitions. This diagonal term captures the fundamental variability of individual predictions and forms the baseline contribution to the overall variance regardless of the cross-validation structure. The second component, $\mathcal{V}_2$, quantifies the covariances between different observations within the same outer partition $\ell$. This within-partition covariance term reflects the correlation structure that emerges when the same training set is used to predict different test observations, capturing dependencies that arise from the shared model-fitting procedure. The third component, $\mathcal{V}_3$, captures the covariances between observations from different outer partitions. This between-partition term reflects more subtle dependencies that can arise from overlapping training sets or from patterns in the data that affect prediction errors across different train-test splits.}

\tr{The weighting coefficients $\varphi_1, \varphi_2, \varphi_3$ are determined by the specific variance estimation form and the number of observations per partition, with the default values being functions of the total sample size $N$.  When $R_O = 1$, there are no pairs of different observations within partitions, causing the within-partition covariance component $\mathcal{V}_2$ to vanish. In this case, the default weights are $\varphi_1 = \frac{1}{N}$ and $\varphi_3 = \frac{-1}{N(N-1)}$. When $R_O = 2$, the situation becomes more complex as both within-partition and between-partition dependencies contribute to the overall variance. The default weights are $\varphi_1 = \frac{3}{2N(N-1)}$, $\varphi_2 = \frac{-1}{2N(N-1)}$, and $\varphi_3 = \frac{-1}{N(N-1)[N(N-1)-2]}$. These weighting schemes reflect the different ways in which the variance decomposition accounts for the correlation structure inherent in the cross-validation procedure.}

\tr{The mean squared error (MSE) estimate combines the  square of the bias term and variance term. This leads to our standardized test statistic that quantifies the evidence for predictive improvement
\begin{equation}
\label{eqn:TN0CV_MSE}
    \mathbb{T}_{R_O} = \frac{\sqrt{L_O R_O} ~\cdot \overline{T}_{R_O}}{\sqrt{\text{MSE}(\overline{T}_{R_O})}} = \frac{\sqrt{L_O R_O} ~\cdot \overline{T}_{R_O}}{\sqrt{\widehat{\mathbb{B}}(\overline{T}_{R_O}) + \widehat{\mathbb{V}}(\overline{T}_{R_O})}}.
\end{equation}
The statistic $\mathbb{T}_{R_O}$ provides a rigorous foundation for statistical comparison between competing model-fitting approaches, properly accounting for the nested structure's impact on both bias and variance.}

% -------------------------------------------------
\section{Numerical Experiments}
\label{sec:NumericalExperiments}
% -------------------------------------------------

\tr{This section presents comprehensive simulation studies designed to evaluate the performance of our proposed testing procedures. We focus on two critical performance metrics: Type I error control and Empirical Power. These simulation studies provide practical guidance for selecting methods that achieve high sensitivity in detecting true signals while maintaining stringent false positive rates in high-dimensional ($P > N$) settings. Throughout our analysis, we define $P$ as the total number of columns in the design matrix $\mathbf{X}$, where $P = P^{\star} + 1$, with $P^{\star}$ representing the number of hypothesized features under investigation and $N$ denoting the sample size.}

\tr{Our simulation framework systematically varies three key parameters to capture realistic analytical scenarios. First, we examine sample sizes $N \in \{50, 100, 150, 200\}$ to include typical ranges in high-dimensional data analysis. Second, we consider the dimensionality ratio $\gamma = P/N$ across a comprehensive range: $\gamma \in \{1 + 1/N, 2, 3, 4, 5, 10, 15, 20, 25, 50, 75, 100\}$, spanning from equi- to ultra-high-dimensional settings. Third, we vary the signal strength through effect size parameters $\xi = 0.025A$ where $A \in \{0, 1, 2, 3, 4, 5\}$, with $\xi = 0$ representing the null hypothesis scenario.}

\tr{On the other hand, the comparative analysis examines 24 distinct test statistics, all based on the Nested Leave-$R_O$-out cross-validation framework. This comprehensive comparison results from systematically combining four methodological choices. We vary the leave-out size $R_O \in \{1, 2\}$ and examine three approaches for quantifying prediction error variability: variance-only (V), squared bias-only (B), and mean squared error (M). Additionally, we compare two penalized regression methods---Ridge (R) and LASSO (L)---each implemented with either adaptive optimal hyperparameters $\widehat{\boldsymbol{\lambda}}$ or averaged optimal parameters $\overline{\lambda}$.}

\begin{figure}[ht]
    \begin{tabular}{cc}
        \includegraphics[width=0.48\linewidth]{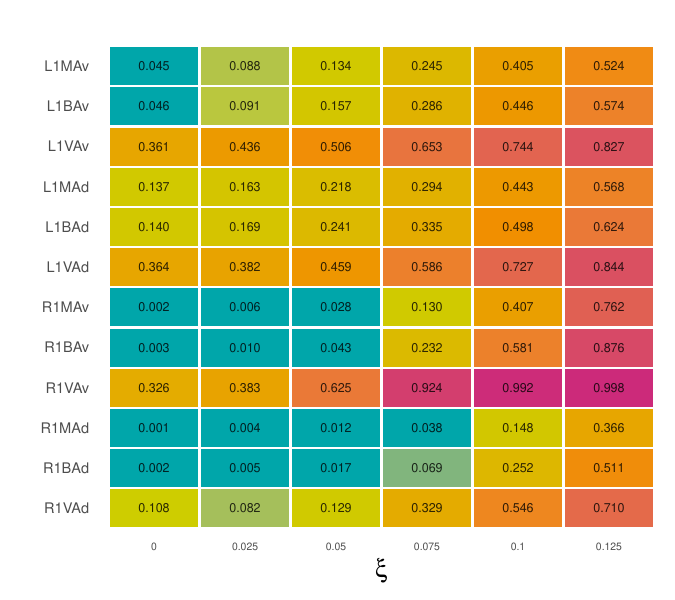}
        & \includegraphics[width=0.48\linewidth]{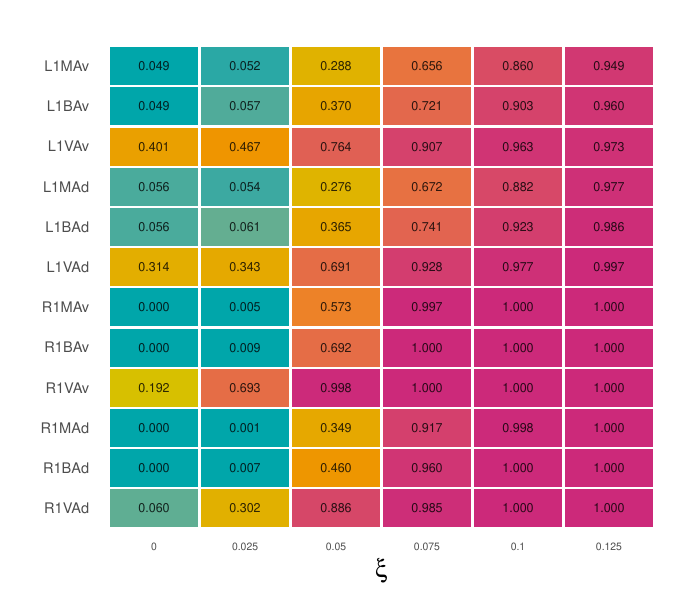}\\
        $N = 50, P = 51$ & $N = 100, P = 101$\\
        \includegraphics[width=0.48\linewidth]{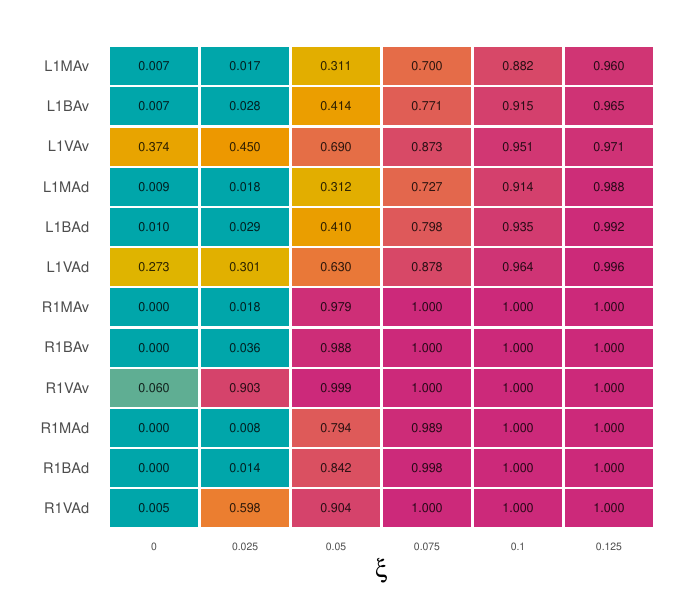}
        & \includegraphics[width=0.48\linewidth]{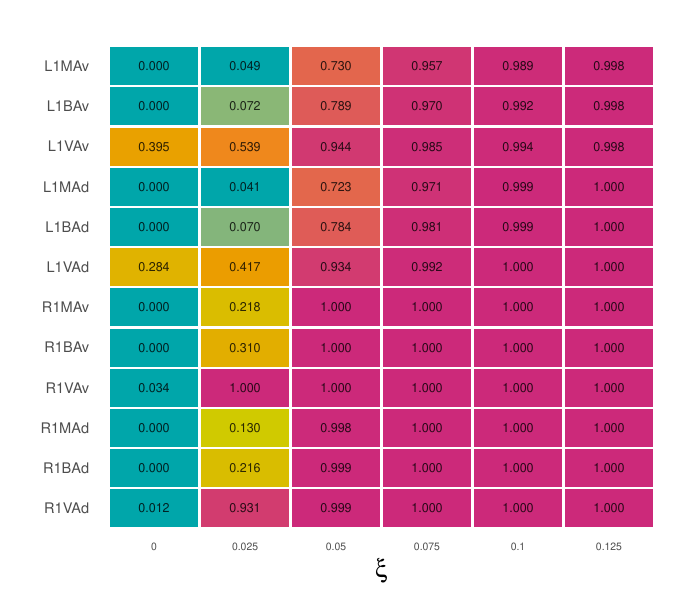}\\
        $N = 150, P = 151$ & $N = 200, P = 201$\\
    \end{tabular}
    \caption{Empirical rejection rates for Nested L1OCV procedures across varying effect sizes ($\xi$). Each panel represents different sample sizes $N $ with $P = N+1$ parameters. The first column ($\xi$ = 0) shows Type I error rates; subsequent columns ($\xi$ = 0.025 to 0.125) show empirical power.}
    \label{fig:RejectionRate_NL1OCV_1}
\end{figure}

\tr{The resulting test statistics using \eqref{eqn:TN0CV_Bias}, \eqref{eqn:TN0CV_Variance} and \eqref{eqn:TN0CV_MSE} follow the general form:
\begin{equation}
    \mathbb{T}_{iR_O\text{V}j} = \frac{\sqrt{N} \cdot \overline{T}_{iR_Oj}}{\sqrt{\widehat{\mathbb{V}}(\overline{T}_{iR_Oj})}}, 
\mathbb{T}_{iR_O\text{B}j} = \frac{\sqrt{N} \cdot\overline{T}_{iR_Oj}}{\sqrt{\widehat{\mathbb{B}}(\overline{T}_{iR_Oj})}}, ~\text{and}~
\mathbb{T}_{iR_O\text{M}j} = \frac{\sqrt{N} \cdot\overline{T}_{iR_Oj}}{\sqrt{\text{MSE}(\overline{T}_{iR_Oj})}}
\end{equation}
where the subscript notation indicates: penalized method ($i \in \{\text{R, L}\}$), leave-out size ($R_O$), standard error component ($\text{V, B, M}$), and hyperparameter selection ($j \in \{\text{Ad, Av}\}$) where Ad refers to adaptive optimal hyperparameters $\widehat{\boldsymbol{\lambda}}$ and Av is for averaged optimal parameters $\overline{\lambda}$.  Optimal statistical testing procedures control Type I error at the nominal level $\alpha = 0.05$ under $H_0$ ($\xi = 0$) while achieving power approaching unity as $\xi$ increases.}

\begin{figure}[ht]
    \begin{tabular}{cc}
        \includegraphics[width=0.48\linewidth]{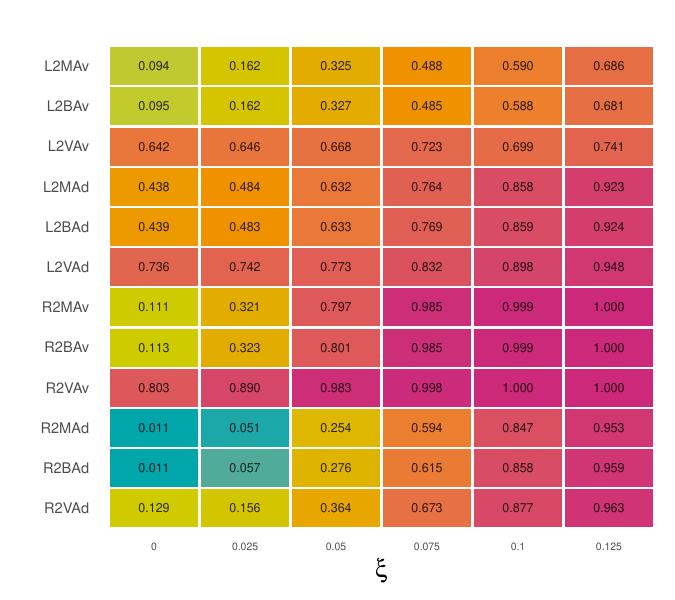}
        & \includegraphics[width=0.48\linewidth]{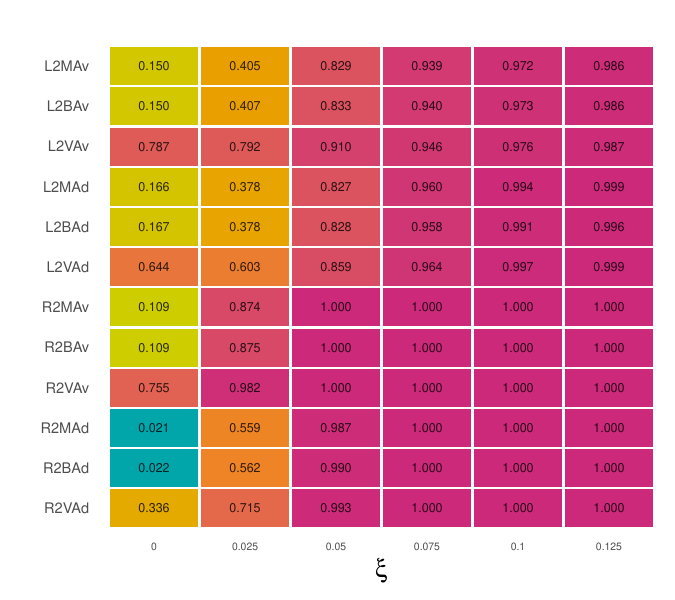}\\
        $N = 50, P = 51$ & $N = 100, P = 101$\\
        \includegraphics[width=0.48\linewidth]{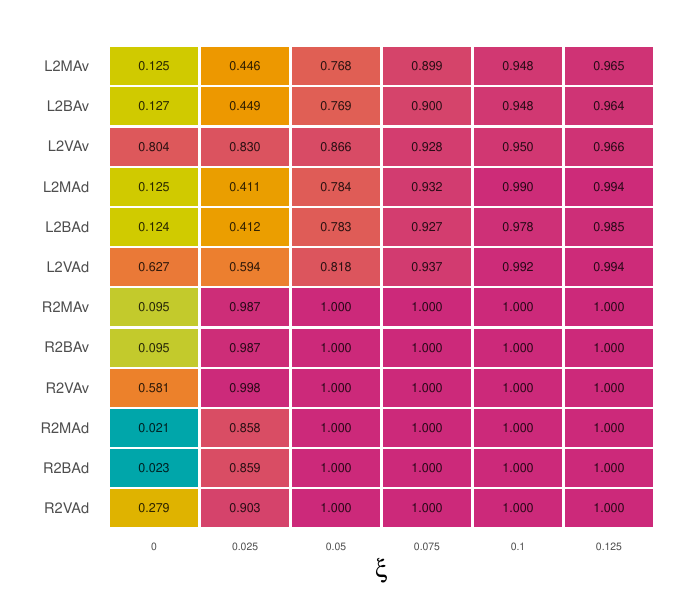}
        & \includegraphics[width=0.48\linewidth]{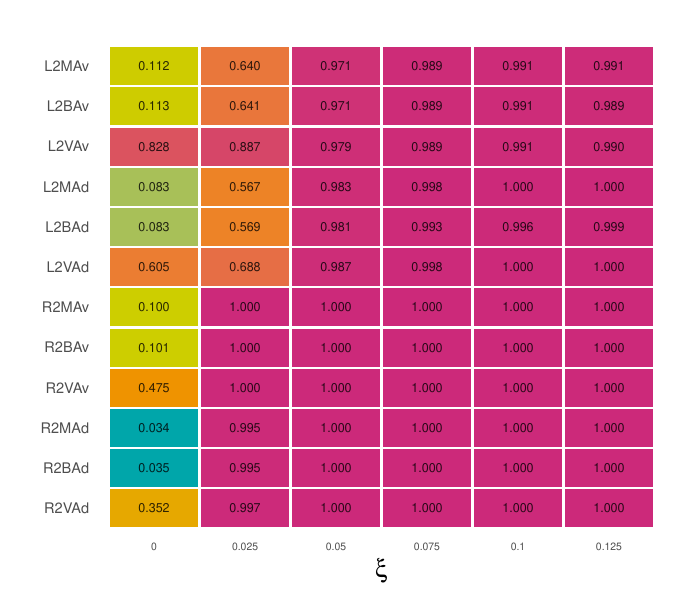}\\
        $N = 150, P = 151$ & $N = 200, P = 201$\\
    \end{tabular}
    \caption{Empirical rejection rates for NL2OCV procedures across varying effect sizes ($\xi$). Each panel represents different sample sizes $N $ with $P = N+1$ parameters. The first column ($\xi$ = 0) shows Type I error rates; subsequent columns ($\xi$ = 0.025 to 0.125) show empirical power.}
    \label{fig:RejectionRate_NL2OCV_1}
\end{figure}

\tr{Rather than presenting all dimensionality configurations, we focus on two representative regimes: equi-dimensional ($P = N + 1$) and ultra-high-dimensional ($P = 100$) settings. Figures \ref{fig:RejectionRate_NL1OCV_1} and \ref{fig:RejectionRate_NL1OCV_100} display the corresponding rejection rates for NL1OCV procedures, while Figures \ref{fig:RejectionRate_NL2OCV_1} and \ref{fig:RejectionRate_NL2OCV_100} show results for NL2OCV procedures.  The rest of the figures are available in the supplementary material.}

% -------------------------------------------------
\subsection{Uncertainty Quantification Approaches in Test Standardization}
\label{sec:UncertaintyQuantification}
% -------------------------------------------------

\begin{figure}[ht]
    \begin{tabular}{cc}
        \includegraphics[width=0.48\linewidth]{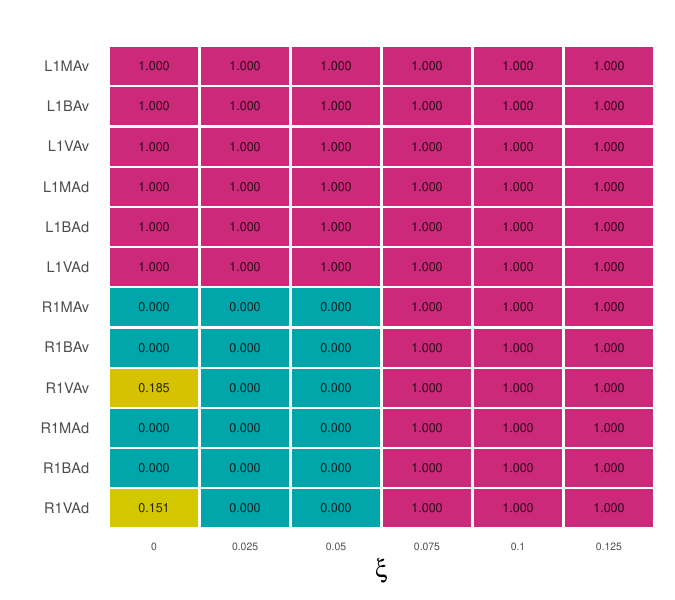}
        & \includegraphics[width=0.48\linewidth]{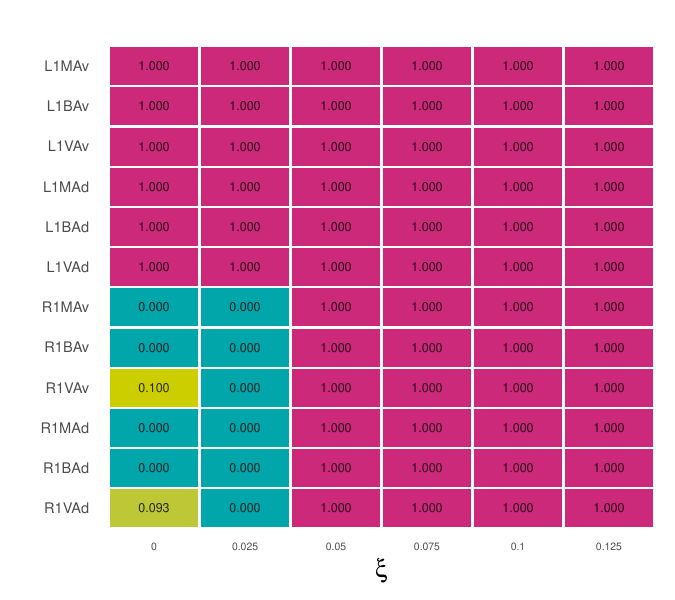}\\
        $N = 50, P = 5000$ & $N = 100, P = 10000$\\
        \includegraphics[width=0.48\linewidth]{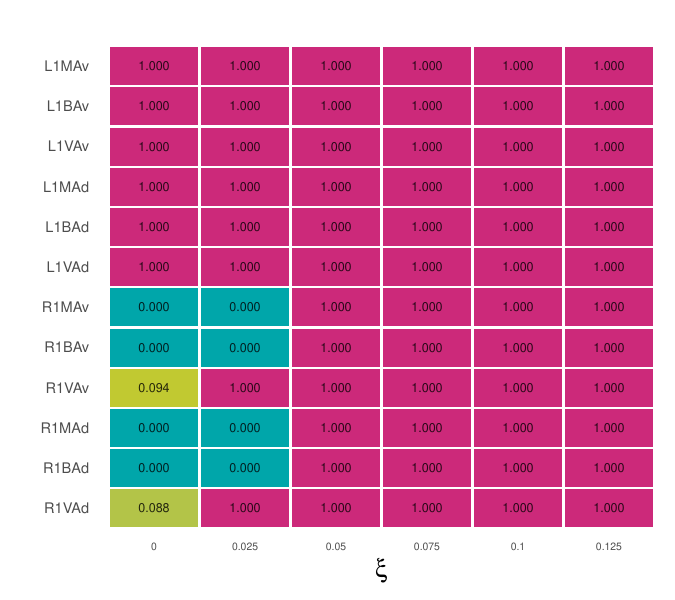}
        & \includegraphics[width=0.48\linewidth]{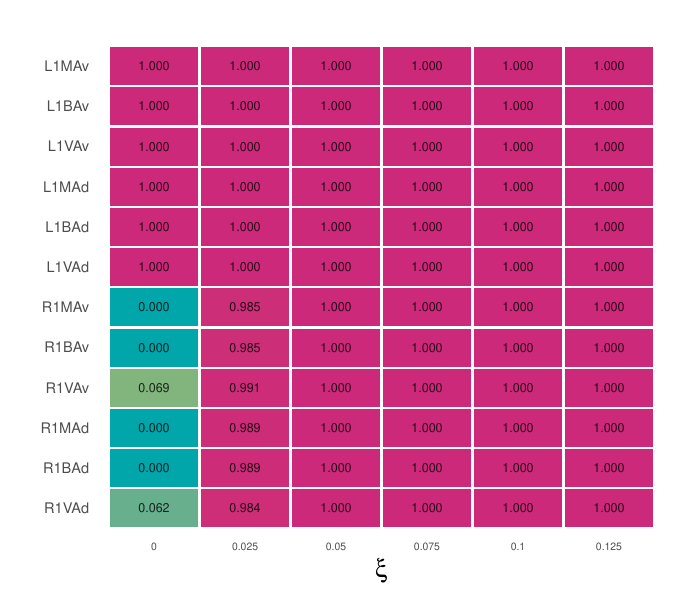}\\
        $N = 150, P = 15000$ & $N = 200, P = 20000$\\
    \end{tabular}
    \caption{Empirical rejection rates for NL1OCV procedures across varying effect sizes ($\xi$). Each panel represents different sample sizes $N $ with $P = 100N$ parameters. The first column ($\xi$ = 0) shows Type I error rates; subsequent columns ($\xi$ = 0.025 to 0.125) show empirical power.}
    \label{fig:RejectionRate_NL1OCV_100}
\end{figure}

\begin{figure}[ht]
    \begin{tabular}{cc}
        \includegraphics[width=0.48\linewidth]{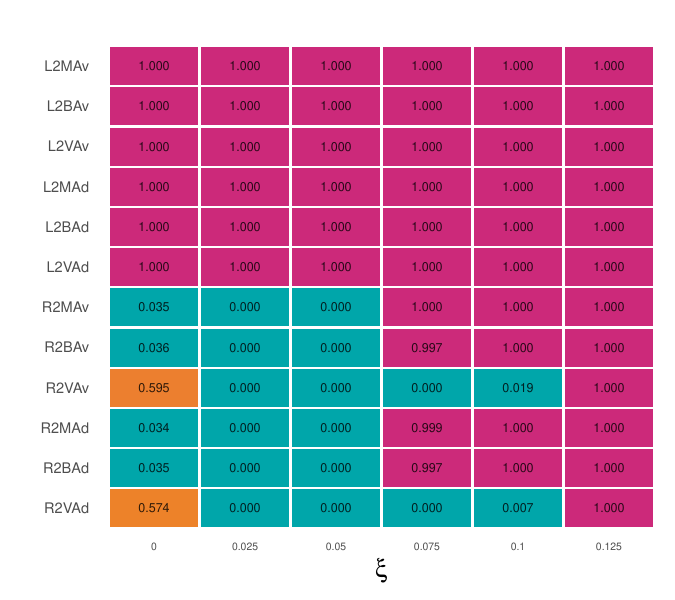}
        & \includegraphics[width=0.48\linewidth]{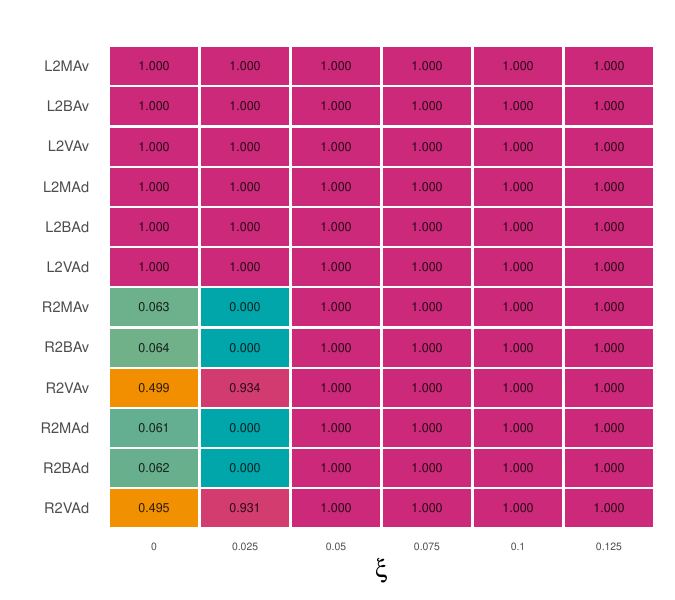}\\
        $N = 50, P = 5000$ & $N = 100, P = 10000$\\
        \includegraphics[width=0.48\linewidth]{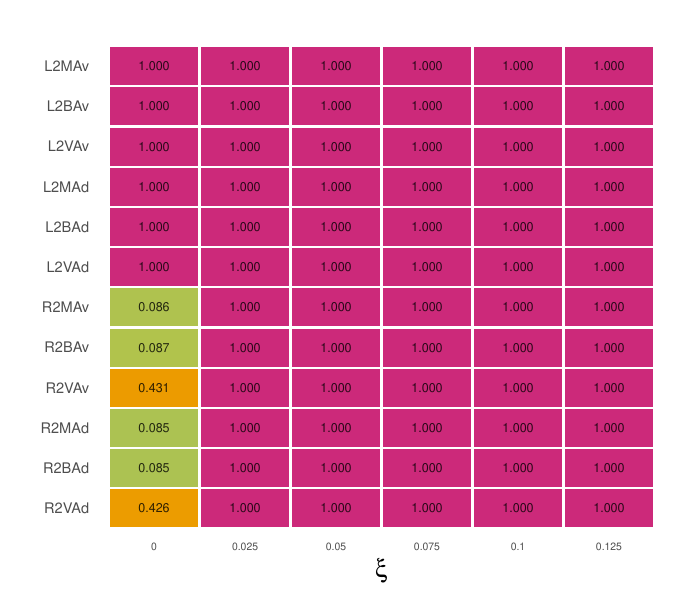}
        & \includegraphics[width=0.48\linewidth]{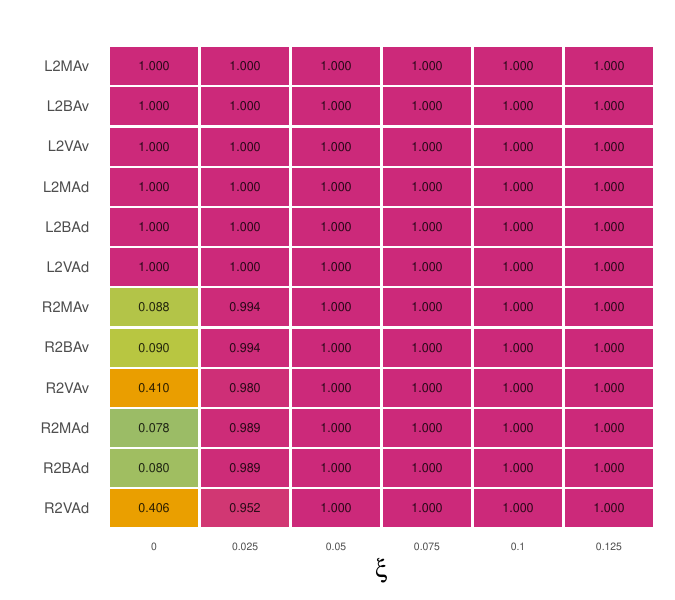}\\
        $N = 150, P = 15000$ & $N = 200, P = 20000$\\
    \end{tabular}
    \caption{Empirical rejection rates for NL2OCV procedures across varying effect sizes ($\xi$). Each panel represents different sample sizes $N $ with $P = 100N$ parameters. The first column ($\xi$ = 0) shows Type I error rates; subsequent columns ($\xi$ = 0.025 to 0.125) show empirical power.}
    \label{fig:RejectionRate_NL2OCV_100}
\end{figure}

\tr{Figures \ref{fig:RejectionRate_NL1OCV_1} to \ref{fig:RejectionRate_NL2OCV_100} reveal patterns in rejection rate behavior that depend critically on the interplay between sample size ($N$), dimensionality ($P$), and effect size ($\xi$).  A striking pattern emerges across all methods: rejection rates increase dramatically as dimensionality grows, even when sample sizes scale proportionally. This dimensional sensitivity suggests that the $P/N$ ratio alone does not fully capture method performance, as procedures become increasingly responsive to departures from the null hypothesis in higher-dimensional settings. When sample size $N$ is large, both Ridge and LASSO procedures exhibit remarkably high power, with proportion of correct rejections approaching $1.00$ even for small effect sizes such as $\xi = 0.025$. This enhanced sensitivity reflects the improved precision available for detecting subtle signals when abundant data are available, demonstrating the fundamental advantage of larger samples in high-dimensional hypothesis testing.}

\tr{Consequently, the uncertainty quantification approaches in the test standardization has profound implications for both Type I error control and power characteristics. Variance-based denominators consistently produce severe Type I error inflation across all dimensional regimes. The core problem is that variance-based denominators systematically underestimate how much the test statistic actually varies in regularized regression settings. This underestimation creates artificially large test statistics, leading to inflated rejection rates even when the null hypothesis is true.  The regularization process introduces additional sources of variability that simple variance estimates fail to capture. This finding has important implications for practitioners who have commonly used variance-based approaches in similar contexts.}

\tr{Conversely, squared bias-based denominators demonstrate superior Type I error control, maintaining rates close to nominal levels across diverse experimental conditions. These denominators provide an appropriate balance between Type I error control and power, making them particularly suitable for reliable statistical inference.  MSE-based denominators, which incorporate both bias and variance components, show similar Type I error control to bias-based approaches in most settings. However, in Figure \ref{fig:RejectionRate_NL1OCV_1} when $P = N + 1$, the test based MSE exhibit slightly lower power than the bias-based test. This reduction in power stems from the slightly larger denominator values that result from including both bias and variance terms, effectively making the test statistic more conservative.}

\tr{Moreover, Ridge and LASSO methods show different rejection rate profiles. Ridge methods demonstrate more conservative behavior, with rejection rates that increase gradually with effect size and show reasonable Type I error control in many configurations when bias-based or MSE-based denominators are employed.  LASSO methods, while achieving exceptional power in high-dimensional settings with rejection rates of 1 across all non-null effect sizes, suffer from severe Type I error inflation in nearly all scenarios examined. Acceptable Type I error control is observed only among some cases of $P = N + 1$ in Figure \ref{fig:RejectionRate_NL1OCV_1}. In all other configurations, LASSO exhibits violations of nominal Type I error rates, making these procedures unsuitable for formal statistical inference in most practical applications.}

\tr{These empirical findings align with some literature demonstrating limitations of LASSO for inference. \citet{lockhart2014significance} demonstrated that LASSO selects variables even under the global null hypothesis ($\boldsymbol{\beta} = \mathbf{0}$), where all associations are purely spurious due to random noise. In addition, \citet{su2017false} proved that true features and null features are always interspersed on the LASSO path, occurring regardless of effect size magnitudes. In their regime of linear sparsity, they derived a sharp asymptotic trade-off showing that achieving adequate power (low false negative rate) necessarily entails substantial false positive rates at any threshold along the LASSO path.}

\tr{Furthermore, the performance of adaptive versus averaged hyperparameter selection strategies shows important dependencies on the cross-validation framework employed. In NL1OCV procedures, averaged hyperparameter selection consistently achieves superior power compared to adaptive selection, particularly in settings where $N \approx P$. This advantage is most pronounced for Ridge methods where the power differences are consistent across effect sizes. However, NL2OCV procedures reveal a critical limitation of averaged hyperparameter selection: consistent Type I error inflation compared to adaptive selection. This inflation affects both Ridge and LASSO methods and occurs across all dimensional regimes examined, making averaged hyperparameter selection unsuitable for NL2OCV frameworks when statistical validity is required. For LASSO specifically, the Type I error inflation under averaged hyperparameter selection is particularly concerning given the theoretical results discussed above. Using a pooled, averaged $\lambda$ value across all train-test splits fails to account for the variability in variable selection behavior across different data partitions. When spurious variables are selected early along the LASSO path \citep{su2017false}, averaged hyperparameter approaches may favor $\lambda$ values that include these false discoveries, since they appear stable across folds despite lacking true signal. Therefore, this mechanism explains why adaptive selection procedures that allow $\lambda$ to vary based on each specific train-test split better control Type I error, even though some inflation persists.}

\begin{figure}[ht]
\scriptsize
    \begin{tabular}{cc}
        R1BAd Type I Error & R1BAd Empirical Power\\
        \includegraphics[width=0.4\linewidth]{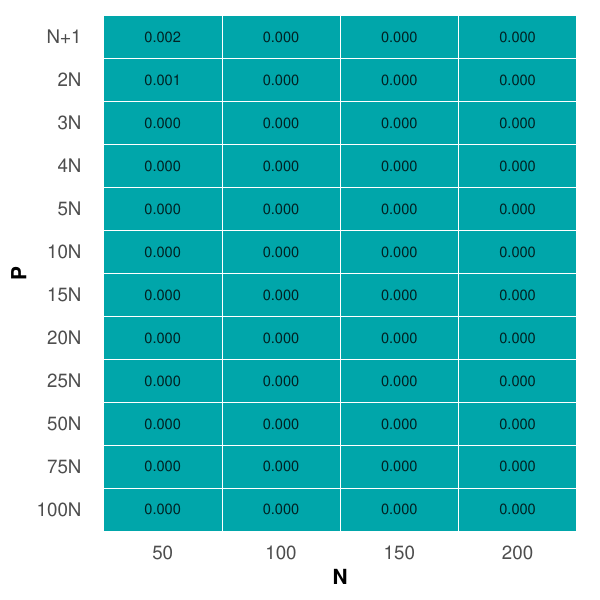}
        & \includegraphics[width=0.4\linewidth]{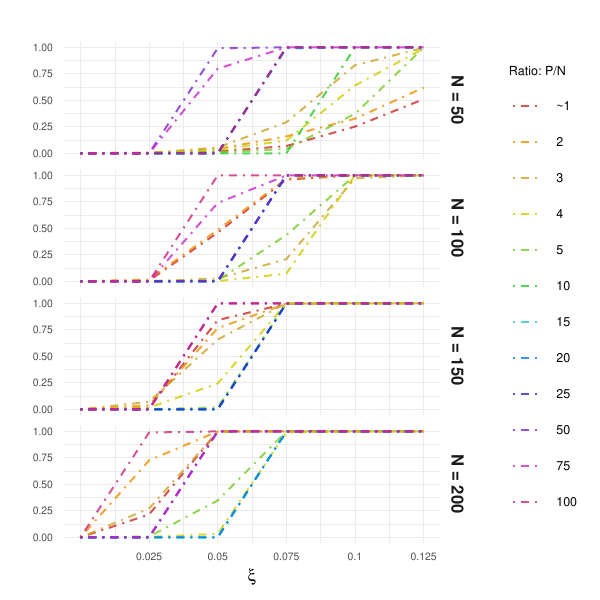}\\
        R1BAv Type I Error & R1BAv Empirical Power\\
        \includegraphics[width=0.4\linewidth]{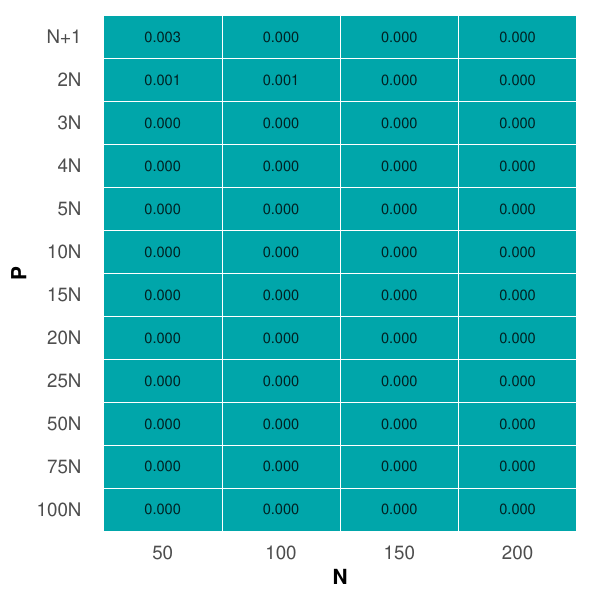}
        & \includegraphics[width=0.4\linewidth]{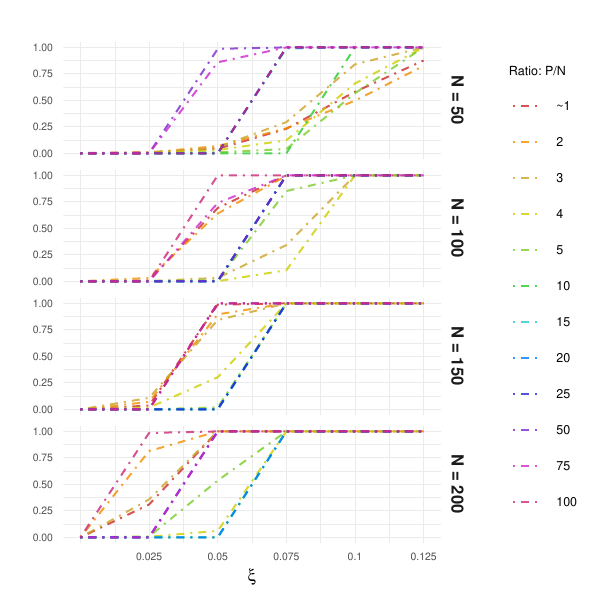}\\
    \end{tabular}
    \caption{Type I Error Rates and Empirical Power for Ridge Regression with Nested Leave-one-out Cross-Validation using Bias-Based Test Statistics}
    \label{fig:NL1OCV_OptimalTest}
\end{figure}

% -------------------------------------------------
\subsection{Optimal Testing Procedure based on NL1OCV and NL2OCV}
% -------------------------------------------------

\tr{Figure \ref{fig:NL1OCV_OptimalTest} demonstrates the superior performance achievable with bias-based test statistics in NL1OCV procedures. Both adaptive and averaged hyperparameter selection strategies can be employed effectively in the NL1OCV framework, with averaged selection showing superior power in $N \approx P$ settings. The power curves demonstrate adequate detection capability across effect sizes while preserving statistical validity, representing the optimal balance for practical applications.}

\begin{figure}[ht]
\scriptsize
    \begin{tabular}{cc}
        R2BAd Type I Error & R2BAd Empirical Power\\
        \includegraphics[width=0.4\linewidth]{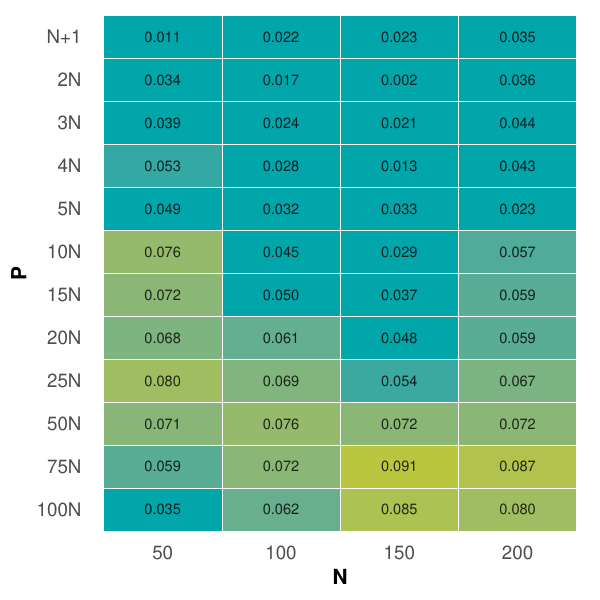}
        & \includegraphics[width=0.4\linewidth]{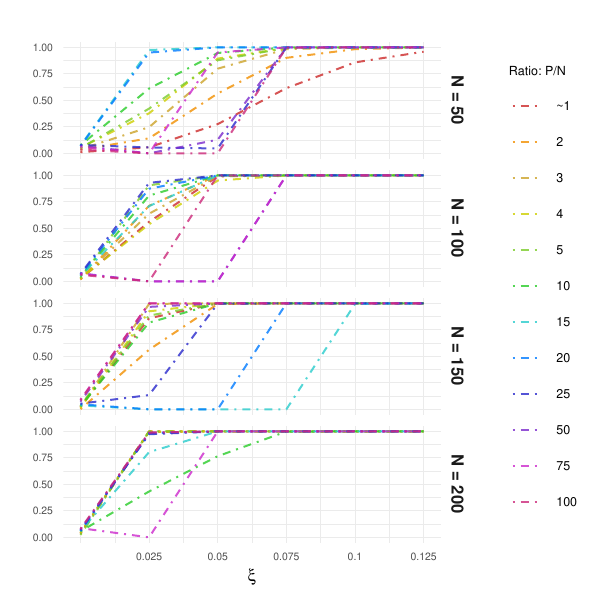}\\
        R2MAd Type I Error & R2MAd Empirical Power\\
        \includegraphics[width=0.4\linewidth]{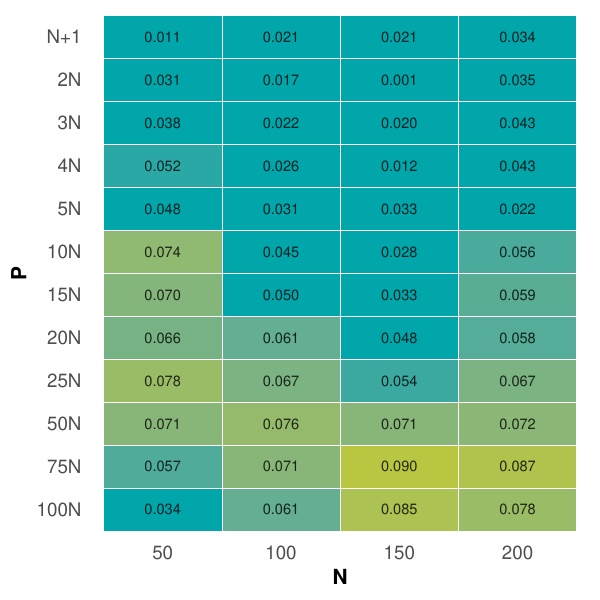}
        & \includegraphics[width=0.4\linewidth]{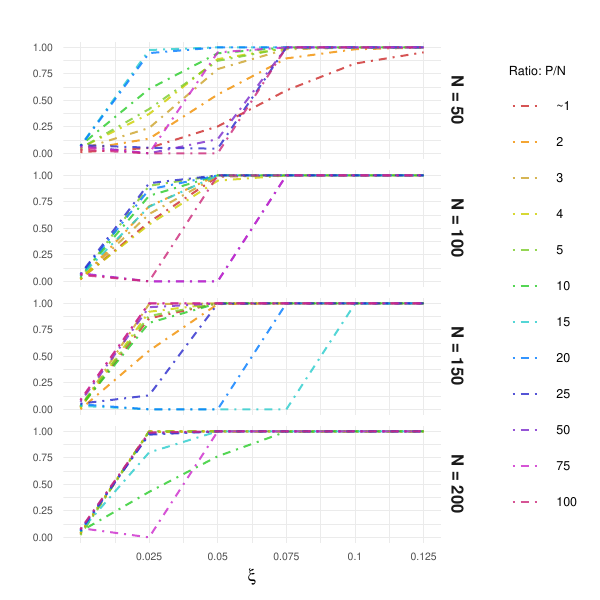}\\
    \end{tabular}
    \caption{Type I Error Rates and Empirical Power for Ridge Regression with Nested Leave-two-out Cross-Validation using Bias-Based Test Statistics}
    \label{fig:NL2OCV_OptimalTest}
\end{figure}

\tr{Notably, the comparison between bias-based and MSE-based denominators within this framework reveals that while both maintain appropriate Type I error control, MSE-based approaches exhibit slightly reduced power, particularly in high-dimensional settings. This power reduction reflects the larger denominator values that result from including variance components alongside bias terms. In scenarios where maximum power is desired while maintaining statistical control, bias-based test statistics emerge as the preferred choice.}

\tr{Figure \ref{fig:NL2OCV_OptimalTest} reveals important distinctions between NL1OCV and NL2OCV frameworks. Unlike NL1OCV procedures where bias-based and MSE-based denominators show meaningful differences in power characteristics, NL2OCV procedures exhibit minimal differences between these two approaches.  This convergence in performance suggests that the additional variability introduced by the NL2OCV  framework may mask the subtle differences between bias-based and MSE-based scaling that are evident in NL1OCV procedures. However, the critical finding is that adaptive hyperparameter selection definitively outperforms averaged selection in the NL2OCV framework.}

\begin{table}[!ht]
\scriptsize
    \centering
    \begin{tabular}{|c|c|c|c|c|c|c|}
    \hline
         & \multicolumn{3}{c}{Computational Time Naive Method (ms)} & \multicolumn{3}{|c|}{Computational Time Proposed Method (ms)} \\ \hline
        $P$ & Median & Min & Max & Median & Min & Max \\ \hline
        201 & 148421.8282 & 148389.0196 & 148454.6368 & 1329.836638 & 925.160833 & 1734.512442 \\ \hline
        400 & 148092.2626 & 147397.6615 & 148786.8638 & 1168.641546 & 1158.531953 & 1178.751139 \\ \hline
        600 & 146441.5275 & 146358.3871 & 146524.6679 & 1255.773528 & 1251.398478 & 1260.148578 \\ \hline
        800 & 146484.1115 & 146482.6008 & 146485.6222 & 1400.328740 & 1377.101075 & 1423.556404 \\ \hline
        1000 & 146615.7335 & 146606.8288 & 146624.6383 & 1560.522754 & 1543.166155 & 1577.879353 \\ \hline
        2000 & 150056.8600 & 150050.3456 & 150063.3743 & 3515.730554 & 3492.938804 & 3538.522303 \\ \hline
        3000 & 149412.1293 & 149204.4119 & 149619.8466 & 3816.778883 & 3794.942402 & 3838.615363 \\ \hline
        4000 & 152319.1055 & 152127.9229 & 152510.2881 & 7097.903168 & 6874.821894 & 7320.984442 \\ \hline
        5000 & 150846.9265 & 150788.7097 & 150905.1433 & 5626.055879 & 5609.996457 & 5642.115300 \\ \hline
        10000 & 159688.6952 & 156545.5276 & 162831.8628 & 11173.18168 & 11107.46783 & 11238.89554 \\ \hline
        15000 & 186762.0351 & 184629.3115 & 188894.7588 & 25954.79571 & 25871.98604 & 26037.60538 \\ \hline
        20000 & 172017.4862 & 168807.3631 & 175227.6092 & 25169.45274 & 24824.47188 & 25514.43359 \\ \hline
    \end{tabular}
\caption{Comparison of the Computational Time (in milliseconds) using the Proposed Computationally Efficient Method versus the Naive Implementation when $N = 200$}
\label{tab:ComputationalTime}
\end{table}

\tr{The exclusive use of adaptive hyperparameters in the proposed NL2OCV methodology ensures reliable Type I error control, addressing the inflation observed with averaged approaches. This represents a crucial methodological consideration for practitioners choosing between cross-validation frameworks. Adaptive selection achieves better Type I error control by allowing $\lambda$ to vary based on each specific train-test split, which accommodates the data-dependent nature of variable selection that averaged approaches cannot capture through a pooled, averaged $\lambda$ value.}

\begin{table}[ht]
\centering
\small
\begin{tabular}{|l|p{10cm}|}
\hline
\textbf{Method} & \textbf{Type I Error Control Key Points}\\ \hline
Variance-based & 
Severe Type I error inflation across all settings; 
Caused by underestimation of test statistic variability\\ \hline
Bias-based & 
Consistent Type I error at nominal $\alpha = 0.05$; Robust across all $P/N$ ratios investigated; Works for both NL1OCV and NL2OCV frameworks
\\ \hline
MSE-based & 
Consistent Type I error at nominal $\alpha = 0.05$; Reduced power vs. bias-based approaches
\\ \hline\hline
\textbf{Feature} & \textbf{Empirical Power Key Points}\\
\hline
Dimensionality & 
Rejection rates increase dramatically with dimensionality; $P/N$ ratio alone insufficient to predict performance
\\ \hline
Sample size & 
Large $N$ yields power $\approx$ 1.00 for small effects ($\xi = 0.025$); Applies to both Ridge and LASSO procedures
\\ \hline
Bias vs. MSE & 
NL1OCV: Bias-based shows superior power; NL2OCV: Minimal difference between approaches
\\ \hline\hline
\textbf{Framework} & \textbf{Hyperparameter Selection Key Points}\\
\hline
NL1OCV & 
Both adaptive and averaged are effective; Averaged superior when $N \approx P$
\\ \hline
NL2OCV & 
Adaptive definitively outperforms averaged; Critical for maintaining Type I error control
\\ \hline\hline
\textbf{Framework} & \textbf{Recommended Procedures}\\
\hline
NL1OCV & 
\textbf{R1BAv} or \textbf{R1BAd} (Ridge with bias-based denominator); Choice depends on dimensionality regime
\\ \hline
NL2OCV & 
\textbf{R2BAd} or \textbf{R2MAd} (Ridge with adaptive hyperparameters); Bias-based preferred for maximum power
\\ \hline
\end{tabular}
\caption{Summary of Key Findings from Numerical Experiments}
\label{tab:ResultsSummary}
\end{table}

\tr{To empirically validate the computational efficiency of our proposed method, we conducted timing experiments comparing the theoretically-derived efficient implementation of both NL1OCV and NL2OCV from Section \ref{sec:MainResults} against a naive algorithmic implementation across various problem dimensions. Table \ref{tab:ComputationalTime} presents results for a realistic scenario with $N = 200$ observations and $P$ ranging from 201 to 20,000 predictors. Unit testing confirmed that both implementations yield identical results, with computational time being the sole difference. The proposed method demonstrates substantial computational advantages, achieving speedups ranging from approximately 100-fold at moderate dimensions to 7-fold at $P$ = 20,000. While the naive implementation requires a relatively constant computational time around 150 seconds regardless of dimension, our efficient approach scales approximately linearly with $P$ but maintains dramatically lower runtime across all scenarios, requiring only 25 seconds even at the highest dimension tested.}

\tr{Table \ref{tab:ResultsSummary} synthesizes the key findings from our comprehensive numerical experiments. Overall, our results strongly recommend Ridge-based methods with bias denominators and adaptive hyperparameter selection as the most reliable approach for high-dimensional predictive performance testing. As discussed in the preceding sections, Ridge demonstrates superior Type I error control compared to LASSO, which suffers from severe inflation due to the interspersed nature of true and null features along its solution path. Adaptive hyperparameter selection is preferred over averaged approaches because it allows $\lambda$ to vary based on each specific train-test split, accounting for the data-dependent nature of variable selection across different data partitions. Finally, bias-based denominators provide the optimal balance of statistical validity and power, while variance-based alternatives exhibit Type I error inflation that renders them unsuitable for inference.}

% -------------------------------------------------
\section{Analysis of RNA Sequencing Data}
\label{sec:RealDataAnalysis}
% -------------------------------------------------

\tr{We analyzed data from the Aging, Dementia, and Traumatic Brain Injury Study, a collaborative effort between the University of Washington, Kaiser Permanente Washington Health Research Institute, and the Allen Institute for Brain Science (https://aging.brain-map.org). This comprehensive dataset includes mRNA expression patterns, Luminex protein quantifications, and immunohistochemistry pathology metrics from $N = 44$ subjects with a documented history of traumatic brain injury (TBI), drawn from the population-based Adult Changes in Thought (ACT) cohort.}

\tr{The transcriptomic data were generated through RNA sequencing of tissue samples isolated from four distinct brain regions: (1) white matter underlying the parietal neocortex (FWM), (2) hippocampus (HIP), (3) neocortex from the posterior superior temporal gyrus (TCx), and (4) the inferior parietal lobule (PCx). The comprehensive mRNA expression profiles ($P = 17,574$ genes) served as predictive features, while protein-level measurements served as response variables. These protein biomarkers included tau and phosphorylated tau variants, A$\beta$ species, $\alpha$-synuclein, inflammatory mediators, neurotrophic factors, and immunohistochemistry pathology metrics. This configuration represents a high-dimensional supervised learning problem, with $N \ll P$.}

% -------------------------------------------------
\subsection{Regional Patterns of Transcriptomic Regulation}
% -------------------------------------------------

\begin{figure}[ht]
    \centering
    \begin{tabular}{cc}
        HIP & PCx\\
        \includegraphics[width=0.5\linewidth]{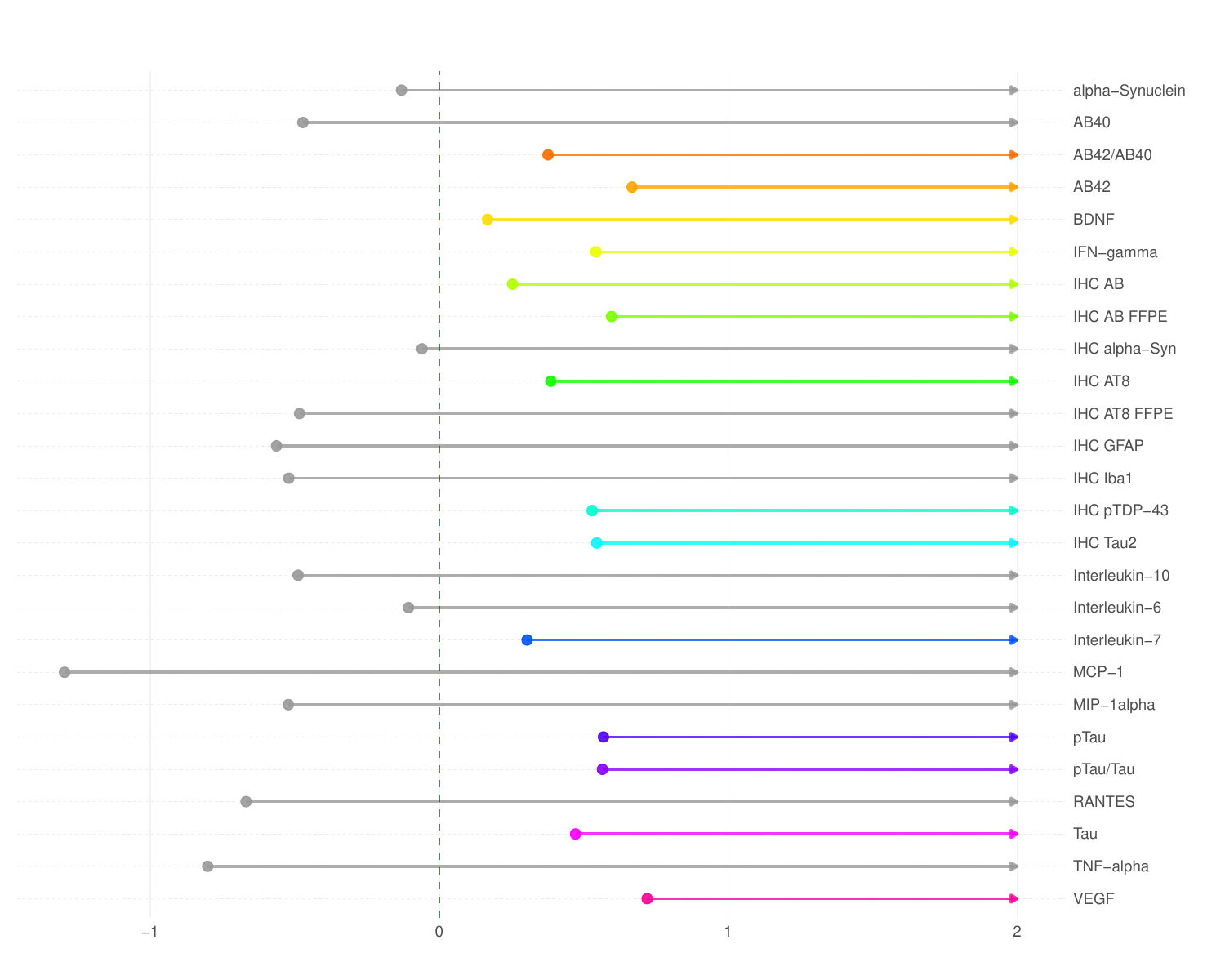} & \includegraphics[width=0.5\linewidth]{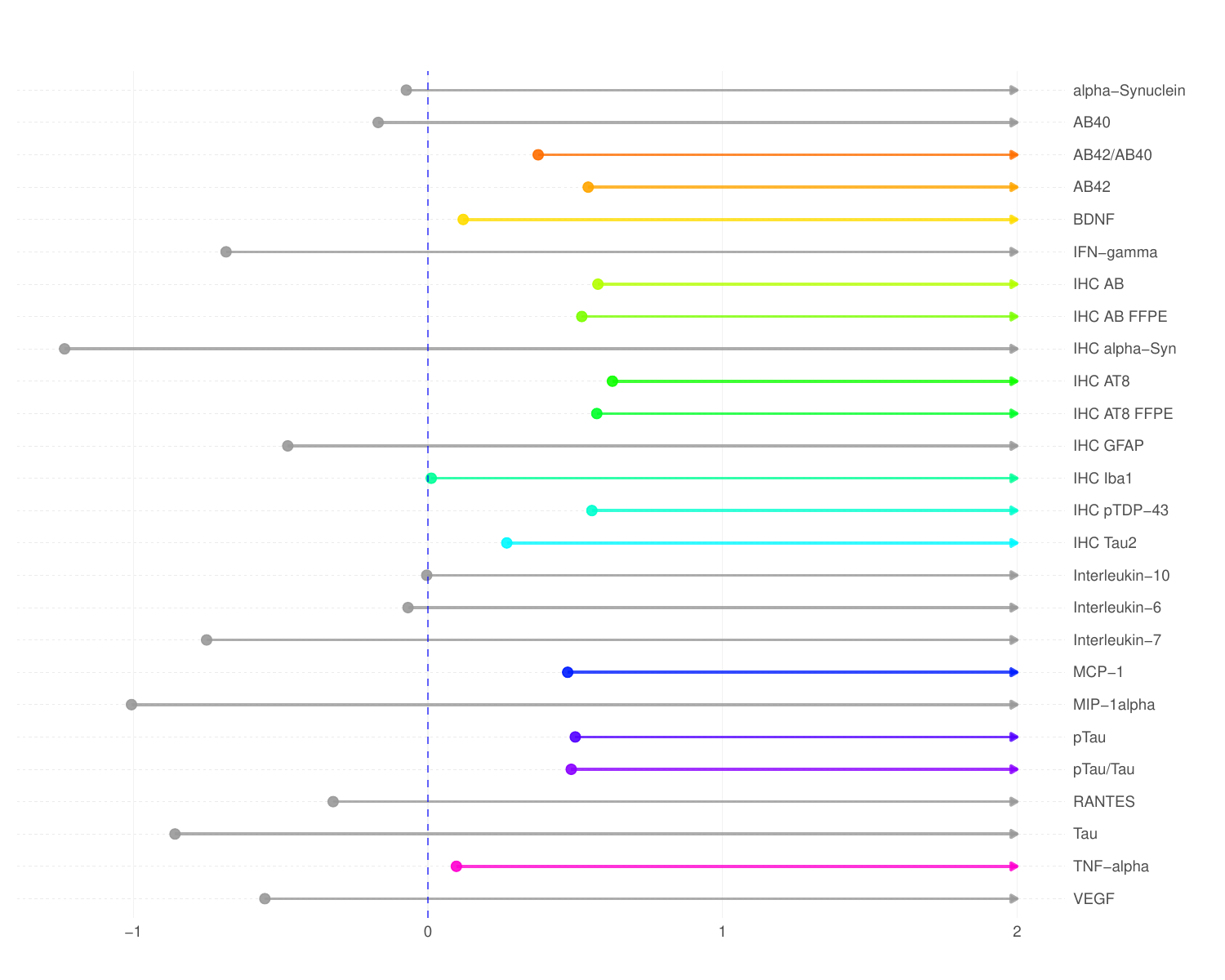} \\
        FWM & TCx\\
        \includegraphics[width=0.5\linewidth]{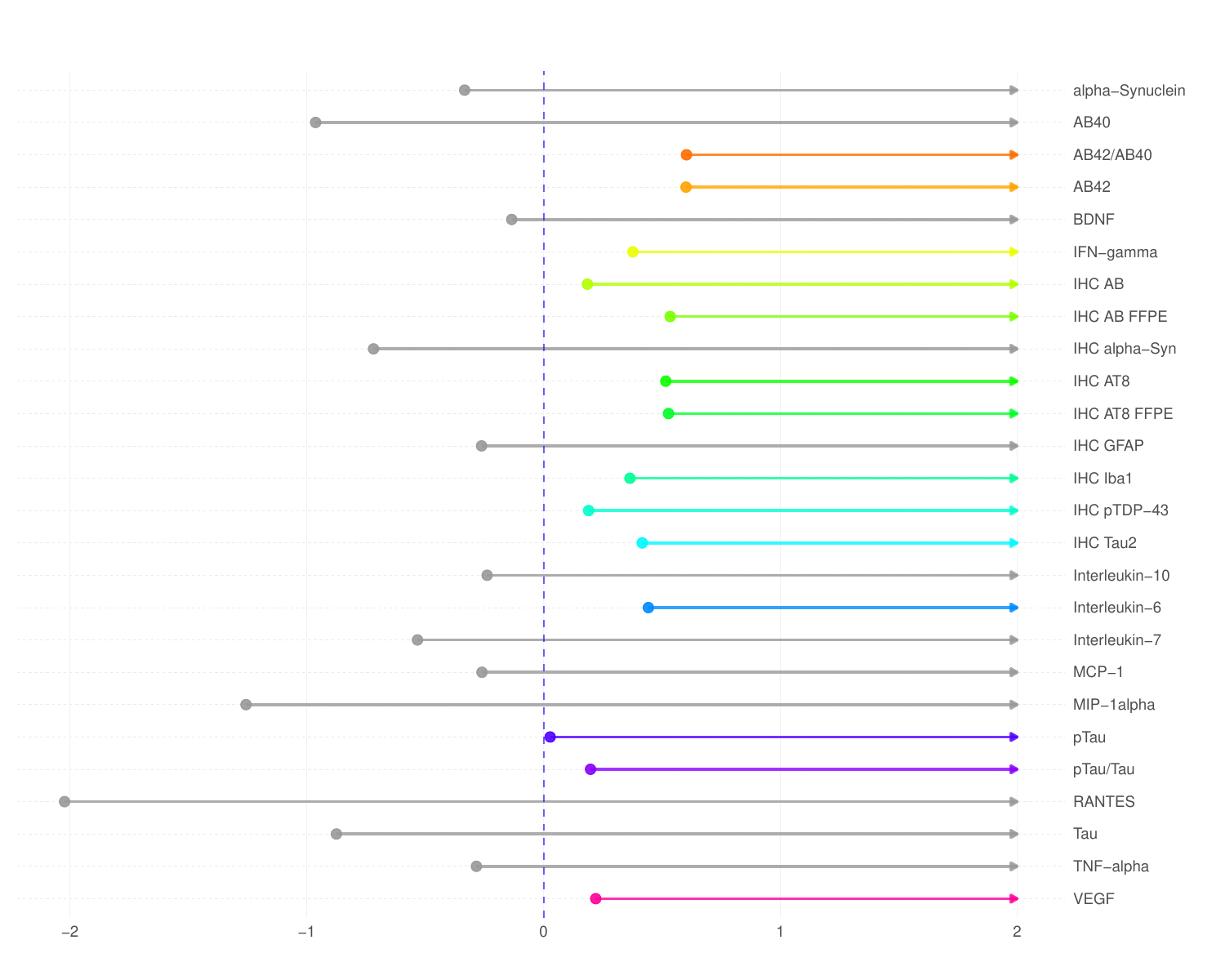} &
        \includegraphics[width=0.5\linewidth]{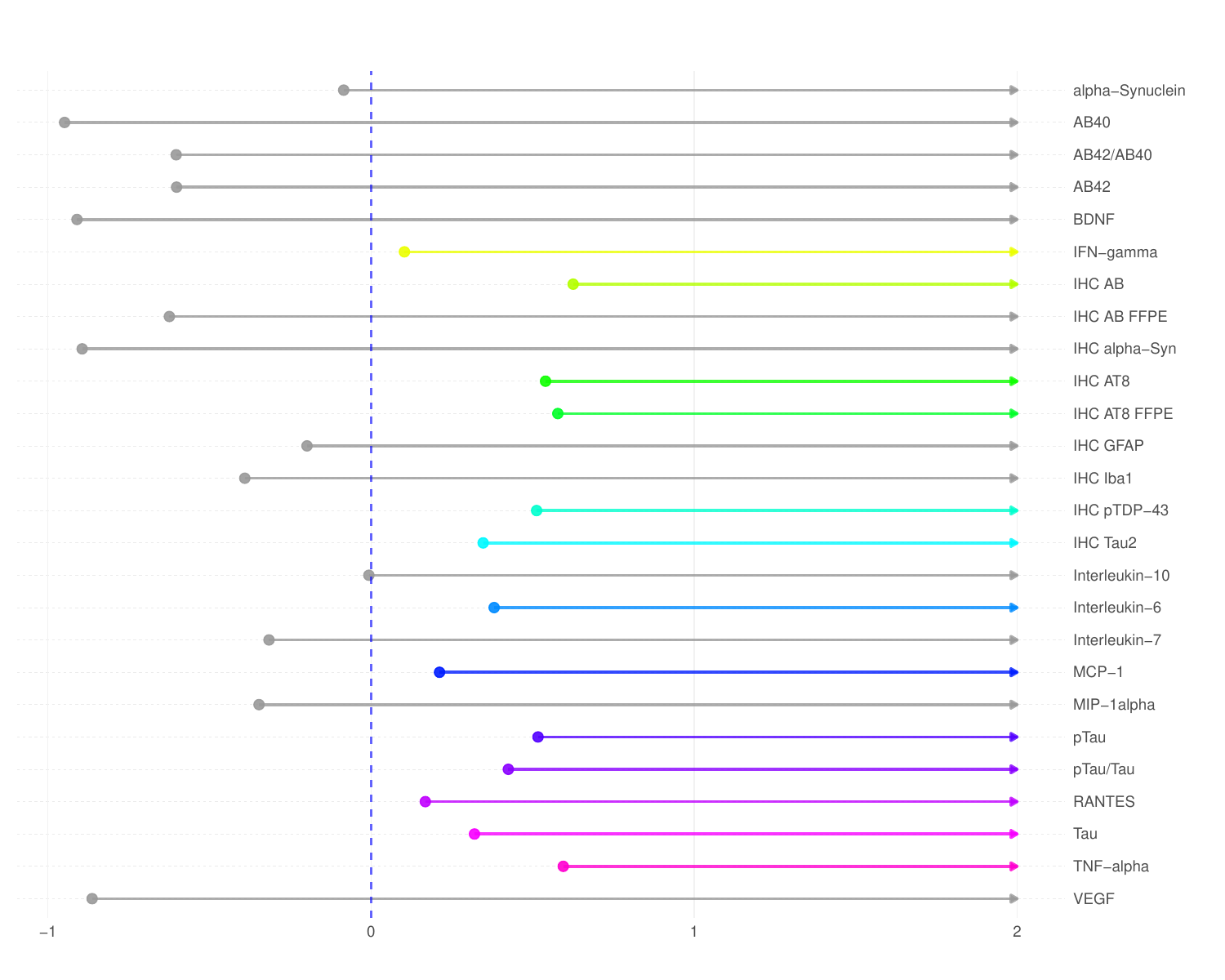} \\
    \end{tabular}
    \caption{Lower bound confidence interval estimates for prediction error improvement when using mRNA expression to predict protein biomarkers across four brain regions. Positive values indicate that transcriptomic data improves prediction accuracy over baseline models. Results shown for white matter underlying parietal neocortex (FWM), hippocampus (HIP), posterior superior temporal gyrus neocortex (TCx), and inferior parietal lobule (PCx).}
    \label{fig:SignificantProteins}
\end{figure}

\tr{In the analysis, we implemented our proposed method described in Section \ref{sec:NumericalExperiments}: Nested L1OCV method with squared difference between inner and outer CV performance to estimate prediction error variability. Both inner and outer CV estimates used averaged hyperparameter values $\overline{\lambda}$. This approach revealed that the mRNA expression profiles showed region-specific patterns of predictive improvement across protein biomarkers, with some proteins demonstrating consistent transcriptomic relationships across multiple brain regions while others showed region-specific associations.  Analysis of Figure \ref{fig:SignificantProteins} reveals distinct regional patterns in the predictive relationship between mRNA expression and protein biomarkers:}

\tr{\textbf{Proteins Significant Across All Four Regions:} Six biomarkers showed consistent significant predictive improvement across all brain regions: IHC A$\beta$, IHC AT8, IHC pTDP-43, IHC Tau2, pTau, and pTau/Tau ratio. These represent transcriptomic regulatory mechanisms with consistent relationships to core neurodegenerative processes across diverse brain regions. The predominance of tau-related biomarkers reflects the fundamental role of tau phosphorylation in neurodegeneration, where pTau-181 represents tau abnormally phosphorylated at threonine-181, playing a critical role in tau aggregation and neurofibrillary tangle formation \citep{karikari2020blood}. The AT8 immunohistochemistry identifies both immature and mature neurofibrillary tangles formed of tau phosphorylated at serine-202 and threonine-205, also detecting dystrophic neurites and glial tangles \citep{goedert1995monoclonal}.}

\tr{\textbf{Proteins Significant in Three Regions:} Five biomarkers demonstrated significance in three of four regions: 
A$\beta$42 (FWM, HIP, PCx), A$\beta$42/A$\beta$40 ratio (FWM, HIP, PCx), IFN-$\gamma$ (HIP, FWM, TCx), IHC A$\beta$ FFPE (PCx, FWM, HIP), and IHC AT8 FFPE (FWM, PCx, TCx). The amyloid markers A$\beta$42 and A$\beta$40 represent beta-amyloid peptides derived from amyloid precursor protein, with the 42-amino acid form being more amyloidogenic and prominently found in amyloid plaques characteristic of Alzheimer's disease \citep{iwatsubo1994visualization}. The IFN-$\gamma$ pattern is particularly notable, showing significance in hippocampus and both white matter regions while being non-significant in parietal cortex. As a prototype pro-inflammatory cytokine critical for innate and adaptive immunity, IFN-gamma activates macrophages and induces MHC class II expression \citep{schroder2004interferon}, suggesting region-specific inflammatory responses in post-traumatic neurodegeneration.}

\tr{\textbf{Proteins Significant in Two Regions:} Seven biomarkers showed paired regional associations: BDNF (PCx, HIP), IHC Iba1 (PCx, FWM), Interleukin-6 (FWM, TCx), MCP-1 (TCx, PCx), total Tau (HIP, TCx), TNF-$\alpha$ (TCx, PCx), and VEGF (FWM, HIP). 
TNF-$\alpha$, a pro-inflammatory cytokine produced primarily by macrophages, regulates inflammatory responses particularly in acute phases \citep{locksley2001tnf}. MCP-1 (CCL2) displays chemotactic activity for monocytes to inflammation sites \citep{deshmane2009monocyte}, while IL-6 serves as a pro-inflammatory cytokine important in acute inflammatory responses and B-cell maturation \citep{tanaka20146}. The neuroplasticity markers BDNF (brain-derived neurotrophic factor) promotes neuronal growth, differentiation, and synapse formation \citep{huang2001neurotrophins}, while VEGF stimulates vasculogenesis and angiogenesis, suggesting region-specific roles in neural repair and vascular remodeling.}

\tr{\textbf{Region-Specific Proteins:} Two biomarkers demonstrated unique regional relationships: RANTES (TCx only) and Interleukin-7 (HIP only). RANTES (CCL5) is a chemokine that recruits immune cells to sites of inflammation and can cross the blood-brain barrier \citep{appay2001rantes}. Its temporal cortex-specific significance suggests region-specific inflammatory recruitment patterns in post-traumatic neurodegeneration. Interleukin-7 is essential for T cell development and survival, and its hippocampus-specific pattern may reflect unique immune regulation requirements in this region critical for learning and memory \citep{lawson2015interleukin}.}

\tr{\textbf{Non-Significant Proteins:} Six biomarkers showed no predictive improvement in any region: MIP-1 $\alpha$, Interleukin-10, IHC GFAP, IHC alpha-Syn, A$\beta$40, and $\alpha$-Synuclein. Notably, IL-10 functions as an anti-inflammatory cytokine that downregulates Th1 cytokines and macrophage activation \citep{couper200810}, while GFAP identifies activated astrocytes, and alpha-synuclein represents a presynaptic terminal protein and major component of Lewy bodies \citep{stefanis2012alpha}. The lack of predictive relationships suggests these proteins may be regulated through non-transcriptomic pathways in chronic post-traumatic states.}

% -------------------------------------------------
\subsection{Clinical Implications}
% -------------------------------------------------

\tr{The identification of transcriptomically-regulated protein biomarkers in chronic TBI presents several translational opportunities for improving patient care and advancing therapeutic development. These findings support the development of multi-modal diagnostic approaches combining transcriptomic and proteomic profiling. The strong predictive relationships between mRNA expression and key neurodegeneration markers suggest that transcriptomic assessment could serve as a complementary diagnostic tool, particularly in cases where protein biomarker detection faces technical limitations or accessibility constraints.}

\tr{The demonstration of transcriptomic predictability for specific biomarkers identifies novel therapeutic entry points. RNA-based interventions, including antisense oligonucleotides targeting tau phosphorylation pathways, represent promising approaches given the universal transcriptomic regulation of tau-related markers across brain regions. Recent clinical trials have demonstrated that tau-targeting antisense oligonucleotides can achieve significant reductions in cerebrospinal fluid tau levels, with sustained effects lasting weeks to months after administration \citep{mummery2023tau}. Similarly, the predictable inflammatory mediator patterns suggest that precision immunomodulatory therapies could be developed based on individual transcriptomic profiles. For clinical trial design, patients exhibiting transcriptomically-predictable biomarker profiles may represent enriched populations more likely to respond to interventions targeting specific molecular pathways. This stratification approach could improve trial efficiency and success rates while advancing personalized medicine in TBI care \citep{papa2008use}.}

\tr{The predictive relationships between gene expression and protein biomarkers enable development of dynamic monitoring strategies. Serial transcriptomic assessment could provide earlier indicators of treatment response or disease progression compared to protein biomarkers alone, which may lag behind transcriptional changes. Transcriptomic monitoring has shown promise in other clinical contexts for tracking disease progression and treatment response, with studies demonstrating that mRNA expression changes can precede protein-level alterations \citep{casamassimi2017transcriptome}. Furthermore, the region-specific transcriptomic signatures may inform prognosis and guide rehabilitation planning. Additionally, the development of point-of-care transcriptomic testing platforms could enhance accessibility and clinical utility, particularly in resource-limited settings. Recent advances in point-of-care molecular diagnostics, accelerated by the COVID-19 pandemic, have demonstrated the feasibility of near-patient transcriptomic testing using simplified PCR platforms \citep{wang2018current}.}

% -------------------------------------------------
\section{Conclusions and Future Work}
\label{sec:Conclusions}
% -------------------------------------------------

\tr{In this study, we introduce a novel framework for high-dimensional predictive performance testing that addresses key computational and methodological challenges in cross-validation. We developed a computationally feasible exhaustive cross-validation method that considers all possible data divisions, eliminating partition dependency and reproducibility issues that plague traditional approaches. We incorporated a nested structure within this exhaustive framework, with hyperparameter optimization conducted in the inner stage and prediction error estimation in the outer stage, ensuring proper separation of model selection and assessment. This comprehensive approach provides the statistical rigor of exhaustive enumeration while remaining computationally tractable for high-dimensional datasets.}

\tr{Based on our comprehensive analysis, we provide specific guidance for practitioners working with regularized regression in high-dimensional settings. We strongly recommend avoiding variance-based denominators entirely, as they produce inflated Type I error rates across most scenarios examined due to unreliable variance estimation. Instead, practitioners should prefer bias-based test statistics for optimal balance of Type I error control and power, particularly in high-dimensional applications where detecting genuine signal is paramount. When more conservative inference is desired, MSE-based denominators should be used, accepting reduced power in exchange for additional statistical conservatism. Regarding hyperparameter selection strategies, for Ridge NL1OCV procedures, averaged hyperparameter selection may be employed when $N \approx P$ to maximize power, while adaptive selection remains suitable across all dimensional regimes. However, for NL2OCV procedures, adaptive hyperparameter selection should be used exclusively due to consistent Type I error inflation with averaged approaches. Additionally, practitioners should exercise extreme caution with LASSO-based procedures, as they exhibit acceptable statistical properties only in very limited circumstances and are unsuitable for formal inference in most practical applications. Instead, Ridge-based approaches should be favored for reliable statistical inference, as they maintain superior statistical properties across dimensional regimes when combined with appropriate estimator for prediction error variability and hyperparameter strategies.}

\tr{Our research points toward several critical areas for future investigation, including validation of transcriptomic biomarkers in larger traumatic brain injury cohorts, development of clinical blood-based assays, and exploration of therapeutic interventions targeting identified transcriptomic pathways. We emphasize the potential for integrating transcriptomic markers with neuroimaging and cognitive assessments to advance precision medicine approaches in TBI care. The convergence of transcriptomic regulation with established neurodegeneration biomarkers provides a foundation for precision medicine approaches in TBI care, offering hope for the millions of individuals worldwide living with long-term consequences of traumatic brain injury. This methodological framework opens new possibilities for reliable statistical inference in high-dimensional biomarker discovery, with implications extending far beyond TBI research to any domain requiring robust predictive performance testing in complex, high-dimensional data environments.}

% -------------------------------------------------
\appendix
% -------------------------------------------------

% -------------------------------------------------
\section{Proof of Lemma 1:}
\label{sec:ProofLemma1}
% -------------------------------------------------

\tr{From Section \ref{sec:MainResults}, we define the index sets $\mathcal{T}_{0\ell} = \{\ell(h)\}_{h=1}^{N_0}$ and $\mathcal{T}_{1\ell} = \{\ell(h)\}_{h=N_0+1}^{N}$, $\ell \in [L]$, as the collection of sampling units assigned to the testing and training data, respectively. For $g \in \{0, 1\}$, the partitioned data $\mathbf{D}_{\mathcal{T}_{g\ell}} = (\mathbf{X}_{\mathcal{T}_{g\ell}}, \mathbf{y}_{\mathcal{T}_{g\ell}})$ can be formulated as $\mathbf{X}_{\mathcal{T}_{g\ell}} = [\mathbf{E}_{\mathcal{T}_{g\ell}}]^\top \mathbf{X}$ and $\mathbf{y}_{\mathcal{T}_{g\ell}} = [\mathbf{E}_{\mathcal{T}_{g\ell}}]^\top \mathbf{y}$, where $\mathbf{E}_{\mathcal{T}_{g\ell}}$ is the $N \times N_g$ matrix whose columns correspond to elementary vectors.  The selection matrices are constructed as follows:
\begin{align}
\label{eqn:E_matrix}
\mathbf{E}_{\mathcal{T}_{0\ell}} &= (\mathbf{e}_{\ell(1)}, \mathbf{e}_{\ell(2)}, \ldots, \mathbf{e}_{\ell(N_0)}), \quad \ell(h) \in \mathcal{T}_{0\ell}, \; h \in \{1, 2, \ldots, N_0\} \\
\mathbf{E}_{\mathcal{T}_{1\ell}} &= (\mathbf{e}_{\ell(N_0+1)}, \mathbf{e}_{\ell(N_0+2)}, \ldots, \mathbf{e}_{\ell(N)}), \quad \ell(h) \in \mathcal{T}_{1\ell}, \; h \in \{N_0+1, \ldots, N\} \nonumber
\end{align}
Each elementary vector $\mathbf{e}_{\ell(h)}$ is an $N \times 1$ column vector containing $(N-1)$ zeros and a single 1 in the $\ell(h)$th position.}

\tr{Using \eqref{eqn:E_matrix}, the following matrix expressions can be defined:
\begin{eqnarray}
\mathbf{r}_{\mathcal{T}_{0\ell}}(\lambda) 
&=& [\mathbf{E}_{\mathcal{T}_{0\ell}}]^\top [\mathbf{I}_{N} - \mathbf{H}(\lambda)] \mathbf{y}  
\label{eqn:r_vector}\\
\mathbf{H}_{\mathcal{T}_{0\ell}}(\lambda)
&=& [\mathbf{E}_{\mathcal{T}_{0\ell}}]^\top \mathbf{H}(\lambda)[\mathbf{E}_{\mathcal{T}_{0\ell}}]
\label{eqn:H_matrix_test}\\
\mathbf{I}_{N_0} - \mathbf{H}_{\mathcal{T}_{0\ell}}(\lambda) 
&=& [\mathbf{E}_{\mathcal{T}_{0\ell}}]^\top [\mathbf{I}_{N} - \mathbf{H}(\lambda)][\mathbf{E}_{\mathcal{T}_{0\ell}}] = [\mathbf{R}_{\mathcal{T}_{0\ell}}]^\top [\mathbf{E}_{\mathcal{T}_{0\ell}}].
\label{eqn:IH_matrix_test}
\end{eqnarray}
Here, $\mathbf{r}_{\mathcal{T}_{0\ell}}(\lambda)$ in \eqref{eqn:r_vector} is the the $N_0 \times 1$ subvector of residuals indexed by $\mathcal{T}_{0\ell}$, and $\mathbf{H}_{\mathcal{T}_{0\ell}}(\lambda)$ in \eqref{eqn:H_matrix_test} is the $N_0 \times N_0$ symmetric matrix whose diagonal elements correspond to those of $\mathbf{H}(\lambda)$ with indices in $\mathcal{T}_{0\ell}$, and whose off-diagonal elements correspond to the $N_0(N_0-1)/2$ off-diagonal elements of $\mathbf{H}(\lambda)$ with indices forming pairwise configurations of $\mathcal{T}_{0\ell}$.  We assume that $\mathbf{I}_{N_0} - \mathbf{H}_{\mathcal{T}_{0\ell}}(\lambda)$ in \eqref{eqn:IH_matrix_test} is invertible.}

\tr{Furthermore, define $\mathbf{q}_{\mathcal{T}_{0\ell}}(\lambda) = \mathbf{y}_{\mathcal{T}_{0\ell}} - \mathbf{X}_{\mathcal{T}_{0\ell}}\widehat{\boldsymbol{\beta}}_{\mathcal{T}_{1\ell}}(\lambda)$ as the $N_0 \times 1$ vector of weighted residuals where the main result of Lemma 1 is based.  More rigorously, this can be expressed as
\begin{equation}
\label{eqn:q_vector}
    \mathbf{q}_{\mathcal{T}_{0\ell}}(\lambda) = [\mathbf{E}_{\mathcal{T}_{0\ell}}]^\top [\mathbf{I}_{N}  - \mathbf{M}_{\mathcal{T}_{1\ell}}(\lambda)] \mathbf{y} = \left[\mathbf{I}_{N_0} - \mathbf{H}_{\mathcal{T}_{0\ell}}(\lambda)\right]^{-1}[\mathbf{U}_{\mathcal{T}_{0\ell}}(\lambda)]^\top\mathbf{y}
\end{equation}
where $\mathbf{M}_{\mathcal{T}_{1\ell}}(\lambda) = \mathbf{X}(\mathbf{X}^\top[\mathbf{E}_{\mathcal{T}_{1\ell}}][\mathbf{E}_{\mathcal{T}_{1\ell}}]^\top \mathbf{X} + \lambda \mathbf{I}_{P})^{\dagger} \mathbf{X}^\top[\mathbf{E}_{\mathcal{T}_{1\ell}}][\mathbf{E}_{\mathcal{T}_{1\ell}}]^\top$ and
\begin{equation*}
    \mathbf{U}_{\mathcal{T}_{0\ell}}(\lambda) 
    = [\mathbf{I}_{N}  - \mathbf{M}_{\mathcal{T}_{1\ell}}(\lambda)]^\top [\mathbf{E}_{\mathcal{T}_{0\ell}}] [\mathbf{E}_{\mathcal{T}_{0\ell}}]^\top [\mathbf{I}_{N} - \mathbf{H}(\lambda)] [\mathbf{E}_{\mathcal{T}_{0\ell}}].
\end{equation*}}

\tr{To simplify $\mathbf{U}_{\mathcal{T}_{0\ell}}(\lambda)$, we examine the block matrix representations of $[\mathbf{I}_{N}  - \mathbf{M}_{\mathcal{T}_{1\ell}}(\lambda)]^\top$, $[\mathbf{E}_{\mathcal{T}_{0\ell}}] [\mathbf{E}_{\mathcal{T}_{0\ell}}]^\top$ and $[\mathbf{I}_{N} - \mathbf{H}(\lambda)] [\mathbf{E}_{\mathcal{T}_{0\ell}}]$.  First, $[\mathbf{I}_{N}  - \mathbf{M}_{\mathcal{T}_{1\ell}}(\lambda)]^\top$ simplifies to
\begin{eqnarray}
\label{eqn:IminusM1}
    \begin{pmatrix}
        \mathbf{I}_{N_0}  &  \underset{N_0 \times N_1}{\mathbf{0}} \\
        - \mathbf{X}_{\mathcal{T}_{1\ell}} (\mathbf{X}^\top_{\mathcal{T}_{1\ell}} \mathbf{X}_{\mathcal{T}_{1\ell}} + \lambda \mathbf{I}_{P})^{\dagger}\mathbf{X}^\top_{\mathcal{T}_{0\ell}}
        &  \mathbf{I}_{N_1} - \mathbf{X}_{\mathcal{T}_{1\ell}} (\mathbf{X}^\top_{\mathcal{T}_{1\ell}} \mathbf{X}_{\mathcal{T}_{1\ell}} + \lambda \mathbf{I}_{P})^{\dagger}\mathbf{X}^\top_{\mathcal{T}_{1\ell}} \\
    \end{pmatrix}.
\end{eqnarray}
Second, multiplying $[\mathbf{E}_{\mathcal{T}_{0\ell}}] [\mathbf{E}_{\mathcal{T}_{0\ell}}]^\top$ and $[\mathbf{I}_{N} - \mathbf{H}(\lambda)] [\mathbf{E}_{\mathcal{T}_{0\ell}}]$ produces 
\begin{equation}
\label{eqn:EEIminusHE}
    \begin{pmatrix}
        \mathbf{I}_{N_0} & \mathbf{0} \\
        \mathbf{0} & \mathbf{0} \\
    \end{pmatrix}
    \begin{pmatrix}
        [\mathbf{E}_{\mathcal{T}_{0\ell}}]^\top [\mathbf{I}_{N} - \mathbf{H}(\lambda)][\mathbf{E}_{\mathcal{T}_{0\ell}}]\\
        - \mathbf{E}^\top_{\mathcal{T}_{1\ell}} \mathbf{H}(\lambda)\mathbf{E}_{\mathcal{T}_{0\ell}}
    \end{pmatrix} = 
    \begin{pmatrix}
        \mathbf{I}_{N_0} - \mathbf{H}_{\mathcal{T}_{0\ell}}(\lambda) \\
        \underset{N_1 \times N_0}{\mathbf{0}}
    \end{pmatrix}.
\end{equation}
Finally, by multiplying \eqref{eqn:IminusM1} and \eqref{eqn:EEIminusHE}, we obtain $\mathbf{U}_{\mathcal{T}_{0\ell}}(\lambda)$ as follows 
\begin{eqnarray}
    \underset{N \times N_0}{\mathbf{U}_{\mathcal{T}_{0\ell}}(\lambda)}
    &=& 
    \label{eqn:U_matrix}
    \begin{pmatrix}
        \mathbf{I}_{N_0} - \mathbf{H}_{\mathcal{T}_{0\ell}}\\
        - \mathbf{X}_{\mathcal{T}_{1\ell}} (\mathbf{X}^\top_{\mathcal{T}_{1\ell}} \mathbf{X}_{\mathcal{T}_{1\ell}} + \lambda \mathbf{I}_{P})^{\dagger}\mathbf{X}^\top_{\mathcal{T}_{0\ell}}[\mathbf{I}_{N_0} - \mathbf{H}_{\mathcal{T}_{0\ell}}]
    \end{pmatrix}.
\end{eqnarray}}

\tr{To simplify \eqref{eqn:U_matrix} further, consider $\mathbf{I}_P = (\mathbf{X}_{\mathcal{T}_{0\ell}}^\top\mathbf{X}_{\mathcal{T}_{0\ell}} + \mathbf{X}_{\mathcal{T}_{1\ell}}^\top\mathbf{X}_{\mathcal{T}_{1\ell}} + \lambda \mathbf{I}_{P})(\mathbf{X}^\top\mathbf{X} + \lambda \mathbf{I}_{P})^{\dagger}$.  Exploiting this property of $\mathbf{I}_P$ leads to the equivalent expressions $\mathbf{I}_P - (\mathbf{X}_{\mathcal{T}_{0\ell}}^\top\mathbf{X}_{\mathcal{T}_{0\ell}})(\mathbf{X}^\top\mathbf{X} + \lambda \mathbf{I}_{P})^{\dagger}$ and $(\mathbf{X}_{\mathcal{T}_{1\ell}}^\top\mathbf{X}_{\mathcal{T}_{1\ell}} + \lambda \mathbf{I}_{P})(\mathbf{X}^\top\mathbf{X} + \lambda \mathbf{I}_{P})^{\dagger}$.  Expanding $\mathbf{X}_{\mathcal{T}_{1\ell}} (\mathbf{X}^\top_{\mathcal{T}_{1\ell}} \mathbf{X}_{\mathcal{T}_{1\ell}} + \lambda \mathbf{I}_{P})^{\dagger}\mathbf{X}^\top_{\mathcal{T}_{0\ell}}[\mathbf{I}_{N_0} - \mathbf{H}_{\mathcal{T}_{0\ell}}]$ in \eqref{eqn:U_matrix} then yields $\mathbf{E}^\top_{\mathcal{T}_{1\ell}} \mathbf{H}(\lambda)\mathbf{E}_{\mathcal{T}_{0\ell}}$.  Therefore,
\begin{equation}
\label{eqn:U_matrix_Simplification}
    \underset{N \times N_0}{\mathbf{U}_{\mathcal{T}_{0\ell}}(\lambda)} = 
    \begin{pmatrix}
        [\mathbf{E}_{\mathcal{T}_{0\ell}}]^\top [\mathbf{I}_{N} - \mathbf{H}(\lambda)][\mathbf{E}_{\mathcal{T}_{0\ell}}]\\
        - \mathbf{E}^\top_{\mathcal{T}_{1\ell}} \mathbf{H}(\lambda)\mathbf{E}_{\mathcal{T}_{0\ell}}
    \end{pmatrix}
    = [\mathbf{I}_{N} - \mathbf{H}(\lambda)]^\top [\mathbf{E}_{\mathcal{T}_{0\ell}}].
\end{equation}}

\tr{From Lemma 1, the $\ell$th term in L$N_0$OCV$^{(1)}(\lambda)$ is $[\mathbf{q}_{\mathcal{T}_{0\ell}}(\lambda)]^\top [\mathbf{q}_{\mathcal{T}_{0\ell}}(\lambda)]$.  Combining \eqref{eqn:q_vector} and \eqref{eqn:U_matrix_Simplification} yields
\begin{eqnarray*}
    \mathbf{q}^\top_{\mathcal{T}_{0\ell}}(\lambda) \mathbf{q}_{\mathcal{T}_{0\ell}}(\lambda) 
    &=& \mathbf{y}^\top [\mathbf{I}_{N}  - \mathbf{M}_{\mathcal{T}_{1\ell}}(\lambda)]^\top [\mathbf{E}_{\mathcal{T}_{0\ell}}][\mathbf{E}_{\mathcal{T}_{0\ell}}]^\top [\mathbf{I}_{N}  - \mathbf{M}_{\mathcal{T}_{1\ell}}(\lambda)] \mathbf{y} \\
    &=& \mathbf{y}^\top [\mathbf{U}_{\mathcal{T}_{0\ell}}(\lambda)] \left[\mathbf{I}_{N_0} - \mathbf{H}_{\mathcal{T}_{0\ell}}(\lambda)\right]^{-2} [\mathbf{U}_{\mathcal{T}_{0\ell}}(\lambda)]^\top\mathbf{y} \\
    &=& \mathbf{y}^\top [\mathbf{I}_{N} - \mathbf{H}(\lambda)]^\top [\mathbf{E}_{\mathcal{T}_{0\ell}}] \left[\mathbf{I}_{N_0} - \mathbf{H}_{\mathcal{T}_{0\ell}}(\lambda)\right]^{-2} [\mathbf{E}_{\mathcal{T}_{0\ell}}]^\top[\mathbf{I}_{N} - \mathbf{H}(\lambda)]\mathbf{y} \\
    &=& \mathbf{y}^\top \mathbf{W}_{\mathcal{T}_{0\ell}}(\lambda) \mathbf{y}
\end{eqnarray*}
where $\mathbf{W}_{\mathcal{T}_{0\ell}}(\lambda) = [\mathbf{I}_{N} - \mathbf{H}(\lambda)]^\top [\mathbf{E}_{\mathcal{T}_{0\ell}}] \left[\mathbf{I}_{N_0} - \mathbf{H}_{\mathcal{T}_{0\ell}}(\lambda)\right]^{-2} [\mathbf{E}_{\mathcal{T}_{0\ell}}]^\top[\mathbf{I}_{N} - \mathbf{H}(\lambda)]$ is an $N \times N$ matrix of weights which is a function of the design matrix $\mathbf{X}$, the testing set indices $\mathcal{T}_{0\ell}$, and the regularization parameter $\lambda \in \boldsymbol{\Lambda}$.  Equivalently, this can be re-expressed as 
\begin{equation*}
    \mathbf{q}^\top_{\mathcal{T}_{0\ell}}(\lambda) \mathbf{q}_{\mathcal{T}_{0\ell}}(\lambda) 
    = \mathbf{r}^\top [\mathbf{E}_{\mathcal{T}_{0\ell}}] \left[\mathbf{I}_{N_0} - \mathbf{H}_{\mathcal{T}_{0\ell}}(\lambda)\right]^{-2} [\mathbf{E}_{\mathcal{T}_{0\ell}}]^\top\mathbf{r} = \mathbf{r}^\top_{\mathcal{T}_{0\ell}} \left[\mathbf{I}_{N_0} - \mathbf{H}_{\mathcal{T}_{0\ell}}(\lambda)\right]^{-2} \mathbf{r}_{\mathcal{T}_{0\ell}}. \hspace{2em} \blacksquare
\end{equation*}}

% -------------------------------------------------
\section{Proof of Corollary 1:}
\label{sec:ProofCorollary1}
% -------------------------------------------------

\tr{Define the centered responses as $\widetilde{Y}_{n} = Y_{n} - \overline{Y}$ for $n \in [N]$, or equivalently $\widetilde{Y}_{\ell(h)} = Y_{\ell(h)} - \overline{Y}$ for $\ell(h) \in [N]$.  To compute L$N_0$OCV$^{(0)}$ in Corollary 1, we need the average squared difference between observed responses and the training sample mean. For the $\ell$th partition, this is
\begin{equation}
    \frac{1}{N_0}\sum\limits_{h = 1}^{N_0} (Y_{\ell(h)} - \overline{Y}_{\mathcal{T}_{1\ell}})^2
    = \frac{1}{N_0}\sum\limits_{h = 1}^{N_0} \widetilde{Y}_{\ell(h)}^2 + \frac{N}{N_0N_1}\left(\frac{1}{N_1} + \frac{1}{N}\right)\left(\sum\limits_{h = 1}^{N_0} \widetilde{Y}_{\ell(h)} \right)^2.
    \label{eqn:Null_PEMean_test}
\end{equation}
The L$N_0$OCV$^{(0)}$ estimator then averages \eqref{eqn:Null_PEMean_test} over all $L$ data partitions:
\begin{equation*}
    \text{L}N_0\text{OCV}^{(0)} = 
    \frac{1}{N_0}\dbinom{N}{N_0}^{-1} \underbrace{\sum\limits_{\ell = 1}^L \sum\limits_{h = 1}^{N_0} \widetilde{Y}_{\ell(h)}^2}_{\mathcal{S}_1} + \frac{1}{N_0}\dbinom{N - 1}{N_0}^{-1}\left(\frac{1}{N_1} + \frac{1}{N}\right)\underbrace{\sum\limits_{\ell = 1}^L \left(\sum\limits_{h = 1}^{N_0} \widetilde{Y}_{\ell(h)} \right)^2}_{\mathcal{S}_2}
\end{equation*}}

\tr{Second, we must count both squared terms $\widetilde{Y}_n^2$ and cross-product terms $\widetilde{Y}_m\widetilde{Y}_n$ where $m \neq n$ for the square of sums $\mathcal{S}_2$. Each squared term $\widetilde{Y}_n^2$ appears $\binom{N-1}{N_0-1}$ times, as before. However, each cross-product $\widetilde{Y}_m\widetilde{Y}_n$ appears in exactly $\binom{N-2}{N_0-2}$ partitions—those where both sampling units $m$ and $n$ are in the test set.  This counting allows us to decompose $\mathcal{S}_2$ as
\begin{equation}
\label{eqn:S2}
\mathcal{S}_2 = \left[\binom{N-1}{N_0-1} - \binom{N-2}{N_0-2}\right] \sum_{n=1}^N \widetilde{Y}_n^2 + \binom{N-2}{N_0-2} \left[\sum_{n=1}^N \widetilde{Y}_n^2 + \sum_{m \neq n} \widetilde{Y}_m\widetilde{Y}_n\right].
\end{equation}
The bracketed second expression in \eqref{eqn:S2} equals $\left(\sum\limits_{n=1}^N \widetilde{Y}_n\right)^2 = 0$ since the centered responses sum to zero. Therefore,
\begin{equation}
\mathcal{S}_2 = \left[\binom{N-1}{N_0-1} - \binom{N-2}{N_0-2}\right] \sum_{n=1}^N \widetilde{Y}_n^2 = \binom{N-2}{N_0-1} \sum_{n=1}^N \widetilde{Y}_n^2,
\end{equation}
where the final equality uses the binomial identity $\binom{N-1}{N_0-1} - \binom{N-2}{N_0-2} = \binom{N-2}{N_0-1}$.}

\tr{Finally, we can express the L$N_0$OCV$^{(0)}$ estimator in terms of the sample variance $S_Y^2$ of the observed response vector. Using the results from our counting analysis, we have
\begin{align*}
    \text{L}N_0\text{OCV}^{(0)} 
    &= \frac{1}{N_0}\binom{N}{N_0}^{-1} \binom{N-1}{N_0-1} \sum_{n=1}^N \widetilde{Y}_n^2 + \frac{1}{N_0}\binom{N - 1}{N_0}^{-1}\left(\frac{1}{N_1} + \frac{1}{N}\right)\binom{N-2}{N_0-1} \sum_{n=1}^N \widetilde{Y}_n^2 \\
    &= \left[\frac{1}{N} + \frac{1}{N-1}\left(\frac{1}{N_1} + \frac{1}{N}\right)\right] \sum_{n = 1}^N (Y_n - \overline{Y})^2 = \left(1 + \frac{1}{N - N_0}\right) S_Y^2. \hspace{3em} \blacksquare
\end{align*}}

% -------------------------------------------------
\section{Proof of Corollary 2:}
\label{sec:ProofCorollary2}
% -------------------------------------------------

\tr{From the decomposition discussed in Section \ref{sec:AsymptoticResultsLN0OCVEstimator}, we have $\text{LN}_0\text{OCV}(\mathcal{M}_1)$
\begin{equation*}
     \underbrace{\frac{1}{N} \sum\limits_{n = 1}^{N} (Y_{n} - f(\mathbf{X}_{n}))^2}_{\mathcal{C}_1} + \underbrace{\frac{1}{L N_0} \sum\limits_{\ell = 1}^{L}\sum\limits_{h = 1}^{N_0} (f(\mathbf{X}_{\ell(h)}) - \widehat{f}(\mathbf{X}_{\ell(h)}, \widehat{\boldsymbol{\beta}}_{\mathcal{T}_{1\ell}}))^2}_{\mathcal{C}_2(\mathcal{M}_1)} + \frac{2}{L N_0} \sum\limits_{\ell = 1}^{L}\sum\limits_{h = 1}^{N_0} \mathcal{C}_{3\ell(h)}(\mathcal{M}_1)
\end{equation*}
where $\mathcal{C}_{3\ell(h)}(\mathcal{M}_1) = (Y_{\ell(h)} - f(\mathbf{X}_{\ell(h)}))(f(\mathbf{X}_{\ell(h)}) - \widehat{f}(\mathbf{X}_{\ell(h)}, \widehat{\boldsymbol{\beta}}_{\mathcal{T}_{1\ell}}))$.  To establish the asymptotic normality of $\text{LN}_0\text{OCV}(\mathcal{M}_1)$, we analyze each component separately and show that the $\mathcal{M}_1$-independent term $\mathcal{C}_1$ determines the limiting distribution, while the $\mathcal{M}_1$-dependent terms $\mathcal{C}_2(\mathcal{M}_1)$ and $\mathcal{C}_3(\mathcal{M}_1)$ vanish in probability at rate $\sqrt{N}$.}

% -------------------------------------------------
\subsection[Analysis of the Irreducible Error Term C1]{Analysis of the Irreducible Error Term \(\mathcal{C}_1\)}
\label{sec:ProofCorollary2_C1}
% -------------------------------------------------

\tr{Let $\text{Err}^{(1)} = \mathbb{E}[(Y_n - f(\mathbf{X}_n))^2]$ and $\sigma^2_{\mathcal{C}_1} = \mathbb{V}[(Y_n - f(\mathbf{X}_n))^2]$ denote the true prediction error and its variance, respectively. The term $\mathcal{C}_1$ represents the empirical average of the irreducible errors and is independent of the model-fitting algorithm $\mathcal{M}_1$. By the independence of observations $\{(\mathbf{X}_n, Y_n)\}_{n = 1}^N$,
\begin{equation*}
    \mathbb{E}(\mathcal{C}_1) = \frac{1}{N} \sum\limits_{n = 1}^{N} \mathbb{E}[(Y_{n} - f(\mathbf{X}_{n}))^2] = \text{Err}^{(1)} ~\text{and}~
    \mathbb{V}(\mathcal{C}_1) = \frac{1}{N^2} \sum\limits_{n = 1}^{N} \mathbb{V}[(Y_{n} - f(\mathbf{X}_{n}))^2] = \frac{\sigma^2_{\mathcal{C}_1}}{N}.
\end{equation*}
Since assumption A.1 ensures $\mathbb{E}(Y_n^4) < \infty$ and $\mathbb{V}(Y_n \mid \mathbf{X}_{n}) \leq \boldsymbol{\Sigma}$ with $\left\|\boldsymbol{\Sigma}\right\|_\infty < \infty$, the sequence ${(Y_n - f(\mathbf{X}_{n}))^2}$ satisfies the conditions for the Central Limit Theorem. Therefore, $\sqrt{N}(\mathcal{C}_1 - \text{Err}^{(1)}) \xrightarrow{d} \mathcal{N}(0, \sigma^2_{\mathcal{C}_1})$ as $N \to \infty$.}

% -------------------------------------------------
\subsection[Analysis of the Excess Error Term C2]{Analysis of the Excess Error Term \(\mathcal{C}_2(\mathcal{M}_1)\)}
\label{sec:ProofCorollary2_C2}
% -------------------------------------------------

\tr{The excess error term $\mathcal{C}_2(\mathcal{M}_1)$ measures how well the fitted predictor $\widehat{f}$ approximates the true function $f$, representing the additional error incurred by using the estimated function instead of the true one. Assumption A.2 specifies the rate at which this estimator improves as training data increases. In particular, it provides bounds on 
\begin{equation*}
    N_1^\delta \mathbb{E}\left[ \left(\widehat{f}(\mathbf{X}_{\ell(h)}, \widehat{\boldsymbol{\beta}}_{\mathcal{T}_{1\ell}}) - f(\mathbf{X}_{\ell(h)})\right)^2 \mid \{\mathbf{X}_{\ell(h)}, Y_{\ell(h)}\}_{h = N_0 + 1}^N\right]^{1/2},
\end{equation*}
indicating that the root MSE of the predictor scales like $N_1^{-\delta}$ where $0.25 < \delta < 0.5$. Squaring both sides yields the mean squared error rate: 
\begin{equation*}
    \mathbb{E}\left[ \left(\widehat{f}(\mathbf{X}_{\ell(h)}, \widehat{\boldsymbol{\beta}}_{\mathcal{T}_{1\ell}}) - f(\mathbf{X}_{\ell(h)})\right)^2 \mid \{\mathbf{X}_{\ell(h)}, Y_{\ell(h)}\}_{h = N_0 + 1}^N\right] = \mathcal{O}(N_1^{-2\delta}).
\end{equation*}
Since $\mathcal{C}_2(\mathcal{M}_1)$ is an average of these squared errors over all partitions and test points, each term has expectation $\mathcal{O}(N_1^{-2\delta})$, giving $\mathbb{E}[\mathcal{C}_2(\mathcal{M}_1)] = \mathcal{O}(N_1^{-2\delta})$. With $N_1 = N - N_0$ and $N_0/N \to 0$, we have $N_1 \asymp N$, so $\mathbb{E}[\mathcal{C}_2(\mathcal{M}_1)] = \mathcal{O}(N^{-2\delta})$. The essential point is that the estimator improves fast enough that when scaled by $\sqrt{N}$, the excess error vanishes: since $\delta > 0.25$ implies $2\delta > 0.5$, we have $\mathbb{E}[\sqrt{N} \mathcal{C}_2(\mathcal{M}_1)] = \mathcal{O}(N^{0.5-2\delta})$ where $0.5 - 2\delta < 0$. Applying Markov's inequality, $\sqrt{N} \mathcal{C}_2(\mathcal{M}_1) \xrightarrow{p} 0$ as $N \to \infty$.}

% -------------------------------------------------
\subsection[Analysis of the Cross-Term C3]{Analysis of the Cross-Term \(\mathcal{C}_3(\mathcal{M}_1)\)}
\label{sec:ProofCorollary2_C3}
% -------------------------------------------------

\tr{The cross-term $\mathcal{C}_3(\mathcal{M}_1)$ represents the interaction between the noise and estimation error. The conditional expectation property yields $\mathbb{E}[Y_{\ell(h)} - f(\mathbf{X}_{\ell(h)}) \mid \mathbf{X}_{\ell(h)}, \widehat{f}(\mathbf{X}_{\ell(h)}, \widehat{\boldsymbol{\beta}}_{\mathcal{T}_{1\ell}}))] = 0$ for two reasons: first, the noise term $Y_{\ell(h)} - f(\mathbf{X}_{\ell(h)})$ has mean zero given $\mathbf{X}_{\ell(h)}$ by the model assumptions, and second, the estimation error $\widehat{f}(\mathbf{X}_{\ell(h)}, \widehat{\boldsymbol{\beta}}_{\mathcal{T}_{1\ell}})$ depends only on training data and is independent of the test observation $Y_{\ell(h)}$. This uncorrelatedness between noise and estimation error implies $\mathbb{E}[\mathcal{C}_3(\mathcal{M}_1) \mid \widehat{f}~] = 0$ across all partitions.  To establish the rate of convergence, we bound the variance of $\mathcal{C}_3(\mathcal{M}_1)$. For each partition $\ell$, define 
\begin{equation*}
    \mathcal{C}_{3\ell} = \sum\limits_{h = 1}^{N_0} \mathcal{C}_{3\ell(h)} = \sum\limits_{h = 1}^{N_0} (Y_{\ell(h)} - f(\mathbf{X}_{\ell(h)}))(f(\mathbf{X}_{\ell(h)}) - \widehat{f}(\mathbf{X}_{\ell(h)}, \widehat{\boldsymbol{\beta}}_{\mathcal{T}_{1\ell}})).
\end{equation*}}

\tr{Applying the Cauchy-Schwarz inequality to bound the variance 
\begin{equation*}
    \mathbb{E} \left[ \left.\mathcal{C}_{3\ell}^2 \right\vert \widehat{f}~\right] \leq N_0\sum\limits_{h = 1}^{N_0} \mathbb{E} \left[(Y_{\ell(h)} - f(\mathbf{X}_{\ell(h)}))^2 \right] \mathbb{E} \left[\left. (f(\mathbf{X}_{\ell(h)}) - \widehat{f}(\mathbf{X}_{\ell(h)}, \widehat{\boldsymbol{\beta}}_{\mathcal{T}_{1\ell}}))^2 \right\vert \widehat{f}~\right]
\end{equation*}
where the $N_0$ factor arises from the cross-products in the expansion of $(\mathcal{C}_{3\ell(1)} + \cdots + \mathcal{C}_{3\ell(N_0)})^2$.  Assumption A.1 ensures $\mathbb{E}[(Y_{\ell(h)} - f(\mathbf{X}_{\ell(h)}))^2] \leq \boldsymbol{\Sigma}$ since $\mathbb{V}(Y_n \mid \mathbf{X}_n) \leq \boldsymbol{\Sigma}$, while Assumption A.2 provides the bound $\mathbb{E}[(f(\mathbf{X}_{\ell(h)}) - \widehat{f}(\mathbf{X}_{\ell(h)}, \widehat{\boldsymbol{\beta}}_{\mathcal{T}_{1\ell}}))^2 \mid \widehat{f}~] \leq \frac{c_{+}^2}{N_1^{2\delta}}$ from the excess risk scaling. Substituting these bounds yields
\begin{equation*}
    \mathbb{E} \left[ \left.\mathcal{C}_{3\ell}^2 \right\vert \widehat{f}~\right] \leq \frac{N_0^2 \boldsymbol{\Sigma} c_+^2}{N_1^{2\delta}} ~\text{and}~\mathbb{V}(\mathcal{C}_3(\mathcal{M}_1)) \leq \frac{N_0^2 \boldsymbol{\Sigma} c_{+}^2}{L^2N_0^2 N_1^{2\delta}} = \frac{\boldsymbol{\Sigma} c_{+}^2}{L N_1^{2\delta}}.
\end{equation*}
Crucially, $L = \binom{N}{N_0}$ grows exponentially when $N_0/N \to 0$, while $N_1^{2\delta}$ grows only polynomially. This exponential-polynomial gap ensures that $\mathbb{V}(\mathcal{C}_3(\mathcal{M}_1)) = o(N^{-1})$, and hence $\mathcal{C}_3(\mathcal{M}_1) = o_p(N^{-0.5})$.}

\tr{To establish the final result, we apply Slutsky's theorem to combine the behavior of all three components. Since $\mathcal{C}_1$ converges in distribution to a normal random variable while $\mathcal{C}_2(\mathcal{M}_1)$ and $\mathcal{C}_3(\mathcal{M}_1)$ both vanish in probability when scaled by $\sqrt{N}$, it follows that
\begin{equation*}
\label{eqn:LN0OCV_Distribution}
    \sqrt{N}(\text{L}N_0\text{OCV}(\mathcal{M}_1) - \text{Err}^{(1)}) \overset{d}{\rightarrow} \mathcal{N}(0, \sigma^2_{\mathcal{C}_1}), ~\text{as} ~N \rightarrow \infty.
\end{equation*}}

\begin{figure}[ht]
    \begin{center}
        \begin{tabular}{c}
        \includegraphics[width=0.8\linewidth]{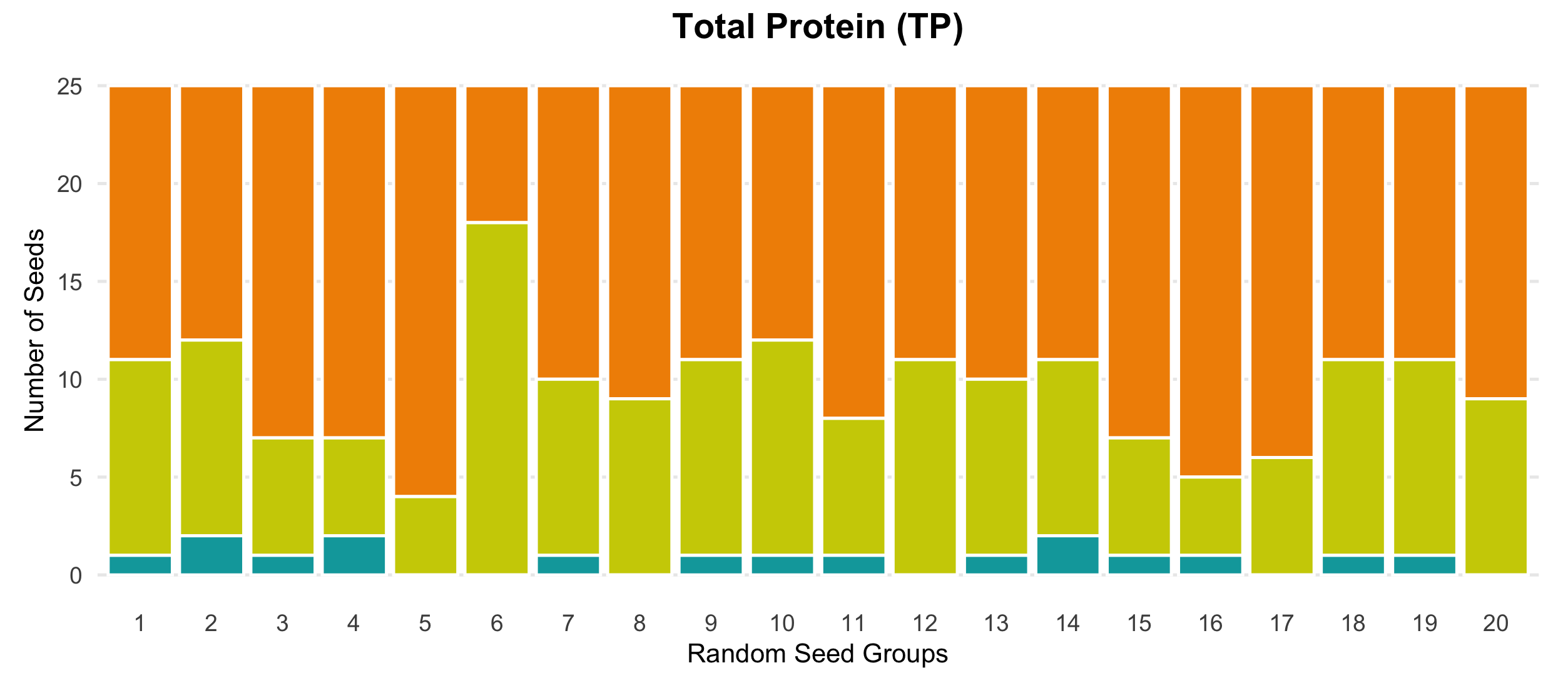}\\
        \includegraphics[width=0.8\linewidth]{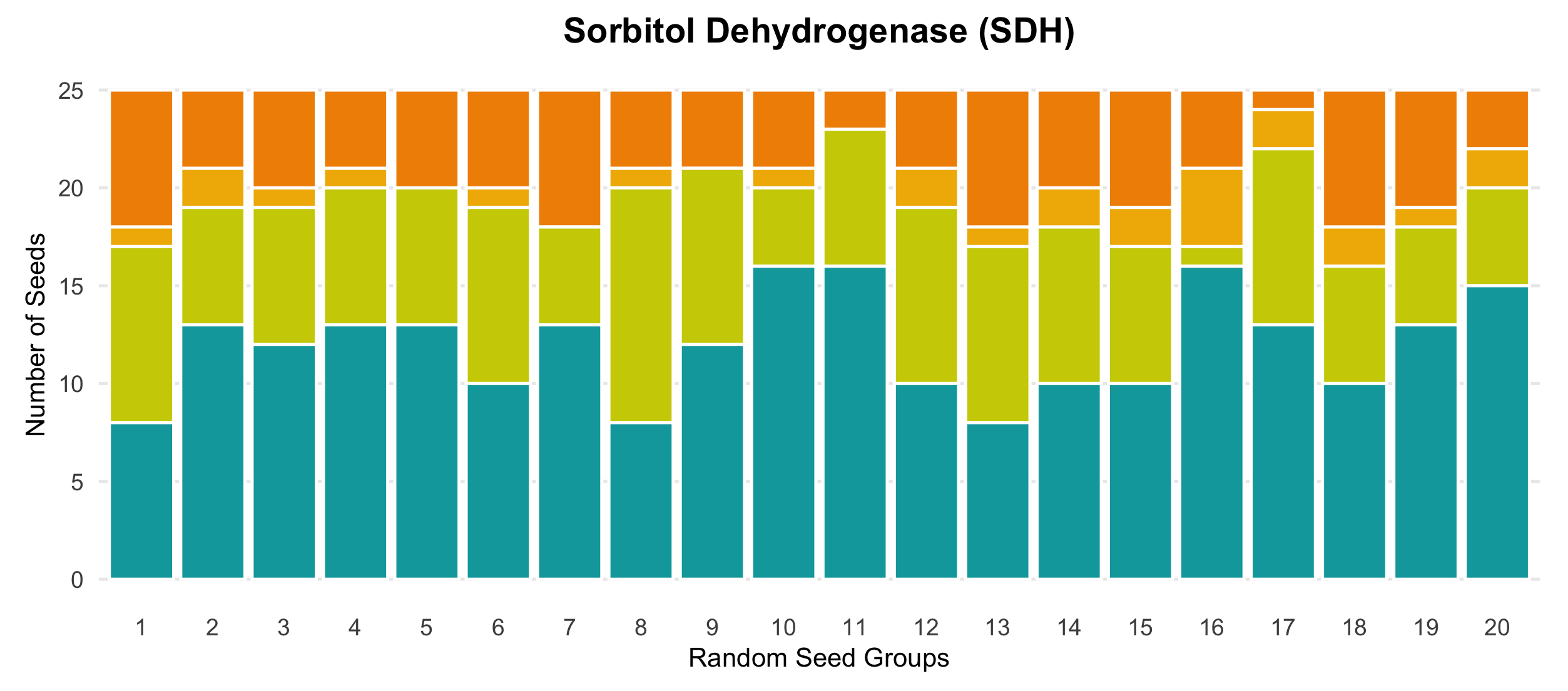}\\
        \includegraphics[width=0.7\linewidth]{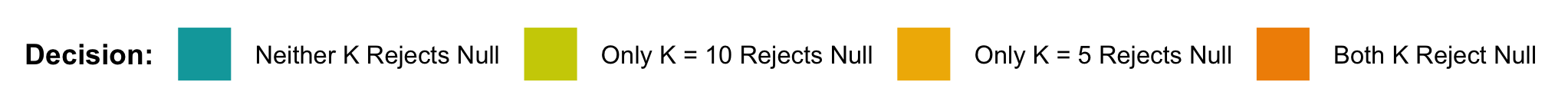}\\
    \end{tabular}
    \end{center}
    \caption{\tr{Reproducibility issues in $K$-fold CV hypothesis testing for ($i$) Total Protein (TP) and ($ii$) Sorbitol Dehydrogenase (SDH). Results from 500 random seeds (8892-9391) grouped into 20 consecutive groups of 25 seeds each. Stacked bars show statistical decision distributions within each group.}}
    \label{fig:LiverToxicityExamples}
\end{figure}

% -------------------------------------------------
\section[Test based on K-fold Cross-Validation]{Test based on \(K\)-fold Cross-Validation}
\label{sec:KFoldCVTest}
% -------------------------------------------------

\tr{Using the liver toxicity data by \citet{bushel2007simultaneous}, we present the detailed methodology for the $K$-fold cross-validation hypothesis test to evaluate whether the 3116 gene expression measurements in $\mathbf{X}$ provide substantial improvement in predicting clinical measurements $\mathbf{y}$ compared to using only the intercept model.}

% -------------------------------------------------
\subsection{Test Statistic Calculation}
\label{sec:KfoldTestStatistic}
% -------------------------------------------------

\tr{For each fold $k \in [K]$, we partition the data into training and testing sets. For every observation in the $k$th testing set, we calculate the prediction errors under both models:
\begin{equation}
\label{eqn:Kfold_TE}
    C_k^{(0)} = \frac{1}{N_k}\sum_{h=1}^{N_k} (Y_{k(h)} - \overline{Y}_{-k})^2 \quad \text{and} \quad C_k^{(1)} = \frac{1}{N_k}\sum_{h=1}^{N_k} (Y_{k(h)} - \mathbf{X}^\top_{k(h)}\widehat{\boldsymbol{\beta}}_{-k}(\widehat{\lambda}))^2
\end{equation}
where $\overline{Y}_{-k}$ is the mean of the training data excluding fold $k$, and $\widehat{\boldsymbol{\beta}}_{-k}(\widehat{\lambda})$ represents the regularized regression coefficients estimated from the training data. This procedure is repeated until all folds serve as testing data.  The paired differences $C_k = C_k^{(0)} - C_k^{(1)}$ capture the improvement in prediction error when using the full model versus the intercept-only model for each fold. The test statistic for the $s$th random seed is then computed as
\begin{equation}
\label{eqn:KfoldCVTestStat}
    T_{\text{KCV}, s} = \displaystyle \frac{\overline{C}}{\sqrt{\displaystyle\frac{S^2_K}{K}}} ~~\text{where}~~\overline{C} = \frac{1}{K}\sum\limits_{k = 1}^K C_k ~~\text{and}~~S^2_K = \frac{1}{K - 1} \sum \limits_{k = 1}^K \left(C_k - \overline{C}\right)^2.
\end{equation}
Under the null hypothesis of no improvement, $T_{\text{KCV}, s}$ follows approximately a $t$-distribution with $K-1$ degrees of freedom. The null hypothesis is rejected if $T_{\text{KCV}, s} > t_{K-1}(\alpha)$, where $t_{K-1}(\alpha)$ is the critical value at significance level $\alpha$.}

% -------------------------------------------------
\subsection{Reproducibility Analysis for the Liver Toxicity Data}
\label{sec:KfoldSystemicReproducibility}
% -------------------------------------------------

\begin{figure}[ht]
    \centering
    \includegraphics[width=0.75\linewidth]{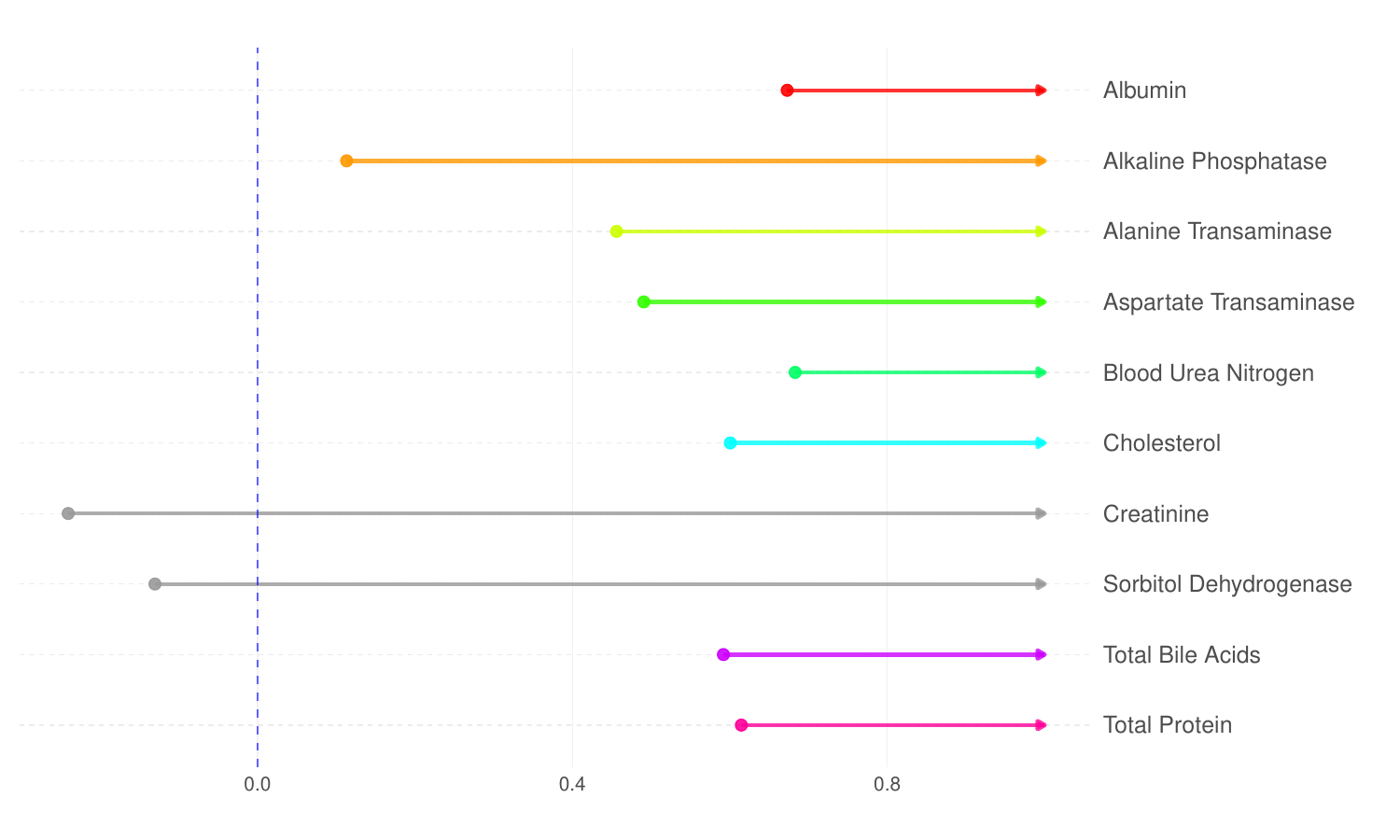}
    \caption{Lower bound confidence interval estimates for prediction error improvement when using gene expression to liver toxicity clinical measurements. Positive values indicate that microarray data improves prediction accuracy over baseline models.}
    \label{fig:SignificantMixOmics}
\end{figure}

\tr{To quantify the instability identified in Section \ref{sec:ReproducibilityCrisis}, we conducted experiments using 500 different random seeds (8892 to 9391) to predict clinical measurements from liver mRNA expression data. We organized these seeds into 20 groups of 25 each and computed test statistics using both $K = 5$ and $K = 10$ fold CV estimators.}

\tr{Figure \ref{fig:LiverToxicityExamples} demonstrates this reproducibility crisis for Total Protein (TP) and Sorbitol Dehydrogenase (SDH). The stacked bars show statistical decision distributions within each group of 25 consecutive seeds, revealing considerable variation in conclusions drawn from identical data using different random partitions. This instability extends systematically across multiple clinical measurements, confirming that the problem represents a fundamental limitation of the $K$-fold CV approach rather than a variable-specific artifact. The results demonstrate that any scientific conclusion drawn from $K$-fold CV hypothesis testing carries substantial risk of irreproducibility, as different random seeds can lead to contradictory statistical decisions even when applied to the same dataset.}

\tr{The fundamental insight of our work is that the reproducibility problem with $K$-fold CV is not a minor implementation detail that can be addressed through careful tuning or averaged away through repeated runs. It represents a systemic failure of the approach for hypothesis testing in high-dimensional settings. Our proposed method, illustrated in Figure \ref{fig:SignificantMixOmics}, addresses this crisis by providing stable, seed-independent inference. Figure \ref{fig:SignificantMixOmics} presents lower bound confidence interval estimates for prediction error improvement when using gene expression data to predict liver toxicity clinical measurements, where positive values indicate that microarray data meaningfully improves prediction accuracy over baseline models. Critically, these intervals represent single, reproducible results for each clinical measurement, in stark contrast to the unstable, seed-dependent conclusions produced by $K$-fold CV shown in Figures \ref{fig:RationaleLiverToxicity} and \ref{fig:LiverToxicityExamples} The consistency of our approach across the ten clinical measurements demonstrates that our method successfully eliminates the reproducibility crisis that plagues standard $K$-fold CV-based inference, providing researchers with reliable, replicable conclusions regardless of arbitrary computational choices.}

% -------------------------------------------------
\bibliographystyle{apalike}
\bibliography{arXiV/references}
% -------------------------------------------------

\end{document}